\documentclass[twocolumn]{aa}  

\usepackage{graphicx}
\usepackage{txfonts}
\usepackage{lscape}
\usepackage{booktabs}
\usepackage{threeparttablex}
\usepackage{wasysym}
\usepackage{mathrsfs}
\usepackage[bookmarks=false,colorlinks=true,linkcolor=black,citecolor=cyan,filecolor=black,urlcolor=cyan]{hyperref}
\usepackage{array}

\newcolumntype{C}[1]{>{\centering\arraybackslash}p{#1}}

\begin{document} 
	
	
	\title{The more the merrier: \emph{SRG}/eROSITA discovers two further galaxies showing X-ray quasi-periodic eruptions}
	
	\author{R. Arcodia\thanks{NASA Einstein fellow}
		\inst{1,2},
        Z. Liu
		\inst{2},
        A. Merloni
		\inst{2},
        A. Malyali
		\inst{2},
        A. Rau
		\inst{2},
        J. Chakraborty
		\inst{1},
        A. Goodwin
		\inst{3},
        D. Buckley
		\inst{4},
        J. Brink
        \inst{4},
        M. Gromadzki
		\inst{5},
        Z. Arzoumanian
		\inst{6},
        J. Buchner
		\inst{2},
        E. Kara
		\inst{1},
        K. Nandra
		\inst{2},
        G. Ponti
		\inst{7,2},
        M. Salvato
		\inst{2},
        G. Anderson
		\inst{3},
        P. Baldini
		\inst{2},
        I. Grotova
		\inst{2},
        M. Krumpe
		\inst{8},
        C. Maitra
		\inst{2},
        J. C. A. Miller-Jones
        \inst{3},
        M. E. Ramos-Ceja
		\inst{2}
	}
	\institute{MIT Kavli Institute for Astrophysics and Space Research, 70 Vassar Street, Cambridge, MA 02139, USA\\
		\email{rarcodia@mit.edu}
		\and
		Max-Planck-Institut f\"ur extraterrestrische Physik (MPE), Gie{\ss}enbachstra{\ss}e 1, 85748 Garching bei M\"unchen, Germany
		\and
		International Centre for Radio Astronomy Research – Curtin University, GPO Box U1987, Perth, WA 6845, Australia
        \and
        South African Astronomical Observatory, P.O. Box 9, Observatory 7935, Cape Town, South Africa
        \and
        Astronomical Observatory, University of Warsaw, Al. Ujazdowskie 4, 00-478 Warszawa, Poland
        \and
        Astrophysics Science Division, NASA Goddard Space Flight Center, 8800 Greenbelt Road, Greenbelt, MD 20771, USA
        \and
        INAF-Osservatorio Astronomico di Brera, Via E. Bianchi 46, I-23807 Merate (LC), Italy
        \and
        Leibniz-Institut f\"ur Astrophysik Potsdam (AIP), An der Sternwarte 16, 14482 Potsdam, Germany
	}
	
	\date{Received ; accepted }
	
	
	\abstract{X-ray quasi-periodic eruptions (QPEs) are a novel addition to the group of extragalactic transients. With only a select number of known sources, and many more models published trying to explain them, we are so far limited in our understanding by small number statistics. In this work, we report the discovery of two further galaxies showing QPEs, hereafter named eRO-QPE3 and eRO-QPE4, with the eROSITA X-ray telescope on board the Spectrum Roentgen Gamma observatory, followed by \emph{XMM-Newton}, \emph{NICER}, \emph{Swift-XRT}, \emph{SALT} ($z=0.024$ and $z=0.044$, respectively), and \emph{ATCA} observations. Among the properties in common with those of known QPEs are: the thermal-like spectral shape in eruption (up to $kT\sim110-120$\,eV) and quiescence ($kT\sim50-90$\,eV) and its evolution during the eruptions (with a harder rise than decay); the lack of strong canonical signatures of active nuclei (from current optical, UV, infrared and radio data); and the low-mass nature of the host galaxies ($\log M_*\approx 9-10$) and their massive central black holes ($\log M_{\rm BH}\approx 5-7$). These discoveries also bring several new insights into the QPE population: i) eRO-QPE3 shows eruptions on top of a decaying quiescence flux, providing further evidence for a connection between QPEs and a preceding tidal disruption event; ii) eRO-QPE3 exhibits the longest recurrence times and faintest peak luminosity of QPEs, compared to the known QPE population, excluding a correlation between the two; iii) we find evidence, for the first time, of a transient component that is harder, albeit much fainter, than the thermal QPE spectrum in eRO-QPE4; and iv) eRO-QPE4 displays the appearance (or significant brightening) of the quiescence disk component after the detection of QPEs, supporting its short-lived nature against a preexisting active galactic nucleus. These new properties further highlight the need to find additional QPE sources to increase the sample size and draw meaningful conclusions about the intrinsic population. Overall, the newly discovered properties (e.g., recent origin and/or transient nature of the quiescent accretion disk; lack of correlation between eruption recurrence timescales and luminosity) are qualitatively consistent with recent models that identify QPEs as extreme mass-ratio inspirals.}
	
	
	\titlerunning{\emph{SRG}/eROSITA discovers two further galaxies showing X-ray quasi-periodic eruptions}
	\authorrunning{R. Arcodia et al.} 
	\maketitle

    \defcitealias{Miniutti+2019:qpe1}{M19}
	\defcitealias{Giustini+2020:qpe2}{G20}
	\defcitealias{Arcodia+2021:eroqpes}{A21}
	
	\section{Introduction}
    \label{sec:intro}

    The advent of wide-field cadenced surveys across the electromagnetic spectrum during the last decade opened a new window to the realm of extra-galactic transients, with the discovery of a few classes of repeaters. Some galactic nuclei have been shown to undergo outbursts recurring on timescales of months to decades in optical and/or X-rays \citep{Payne+2021:14ko,Wevers+2023:ptde,Liu+2023:0456,Malyali+2023:rosat}, others over tens of days \citep{Evans+2023:swift,Guolo+2023:swift} and even down to surprisingly fast recurrences of a few hours, such as the X-ray quasi-periodic eruptions \citep[QPEs;][hereafter \citetalias{Miniutti+2019:qpe1,Giustini+2020:qpe2,Arcodia+2021:eroqpes}]{Miniutti+2019:qpe1,Giustini+2020:qpe2,Arcodia+2021:eroqpes}. Whether all these flavors are connected within a single recipe remains to be seen, although the leading models of each subclass appear to relate somehow to the dynamics of stellar objects in galactic nuclei \citep[e.g.,][]{Linial+2023:repeat}.

    In particular, QPEs are sharp X-ray bursts that last a few hours and repeat in a quasi-periodic manner every several hours to about a day. So far, these peculiar bursts have been observed from massive black holes (MBHs) of $M_{\rm BH}\approx10^{5-6.7}\,M_{\astrosun}$ (\citealp{Shu+2017:Rx}; \citetalias{Miniutti+2019:qpe1,Giustini+2020:qpe2}; \citetalias{Arcodia+2021:eroqpes};\citealp{Wevers+2022:hosts}) residing in the nuclei of nearby low-mass galaxies \citepalias[e.g., stellar masses of $M_{*}\approx1-3\times 10^9\,M_{\astrosun}$;][]{Arcodia+2021:eroqpes}. To date, no associated optical, UV, IR or radio flares have been observed, although most of the current multiwavelength photometry is likely contaminated or dominated by either the host galaxy emission or that of the accretion disk. In fact, when in quiescence in between QPEs, most sources are still detected in X-rays, with a soft spectrum reminiscent of the Wien tail of a radiatively efficient accretion disk (with a peak temperature $kT \sim40-70\,$eV; \citetalias{Miniutti+2019:qpe1,Giustini+2020:qpe2,Arcodia+2021:eroqpes}). Eruptions reach a soft X-ray luminosity of $\sim10^{42}-10^{43}$\,erg\,s$^{-1}$ at their peak, $\sim10-100$ times brighter than the quiescence level, with a spectrum that is hotter when brighter and remains soft and thermal in shape (with a peak temperature $kT \sim100-200\,$eV; \citetalias{Miniutti+2019:qpe1,Giustini+2020:qpe2,Arcodia+2021:eroqpes}). While the timing properties show significant diversity in how regular the eruption arrival times are \citepalias{Miniutti+2019:qpe1,Giustini+2020:qpe2,Arcodia+2021:eroqpes}, all eruptions from all the different QPE sources seem to follow the same spectral evolution: bursts progress with a harder rise than decay at the same X-ray count rate or, modeling the emission as a black body, with a hotter rise than decay at the same luminosity (\citealp{Arcodia+2022:ero1_timing,Miniutti+2023:gsnrebr}; Giustini et al., in prep.).
    
    Here, we consider secure QPE sources GSN\,069 \citepalias{Miniutti+2019:qpe1}, RX J1301.9+2747 (\citealp{Sun+2013:rxj}; \citetalias{Giustini+2020:qpe2}), eRO-QPE1 and eRO-QPE2  \citepalias{Arcodia+2021:eroqpes}. In addition, the QPE candidate XMMSL1 J024916.6-041244 \citep{Chakraborty+2021:qpe5cand} showed remarkably similar spectral properties but only 1.5 eruptions, impeding its secure classification to date. Recently, the start of a flare consistent, in terms of spectral and timing properties, with those of eRO-QPE1 was seen in an optical tidal disruption event (TDE) ``Tormund'' \citep{Quintin+2023:tormund}; however, the lack of repetition to date makes its classification still ambiguous. Further, the source Swift J023017.0+283603 shows repeating soft X-ray outbursts every $\sim21\,$d \citep{Evans+2023:swift,Guolo+2023:swift}, although its opposite asymmetry in the burst shape and opposite spectral evolution during the bursts (with a harder decay than rise) does not allow a secure association with QPEs for the time being.

    The origin of QPEs is still actively debated, with some models proposing some kind of accretion disk instability \citep{Raj+2021:qpemodel,Sniegowska+2023:qpemodel,Kaur+2023:qpemodel,Pan+2023:qpemodel}, whilst most suggest that QPEs are triggered by a binary system including a central MBH and a much smaller body orbiting it \citep{King2020:qpemodel,Sukova+2021:qpemodel,Xian+2021:qpemodel,Zhao+2022:qpemodel,Wang+2022:qpemodel,Metzger+2022:qpemodel,King+2022:qpemodel,Krolik+2022:qpemodel,Linial+2023:qpemodel,Lu+2023:qpemodel,Franchini+2023:qpemodel,Tagawa+2023:qpemodel,Linial+2023:qpemodel2}. For instance, most of the observational properties can be qualitatively reproduced by shocks caused by the interaction between the smaller orbiter and the accretion flow around the MBH \citep{Xian+2021:qpemodel,Franchini+2023:qpemodel,Tagawa+2023:qpemodel,Linial+2023:qpemodel2}. In this framework, the disk is pierced by the orbiter once or twice per orbit, and shocks are caused by an initially optically thick cloud of gas ejected by the collisions \citep{Linial+2023:qpemodel2}. Orbital precession and Lense-Thirring precession of the disk would provide the required departure from exact periodicity \citep{Franchini+2023:qpemodel}. A requirement, if not a prediction, for the collisions model is that the accretion flow around these MBHs is not an extended flow typical of active galactic nuclei (AGN), but rather a more compact density distribution. Otherwise the orbiter would sink into the disk plane and/or be ablated by the collision in a more extended disk. This led the latest models to suggest that the disk originates from a TDE, whose role in the QPE emission is solely to throw gas at the preexisting extreme mass ratio inspiral (EMRI) orbiters \citep{Franchini+2023:qpemodel,Linial+2023:qpemodel2}. Interestingly, this theoretical connection between QPEs and a preexisting behavior akin to that of TDEs was previously suggested based on observational properties alone: both GSN 069 \citep{Shu+2018:gsn,Sheng+2021:tdegsn,Miniutti+2023:gsnrebr} and the two candidate QPEs \citep{Chakraborty+2021:qpe5cand,Quintin+2023:tormund} show evidence of past TDEs prior to the onset of QPE (or candidate QPE) behavior.	
		
	Clearly, at this stage it is fundamental to find more QPEs to draw any significant conclusions on the population and its physical origin. After the serendipitous discovery of the first QPE emitting source \citepalias{Miniutti+2019:qpe1}, QPEs were either found through dedicated searches in the X-ray archives (\citetalias{Giustini+2020:qpe2}; \citealp{Chakraborty+2021:qpe5cand,Quintin+2023:tormund}) or with a blind search within the live data stream of the eROSITA X-ray telescope \citepalias{Arcodia+2021:eroqpes}. As much as the former method is important to make sure no interesting source was overlooked, wide-area surveys in X-rays provide the only way to systematically detect new QPE sources as they happen in the sky. In \citetalias{Arcodia+2021:eroqpes}, we reported results from the first two eROSITA all-sky surveys \citep{Merloni+2024:erass}. Here, we report on two new discoveries based on the subsequent eROSITA surveys.
	
	\section{Data processing and analysis}
	\label{sec:data}

    \subsection{X-ray spectral analysis}

    Spectral analysis is performed with the Bayesian X-ray Analysis software (BXA) version 4.0.7 \citep{Buchner+2014:BXA}, which connects the nested sampling algorithm UltraNest \citep{Buchner2019:mlf, Buchner2021:ultranest} with the fitting environment XSPEC version 12.13.0c \citep{Arnaud+1996:xspec}, in its Python version PyXspec\footnote{\href{https://heasarc.gsfc.nasa.gov/docs/xanadu/xspec/python/html/index.html}{Link to PyXspec}}. The continuum model adopted is absorbed by Galactic column density from HI4PI \citep{HI4PI+2016:HI4PI} and redshifted to rest-frame using spectroscopic redshifts. These are $z=0.024$ for eRO-QPE3 and $z=0.044$ for eRO-QPE4 (more details in Sect.~\ref{sec:optical}). We quote median and 16th and 84th percentiles ($\sim1\sigma$) uncertainties from fit posteriors, unless otherwise stated, for fit parameters, flux and luminosity. The bolometric luminosity (labeled with ``bol'', e.g., $L_{disk,bol}$) is obtained integrating the adopted source model between 0.001 and 100\,keV. For non-detections, we quote $\sim1\sigma$ ($\sim3\sigma$) upper limits using the 84th (99th) percentiles of the fit posteriors, unless otherwise stated. eROSITA source plus background spectra (Sect.~\ref{sec:erosita}) were fit including a model component for the background, which was determined via a principal component analysis \citep[e.g.,][]{Simmonds+2018:pca} from a large sample of eROSITA background spectra \citep[e.g.,][]{Liu+2022efeds}. \emph{XMM-Newton} spectra (Sect.~\ref{sec:xmm}) were instead fit using \texttt{wstat}, namely XSPEC implementation of the Cash statistic \citep{Cash1979:cstat}, given the good counts statistics in both source and background spectra. In plots (e.g., Appendix~\ref{sec:erass_spec}), data are rebinned, with uncertainty on the summed counts ($CTS_{\rm tot}$) computed as $1+\sqrt{CTS_{\rm tot}+0.75}$, only for visualization purposes.

    As discussed in \citet{Miniutti+2023:gsnreapp,Miniutti+2023:gsnrebr}, we interpret the QPE emission as thermal-like and we use a simple model black body (\texttt{zbbody} in XSPEC). This is a rather model-independent interpretation given the observed spectral shape, with consequences only on the QPE's bolometric emission. The soft X-ray emission in quiescence, namely in between the eruptions, is often detected in QPE-sources and interpreted as the inner regions of a radiatively efficient accretion disk. Therefore, we model the quiescence, if detected, with a \texttt{diskbb} in XSPEC\footnote{We refer to this model (\texttt{zashift(diskbb)} in XSPEC) as disk or \texttt{diskbb} throughout the paper.}. This component is held fixed whilst fitting of the eruption, therefore QPE fluxes are integrated only under the QPE model. For spectra with good counts statistics (e.g., \emph{XMM-Newton} spectra) we allow the quiescence parameters free to vary within the 10th and 90th percentiles of the posterior distributions of the quiescence-only spectral fits. In this way, the spectral fit parameters and fluxes of the eruptions are marginalized over the uncertainties of the quiescence model. For spectra with lower counts statistics (e.g., eRASS spectra) we freeze the quiescence model in the QPE fits. Additional spectral components are added, if needed to model residuals, and discussed individually in the following Sections. The best-fit model among the ones adopted is selected by inspection of the residuals and by comparing the logarithmic Bayesian evidence (\emph{Z}), adopting the model with the highest value. In general, these hard residuals, if present, are here interpreted as thermal Comptonization of the disk emission. We use either a simple phenomenological powerlaw (\texttt{zpowerlw} in XSPEC) or a more physically motivated model \citep[e.g., \texttt{nthComp} in XSPEC;][]{Zdziarski+1996:nthcomp,Zycki+1999:nthcomp}, depending on the counts statistics in the spectra. We note that alternative models may also be used in future work, although we make the simple assumption that disk residuals are due to Comptonization of the disk photons. This is for the sake of simplicity and it is motivated by its nearly ubiquitous presence in accretion flows around black holes.

    \begin{figure*}[tb]
		\centering
		\includegraphics[width=0.32\textwidth]{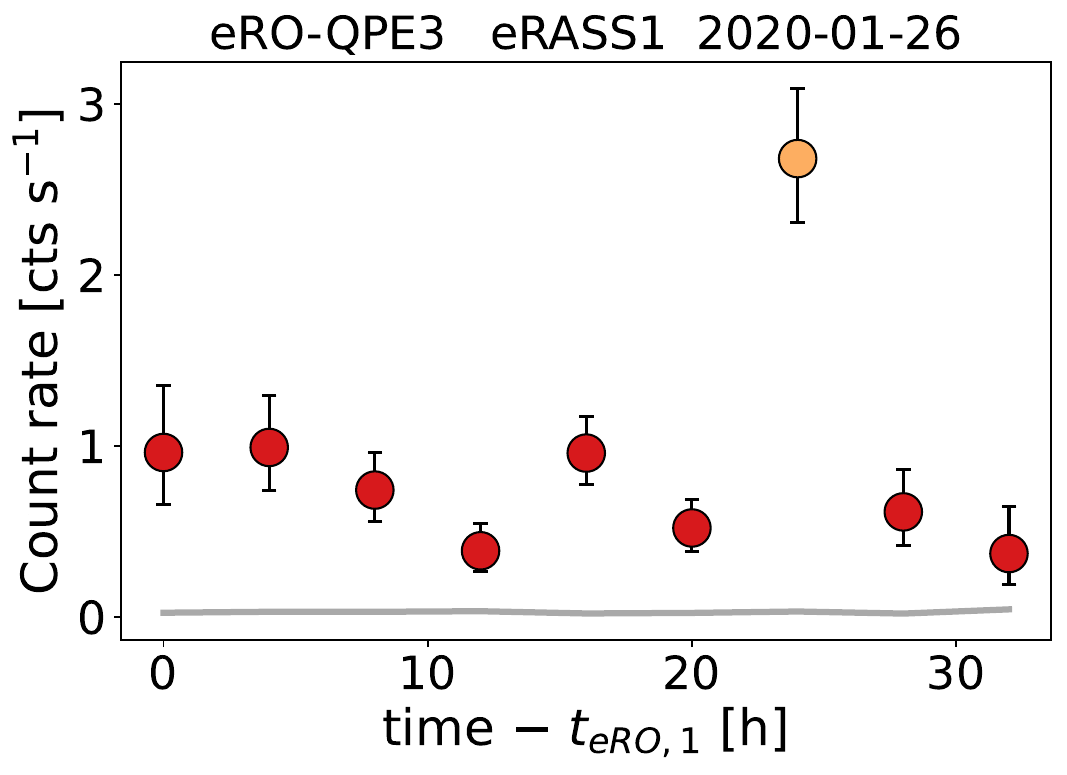}
		\includegraphics[width=0.333\textwidth]{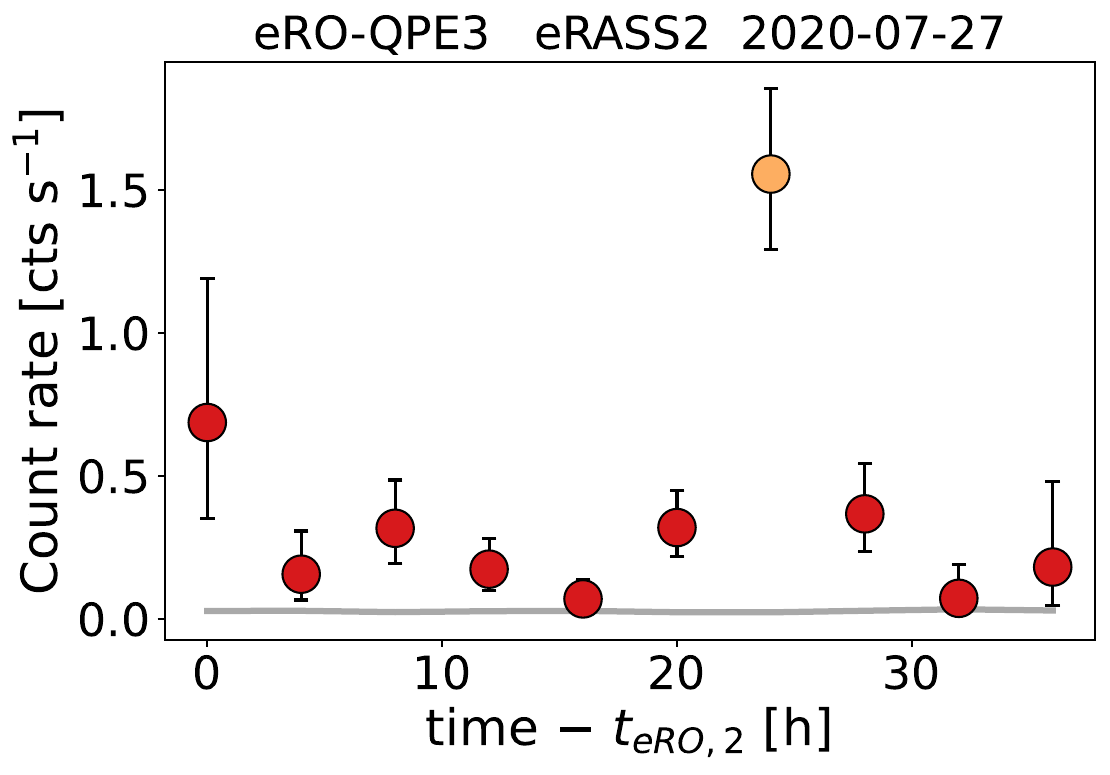}		
        \includegraphics[width=0.336\textwidth]{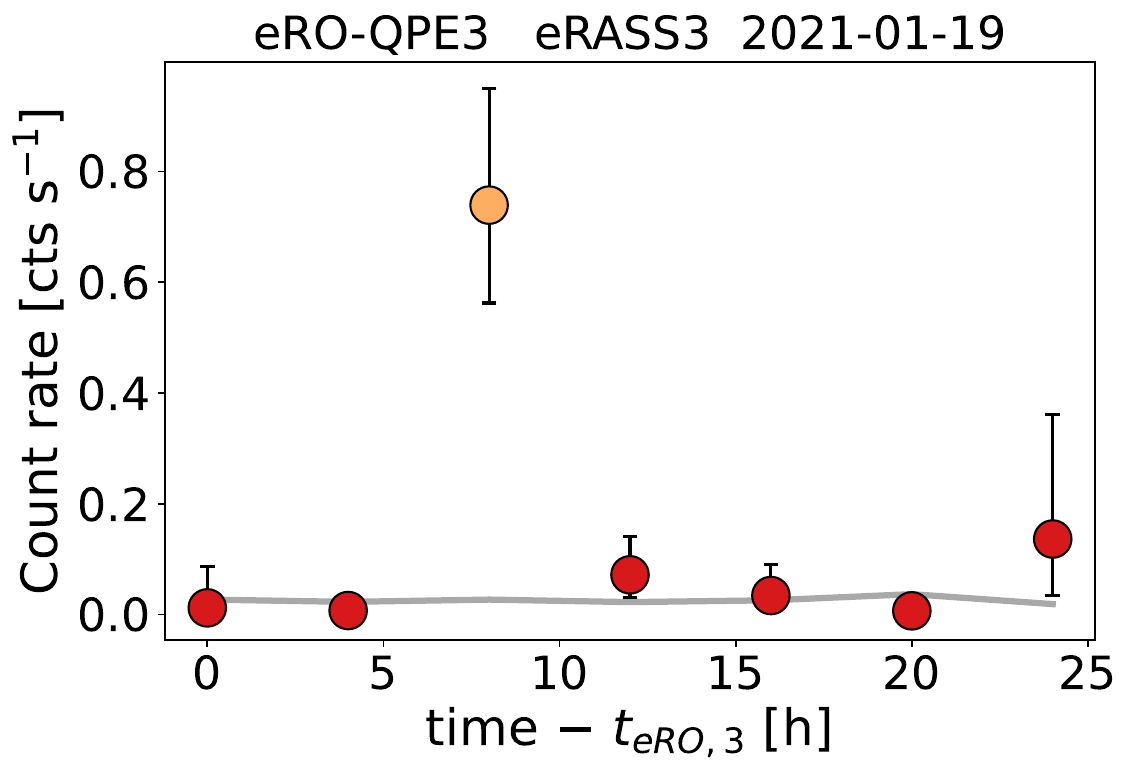}
        \includegraphics[width=0.33\textwidth]{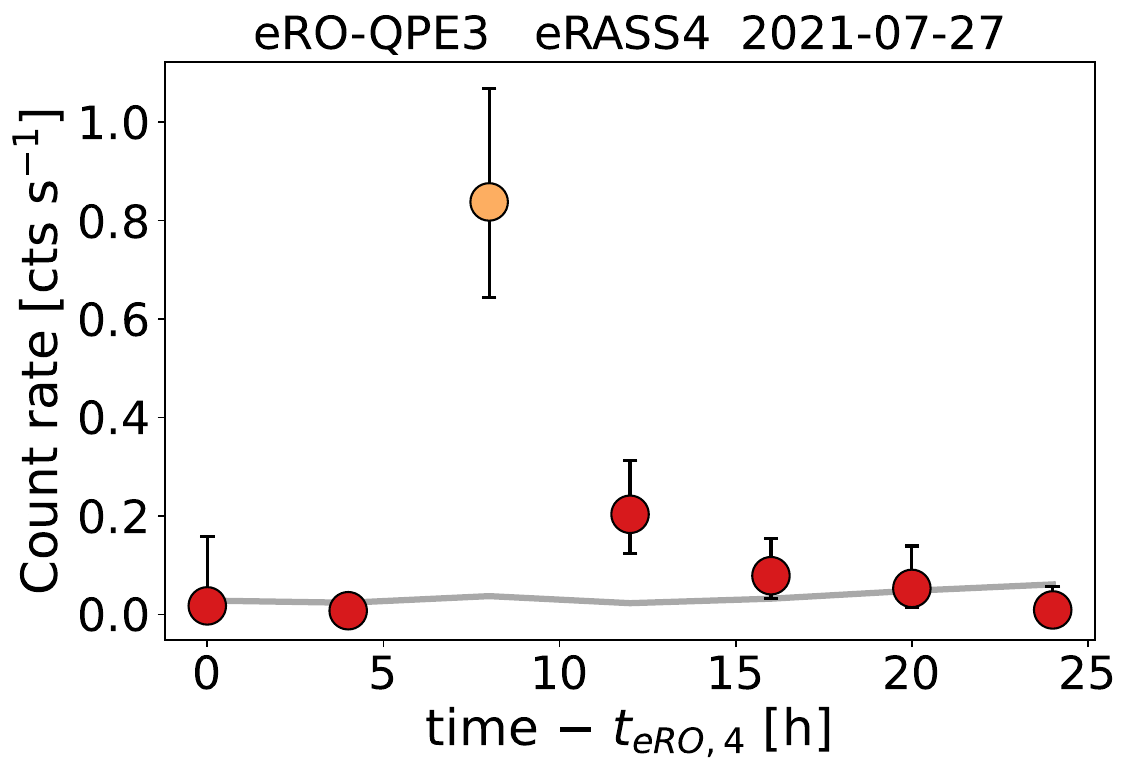}
        \includegraphics[width=0.33\textwidth]{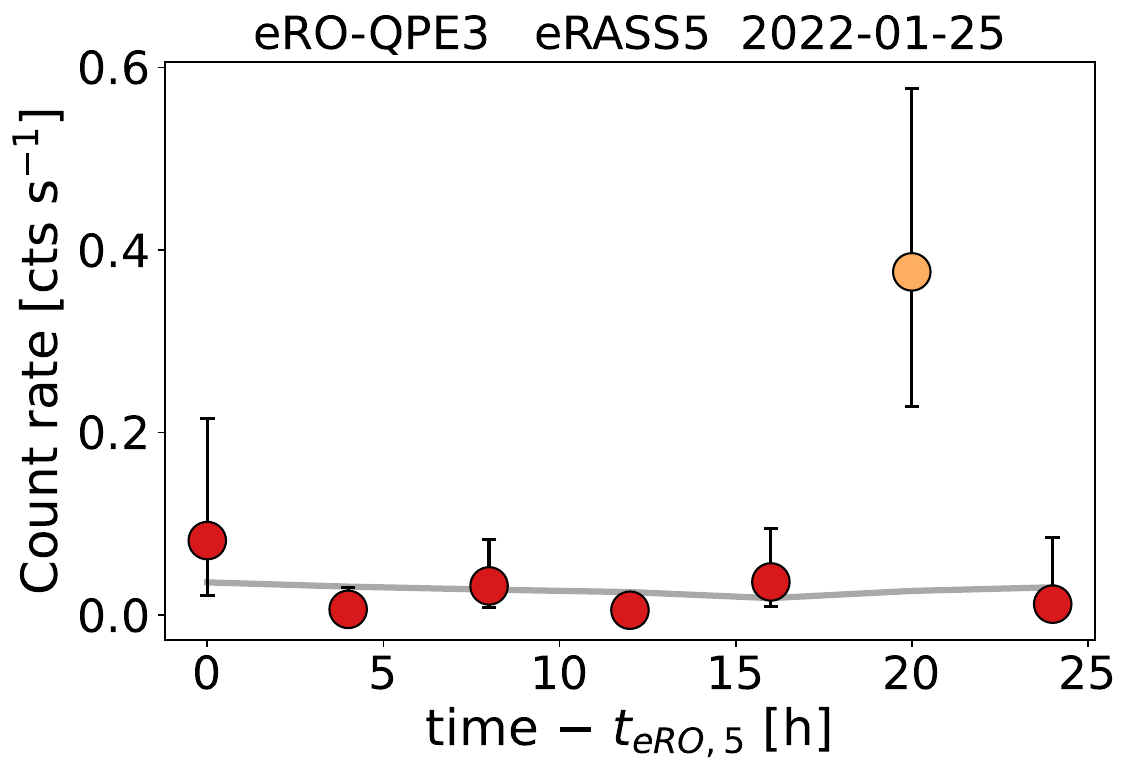}
        \caption{eRASS1-5 light curves of eRO-QPE3 in the $0.2-2.0\,$keV band, each scaled by a reference starting time (MJD $\sim$58874.111, $\sim$59057.576, $\sim$59233.903, $\sim$59422.701 and $\sim$59604.611 from $t_{\rm eRO,1}$ to $t_{\rm eRO,5}$, respectively). Orange data points highlight the putative detection of eruptions, whilst the gray line connects the background level across the eROdays in each eRASS.}
		\label{fig:eRO3_erass_lc}
	\end{figure*}
    	
	\subsection{eROSITA}
    \label{sec:erosita}
	
	eROSITA \citep{Predehl+2021:eROSITA} is the soft X-ray instrument aboard the Spectrum-Roentgen-Gamma (\emph{SRG}) mission \citep{Sunyaev+2021:SRG}. On 13 December 2019 it started observing in survey mode and to this date it has completed four (and started the fifth) of the foreseen eight all-sky surveys (eRASS1-8), each of which is completed in six months. In each survey, every source in the sky is observed for $\sim40\,$s every $\sim4\,$h (i.e., a so-called eROday) for a total number of times within a single eRASS that depends on its location in the sky: around six times on the ecliptic plane and increasing toward higher (ecliptic) latitudes. We have developed an algorithm to look for significant and repeated high-amplitude variability in the eROSITA sources. Light curves are systematically extracted by the eROSITA team with the \texttt{srctool} task of the eROSITA Science Analysis Software System \citep[eSASS;][]{Brunner+2022:eSASS} from event files version 020. In a nutshell, we select sources showing high-amplitude variability within eROdays of the single eRASS, or across multiple eRASS. After excluding secure Galactic objects using {\it Gaia} DR3 \citep{GAIA+2016:gaia,GAIA+2021:edr3}, we visually inspect their X-ray and multiwavelength properties, although only the former (truly alternating variability pattern and a soft spectrum) are used to consider a source as QPE candidate. A first version of this method was described in \citetalias{Arcodia+2021:eroqpes} and we present more details in a companion paper \citep{Arcodia+2024:rates} presenting the calculation of intrinsic volumetric rates based on the eROSITA QPE search method.
     
    \subsubsection{eRO-QPE3}
	
	The source eRASSt J140053-284557 (hereafter eRO-QPE3) is located at the astrometrically corrected X-ray position of (RAJ2000, DECJ2000)=(210.2222, -28.7665), with a total 1$\sigma$ positional uncertainty (including a systematic error of $\sim1.5$") of 1.6", based on the latest-available internal catalog from the first four surveys (eRASS:4). The source is detected in eRASS1 with the identifier 1eRASS J140053.3-284558 \citep{Merloni+2024:erass}. It was also independently found in a search for TDEs, as a new, relative to the archival upper limit, soft eRASS1 source (Grotova et al, in prep). During eRASS1 the source was detected in each eROday with a compatible flux level, with the exception of one eROday showing count rate higher by a factor $\sim2-3$ (top left of Fig.~\ref{fig:eRO3_erass_lc}). However, this intra-eROday variability was not deemed significant by the flare searching code. eRO-QPE3 then triggered an alert from the QPE-candidates searcher in eRASS2, eRASS3 and eRASS4. In eRASS2, the source was fainter compared to eRASS1 (in terms of average count rate), while still detected at all eROdays with an overall constant flux with the exception of one (top middle panel in Fig.~\ref{fig:eRO3_erass_lc}). In eRASS3-4-5, nearly all the counts were detected in a single eROday, with most of (or all, depending on the eRASS) the remaining eROdays consistent with background (see Fig.~\ref{fig:eRO3_erass_lc}).

    \begin{figure}[tb]
		\centering
        \includegraphics[width=0.8\columnwidth]{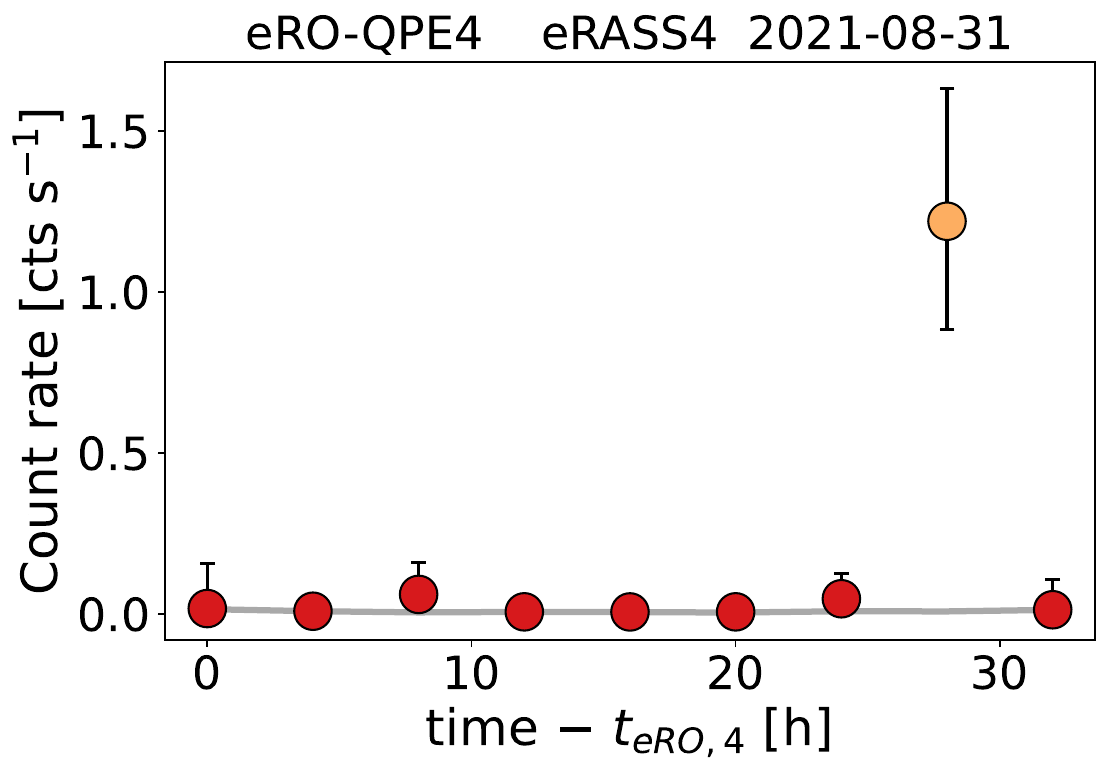}
		\caption{Same as Fig.~\ref{fig:eRO3_erass_lc}, but for eRASS4 data of eRO-QPE4. The reference starting time is $MJD\sim59457.807$. eRO-QPE4 was undetected in the previous eRASS1-3 survey.}
		\label{fig:eRO4_erass_lc}
	\end{figure}

    \begin{figure*}[t]
		\centering
		\includegraphics[width=0.29\textwidth]{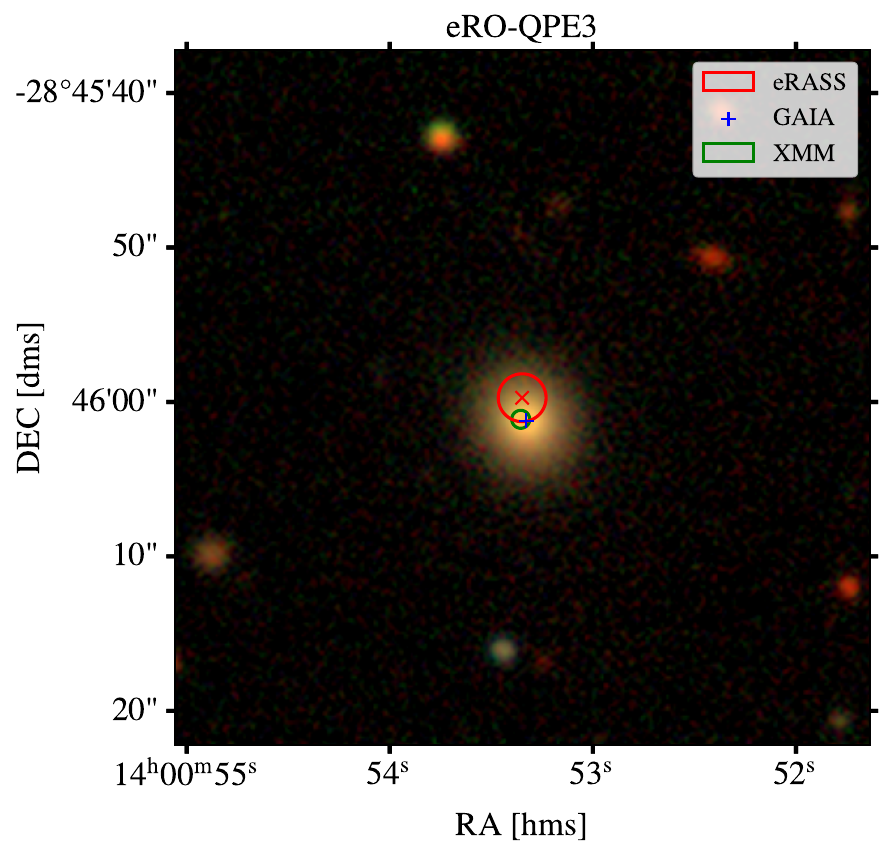}
        \includegraphics[width=0.7\textwidth]{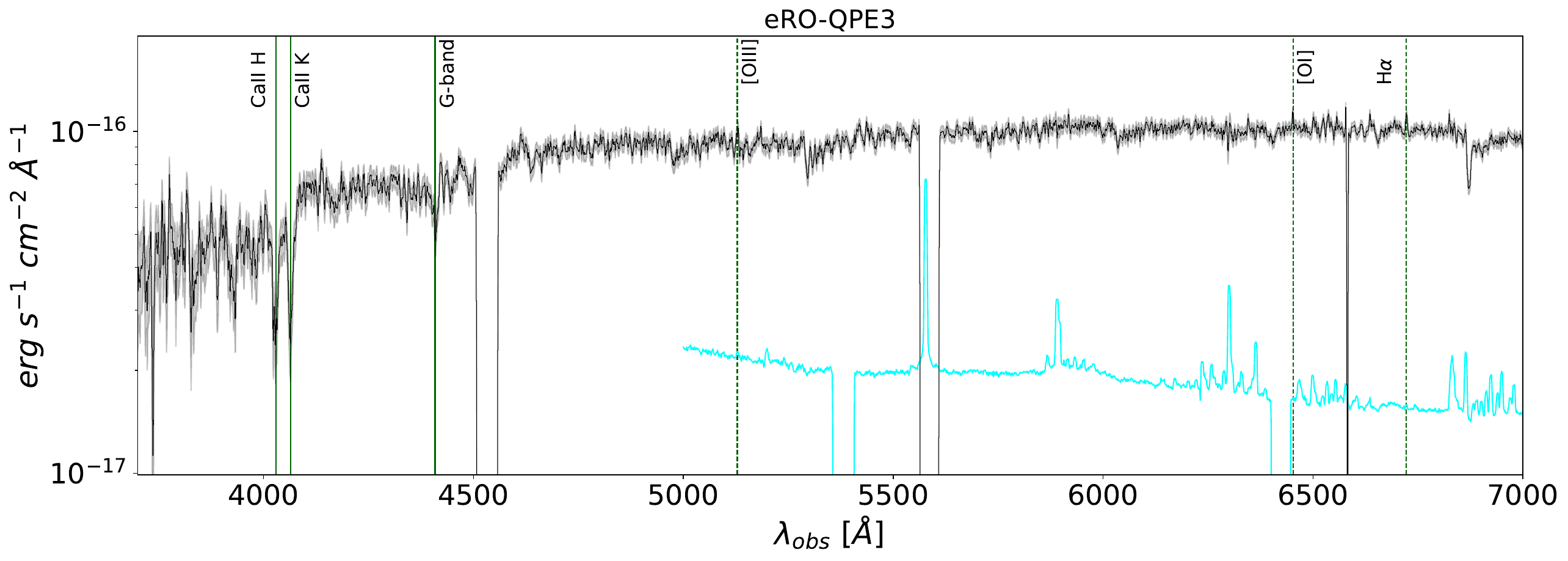}
		\includegraphics[width=0.29\textwidth]{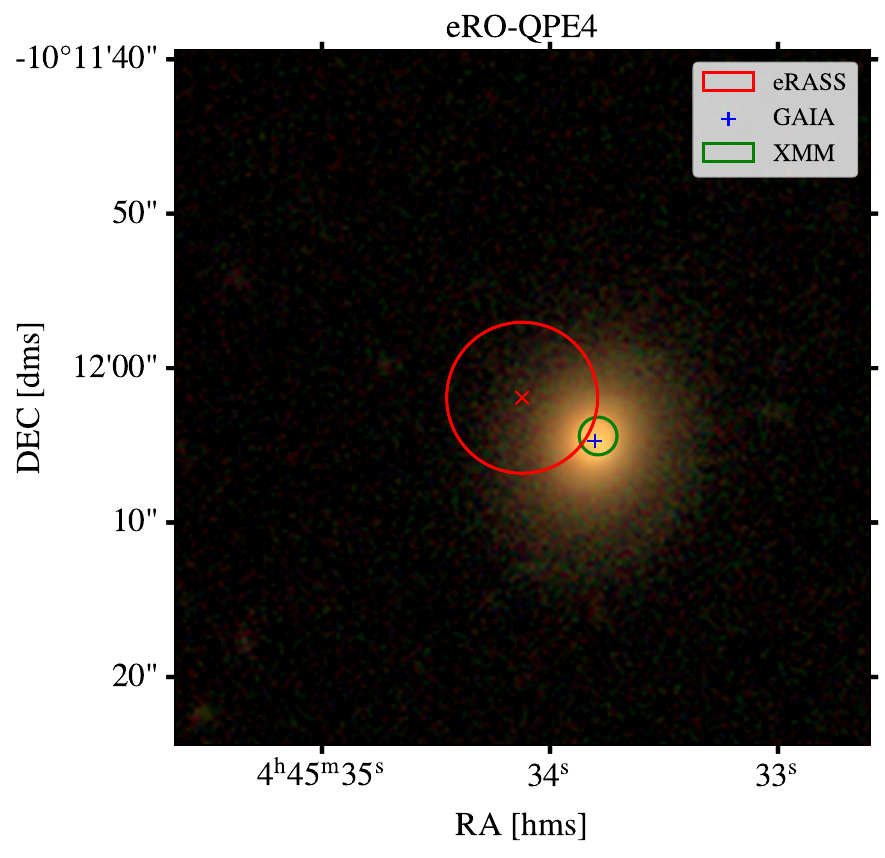}
        \includegraphics[width=0.7\textwidth]{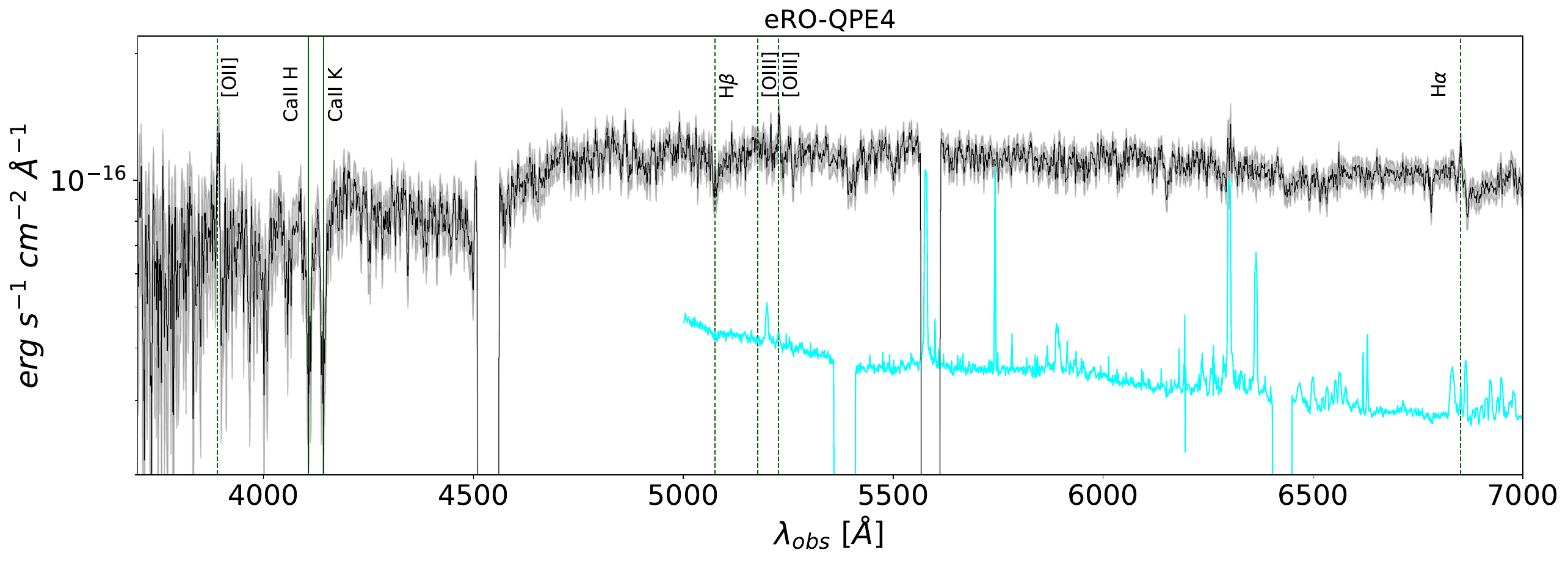}
    	\caption{\emph{Left panels:} $45$"$\times\,45$" cutout of the DESI Legacy Imaging Surveys Data Release 10 [Legacy Surveys / D. Lang (Perimeter Institute)] with the X-ray 1$\sigma$ position circles in red (eROSITA) and green (XMM-Newton), for eRO-QPE3 (\emph{top}) and eRO-QPE4 (\emph{bottom}). \emph{Right panels:} SALT optical spectra of eRO-QPE3 (\emph{top}) and eRO-QPE4 (\emph{bottom}), with inferred spec-z of $\sim0.024$ and $\sim0.044$, respectively. The sky spectrum of the red spectral exposure is shown in cyan in arbitrary units in both subpanels.}
		\label{fig:opt_image}
	\end{figure*}

    Details on the spectral fits performed on eROSITA data of eRO-QPE3 are reported in Appendix~\ref{sec:erass_spec} and Table~\ref{tab:erass1_spec}. The eRASS1 spectrum in quiescence was modeled with a \texttt{diskbb} plus \texttt{zpowerlw} (top panel of Fig.~\ref{fig:eRO3_erass1_spec}). The median (and 16th, 84th percentiles) peak temperature of the disk model is $kT_{\rm in}=kT_{\rm disk}=100_{-20}^{+18}\,$eV, with a rest-frame flux of $F_{\rm 0.2-2.0\,keV} = 1.4_{-0.6}^{+0.4} \times 10^{-12}\,$erg\,s$^{-1}$\,cm$^{-2}$. Here, we consider the orange point in the top-left panel of Fig.~\ref{fig:eRO3_erass_lc} as a bright state part of an eruption. However, we do not exclude the possibility that in eRASS1 no QPEs were present, as the total eRASS1 spectrum can be also modeled with the quiescence model alone (see Appendix.~\ref{sec:erass_spec} and Fig.~\ref{fig:eRO3_erass1_spec_noqpe}). The quiescence model was held fixed during the putative eruption and a black body component was added (bottom panel of Fig.~\ref{fig:eRO3_erass1_spec}). The median (and 16th, 84th percentiles) peak temperature of the QPE component is $kT=kT_{\rm QPE}=122_{-15}^{+19}\,$eV and its flux is $F_{\rm 0.2-2.0\,keV} = (3.0 \pm 0.7) \times 10^{-12}\,$erg\,s$^{-1}$\,cm$^{-2}$. The median temperature is harder than the quiescence disk's peak temperature, but compatible within uncertainties. This confirms that the contrast between the putative quiescence and the QPE, if present, is low during eRASS1. The total flux during the bright eROday, quiescence included, in the observed X-ray band is $F_{\rm 0.2-8.0\,keV} = 4.9_{-0.6}^{+0.7} \times 10^{-12}\,$erg\,s$^{-1}$\,cm$^{-2}$. The eRASS2 spectrum in quiescence is fainter than in eRASS1 (see top panel of Fig.~\ref{fig:eRO3_erass2_spec} and Fig.~\ref{fig:eRO3_erass_lc}). The median (and related 16th, 84th percentiles) temperature of the disk is $kT_{\rm disk}=89_{-60}^{+42}\,$eV, with a flux $F_{\rm 0.2-2.0\,keV} = 3.12_{-3.11}^{+1.51} \times 10^{-13}\,$erg\,s$^{-1}$\,cm$^{-2}$. The total flux in the observed X-ray band is $F_{\rm  0.2-8.0\,keV} = 5.8_{-2.9}^{+2.7} \times 10^{-13}\,$erg\,s$^{-1}$\,cm$^{-2}$. As done for eRASS1, this quiescence model is held fixed during the brighter eROday (the orange point in the top-medium panel of Fig.~\ref{fig:eRO3_erass_lc}) and a black body component is added (bottom panel of Fig.~\ref{fig:eRO3_erass2_spec}). The median (and 16th, 84th percentiles) peak temperature of the QPE component is $kT_{\rm QPE}=82_{-9}^{+10}\,$eV and its flux is $F_{\rm 0.2-2.0\,keV} = 3.2_{-0.7}^{+1.0} \times 10^{-12}\,$erg\,s$^{-1}$\,cm$^{-2}$. The median temperature is colder compared to the bright eROday in eRASS1, despite the flux being compatible. However, given the short $\sim40$s exposure of an eROday compared to the typical QPE duration ($0.5-7\,$h, \citetalias{Arcodia+2021:eroqpes}), eRASS data catch the eruption at different phases and they must be considered as lower limits of the true QPE peaks. The total flux during the bright eROday, quiescence included, in the observed X-ray band is $F_{\rm 0.2-8.0\,keV} = 3.9_{-0.8}^{+1.0} \times 10^{-12}\,$erg\,s$^{-1}$\,cm$^{-2}$. Therefore, despite the decreasing quiescence level, the flux in the bright eROday is compatible between eRASS1 and eRASS2. In eRASS3, eRASS4 and eRASS5, most of the signal is observed during a single eROday, therefore no quiescent state is detected. We adopt a disk model to compute upper limits (Table~\ref{tab:erass1_spec}) and we fit the bright state with a black body QPE component only. In eRASS5, the fit is compatible with background (within $3\sigma$, Fig.~\ref{fig:eRO3_erass5_spec_burst}) even in the putative bright state. We report the fit values in Table~\ref{tab:erass1_spec} for completeness.
    
    \subsubsection{eRO-QPE4}

    The source eRASSt J044534-101201 (hereafter eRO-QPE4) is located at the astrometrically corrected X-ray position of (RAJ2000, DECJ2000)=(71.3921, -10.2005), with a total 1$\sigma$ positional uncertainty (including a systematic error of $\sim1.5$") of 4.9", based on the latest-available internal catalog from the first four surveys (eRASS:4). eRO-QPE4 was undetected in the first three eRASS, with a $1\sigma$ ($3\sigma$) upper limit of $F_{\rm 0.2-2.0\,keV}<3.8 \times 10^{-15}$\,erg\,s$^{-1}$\,cm$^{-2}$ ($<2.8 \times 10^{-14}$\,erg\,s$^{-1}$\,cm$^{-2}$). In eRASS4, most of the X-ray signal was observed in a single eROday (see Fig.~\ref{fig:eRO4_erass_lc}). Adopting a black body model for the QPE emission (a behavior in the source confirmed with follow-up observations), we obtain $kT_{\rm QPE}=93_{-14}^{+18}\,$eV and a flux $F_{\rm 0.2-2.0\,keV} = 3.4^{+1.7}_{-1.1} \times 10^{-12}$\,erg\,s$^{-1}$\,cm$^{-2}$. Excluding this eROday, the source is undetected with a $1\sigma$ upper limit of $F_{\rm 0.2-2.0\,keV}<6.9 \times 10^{-14}$\,erg\,s$^{-1}$\,cm$^{-2}$, using a \texttt{diskbb} model. The position of eRO-QPE4 was not observed in the partial eRASS5 survey.

    \subsection{Optical identification}
    \label{sec:optical}

    \subsubsection{eRO-QPE3}
	
	eRO-QPE3 is associated with the galaxy 2MASS 14005331-2846012 at (RA, Dec) =  (14:00:53.315, -28:46:01.26). We show the Legacy Survey DR10 optical image in the top left panel of Fig.~\ref{fig:opt_image}, centered at the X-ray coordinates. The eROSITA position and 1$\sigma$ accuracy (based on the cumulative eRASS:4 image) are shown with a red cross and circle, respectively, whilst in green we show the analog from the deeper, hence more accurate, \emph{XMM-Newton} observation. Similarly to other QPE sources, eRO-QPE3 is consistent with the nucleus position, given the available positional accuracy \citepalias{Miniutti+2019:qpe1,Giustini+2020:qpe2,Arcodia+2021:eroqpes}. 2MASS 14005331-2846012 was observed in the \textit{grizJHK} bands with the seven-channel imager GROND (Gamma-Ray Burst Optical/Near-Infrared Detector; \citealp{Greiner:2008}) at the MPG 2.2\,m telescope at La Silla Observatory (Chile) on 2020 February 27th. The data were reduced with the standard GROND pipeline \citep{Kruehler:2008}, which performs the bias and flat-field corrections, image stacking and astrometric calibration. The photometric calibration was achieved against PanSTARRS (\textit{griz}) and 2MASS (\textit{JHK}; \citealp{Skrutskie:2006}). The observed Kron magnitudes in the AB system are $18.37 \pm 0.06$, $17.60 \pm 0.05$, $17.29 \pm 0.07$,  $17.05 \pm 0.05$,  $16.82 \pm 0.11$,  $16.60 \pm 0.13$ and  $16.95 \pm 0.14$\,mag, respectively.
		
	We instigated optical spectroscopic follow-up with the Robert Stobie Spectrograph (RSS, \citealp{Burgh+2003:RSS}) on the Southern African Large Telescope (SALT, \citealp{Buckley+2006:SALT}) on the night of April 24 2022 (top right panel of Fig.~\ref{fig:opt_image}). Data were reduced following the procedure outlined in \citetalias{Arcodia+2021:eroqpes}. The host galaxy shows a seemingly inactive spectrum, with $z=0.024$ based on Calcium H and K absorption lines (top right panel of Fig.~\ref{fig:opt_image}). We fit the SALT optical spectrum with \texttt{Firefly}, which is a fitting code for deriving the stellar population properties of galaxy spectra \citep{Wilkinson+2017:firefly,Maraston+2020:firefly}. SALT spectra are not calibrated to absolute values \citep{Buckley+2018:salt}, therefore we first renormalized the spectrum flux using optical photometry from the GROND data using the \texttt{Specutil} Python package \citep{Earl+2023:specutils}. \texttt{Firefly} was run twice using two different stellar population models \citep{Maraston+2011:firefly}. The run with the population model \texttt{ELODIE} \citep{Maraston+2011:firefly} was adopted to estimate the mean values for stellar-mass ($M_{*}$) and star formation rate ($SFR$). We added in quadrature to the measured statistical errors a systematic error to account for the choice between the two stellar population models, which was computed from the difference between the two mean values. The stellar mass inferred with \texttt{Firefly} for eRO-QPE3 is $M_* = 2.56^{+0.24}_{-1.40}\times10^9\,M_{\astrosun}$, whist SFR is $0.20^{+0.02}_{-0.14}\,M_{\astrosun}\,$yr$^{-1}$. The low $M_*$ value is in line with the very compact nature of the host galaxy (Fig.~\ref{fig:opt_image}, top), and is in line with those inferred for eRO-QPE1 and eRO-QPE2 \citepalias{Arcodia+2021:eroqpes}, which is remarkable given the blind nature of our search. Using scaling relations between black hole and total stellar mass of the galaxy \citep{Reines+2015:mstar}, we obtain $M_{\rm BH}=5.3^{+0.7}_{-3.5}\times10^6\,M_{\astrosun}$. Although these scaling relations are not necessarily well-calibrated at low masses, the relation with bulge stellar mass has been shown to hold sufficiently well at lower masses too \citep{Schutte+2019:mbhmstar}. We performed SED photometry fitting to have an independent estimate of $M_*$. We collated Legacy Survey DR10 photometry with \texttt{RainbowLasso}\footnote{\href{https://github.com/JohannesBuchner/RainbowLasso}{https://github.com/JohannesBuchner/RainbowLasso}} and fit the SED with Genuine Retrieval of AGN Host Stellar Population (GRAHSP; Buchner et al., in prep.). The model used in GRAHSP includes AGN components (continuum, emission lines, torus) and galaxy components, both attenuated by dust and redshifted. We obtain $M_* = 0.74^{+0.41}_{-0.29}\times10^9\,M_{\astrosun}$, which is slightly lower than the \texttt{Firefly} estimate. This would correspond \citep{Reines+2015:mstar} to a black hole mass of $M_{\rm BH}=0.93^{+0.79}_{-0.47}\times10^6\,M_{\astrosun}$.
 
    No significant emission lines are apparent in the spectrum (top right panel of Fig.~\ref{fig:opt_image}), although subtracting the continuum model from \texttt{Firefly} the presence of faint narrow H$\alpha$, [\ion{O}{iii}] $\lambda$5007 and \ion{O}{i} $\lambda$6302 lines, and very faint $H\beta$ can be inferred. Since we are mainly interested in computing line flux ratios for narrow-lines diagnostics, we computed them from the continuum-subtracted spectrum. We used the \texttt{lmfit} package \citep{Newville+2023:lmfit} and adopted a polynomial model plus Gaussians for emission lines. The algorithm fits a narrow line to H$\beta$, both [\ion{O}{iii}] lines, \ion{O}{i} $\lambda$6302 and H$\alpha$, whilst no narrow [\ion{N}{ii}] lines are visible. We obtain from this $\log([\ion{O}{iii}]/{\rm H}\beta) \sim 0.42$ and $\log([\ion{N}{ii}]/{\rm H}\alpha) \lesssim - 0.87$, after tentatively computing an upper limit on the [\ion{N}{ii}] $\lambda$6549 line by imposing the presence of a narrow line. These values would place eRO-QPE3 in the star-forming region of narrow lines classifications, whilst using the $\log(\ion{O}{i}/{\rm H}\alpha) = -0.45$ the galaxy is instead classified as a LINER \citep{Kewley+2006:bpt}. 
    Given the limited resolution and tentative detection of some narrow lines, we defer a conclusive classification to future work. However, this result confirms that for these nuclear transients narrow lines classifications are ambiguous and have to be interpreted with care, particularly so since they trace past nuclear activity potentially unrelated to the current transient activity. We obtained an independent estimate of the black hole mass by fitting the SALT spectrum with the Penalized PiXel-Fitting method \citep[pPXF;][]{Cappellari+2004:ppxf,Cappellari+2017:ppxf,Cappellari2023:ppxf}. The fit velocity dispersion is $\sim 204\,$km\,s$^{-1}$. Estimating an instrumental broadening of $6\AA$ from the narrowest arc lines for the grating angle settings adopted, the instrumental velocity dispersion computed around the median wavelength $\sim 150\,$km\,s$^{-1}$. Subtracting this dispersion in quadrature and summing in quadrature the template dispersion of $\sim83\,$km\,s$^{-1}$, we obtain a corrected velocity dispersion of $\sim152\,$km\,s$^{-1}$. Using the scaling relation from \citet{Gultekin+2009:msigma}, we infer $M_{\rm BH}\sim4\times10^7\,M_{\astrosun}$, which is significantly larger than the estimate obtained from the stellar mass. Given the good agreement of the $\log M_* \sim 9$ estimate from both spectroscopy and photometry fitting, we adopt the $M_{\rm BH}$ estimate from stellar mass as reference for eRO-QPE3 at this stage (thus in the range $M_{\rm BH}\sim (0.9-5.3)\times10^6\,M_{\astrosun}$). Further analysis with spectra obtained at higher resolution is needed for a more robust narrow-line classification and black hole mass estimate from velocity dispersion.
	
	\subsubsection{eRO-QPE4}
	
	eRO-QPE4 was associated with the galaxy 2MASS 04453380-1012047 at (RA, Dec) =  (04:45:33.80, -10:12:04.74). We show the 1$\sigma$ eROSITA positional accuracy (based on the cumulative eRASS:4 image) in the bottom left panel of Fig.~\ref{fig:opt_image} with a red circle, whilst in green the more accurate \emph{XMM-Newton} circle. We took an optical spectrum with RSS on SALT on February 23 2023 (bottom right panel of Fig.~\ref{fig:opt_image}), data were reduced as reported in \citetalias{Arcodia+2021:eroqpes}. Similarly to the other eRO-QPEs, the galaxy does not show prominent emission lines and a redshift of 0.0437 is estimated from Calcium absorption lines (bottom right panel of Fig.~\ref{fig:opt_image}). We fit the salt spectrum with \texttt{Firefly}, after renormalizing it using optical photometry from VEXAS DR2 catalogs \citep{Khramtsov+2021:vexas}, which are also consistent with SkyMapper petrosian magnitudes \citep{Wolf+2018:skymapper}. We refer to the eRO-QPE3 fit for more details on the stellar population models and the related systematic uncertainties. The stellar mass inferred with \texttt{Firefly} for eRO-QPE4 is $M_* = 1.6^{+0.7}_{-0.6}\times10^{10}\,M_{\astrosun}$, whist SFR is $2.26^{+1.06}_{-2.21}\,M_{\astrosun}\,$yr$^{-1}$. Using \citet{Reines+2015:mstar}, we obtain $M_{\rm BH}=6.8^{+4.8}_{-3.2}\times10^7\,M_{\astrosun}$. This is the highest $M_*$ estimate inferred from any of the eROSITA QPEs \citepalias{Arcodia+2021:eroqpes} and the highest black hole mass ever inferred for a QPE source \citep{Wevers+2022:hosts}. With the GRAHSP SED fitting we obtain $M_* = 0.60^{+0.29}_{-0.12}\times10^{10}\,M_{\astrosun}$, which is slightly lower than the \texttt{Firefly} estimate, even including uncertainties, but still much higher than what is estimated for eRO-QPE3 with the same method. This corresponds \citep{Reines+2015:mstar} to $M_{\rm BH}=1.74^{+1.28}_{-0.47}\times10^7\,M_{\astrosun}$. From the SALT spectrum we can only identify faint narrow H$\alpha$, [\ion{O}{ii}] and [\ion{O}{iii}] $\lambda$5007 in emission, while no H$\beta$ nor [\ion{N}{ii}] lines are visible, even from the residual spectrum provided by \texttt{Firefly} after subtracting the continuum. Hence, we are unable to securely classify the galaxy at this stage, although in similarity with other eROSITA QPEs a strong preexisting AGN is disfavored. We obtained a corrected velocity dispersion of $\sim133\,$km\,s$^{-1}$ with pPXF. Using the scaling relation from \citet{Gultekin+2009:msigma}, we infer $M_{\rm BH}\sim2.4\times10^7\,M_{\astrosun}$, which is reasonably consistent with that obtained via the $M_*$ estimate. Hence, for eRO-QPE4 all diagnostics used point to $M_{\rm BH}$ in the range $\sim(1.7-6.8)\times10^7\,M_{\astrosun}$.

    \begin{figure*}[tb]
		\centering
		\includegraphics[width=\textwidth]{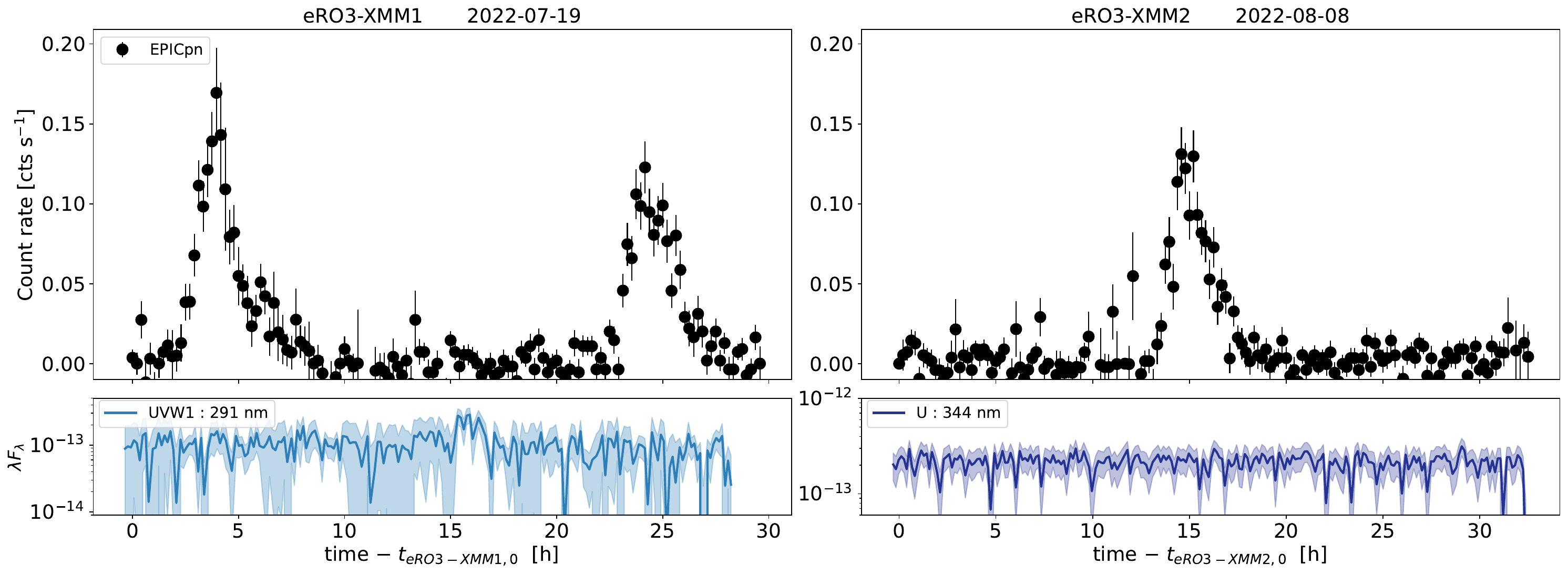}
		\caption{\emph{XMM-Newton} $0.2-10.0\,$keV EPICpn light curve of eRO-QPE3, of the first (eRO3-XMM1, \emph{left}) and second observation (eRO3-XMM2, \emph{right}). The two eruptions in eRO3-XMM1 are separated by $\sim20.4$\,h. Here, $t_{\rm eRO3-XMM1,0}$ corresponds to MJD\,$\sim59779.799$ and $t_{\rm eRO3-XMM2,0}$ to $\sim59799.745$.}
		\label{fig:xmm_lc}
	\end{figure*}

    \begin{figure}[tb]
    	\centering
        \includegraphics[width=0.99\columnwidth]{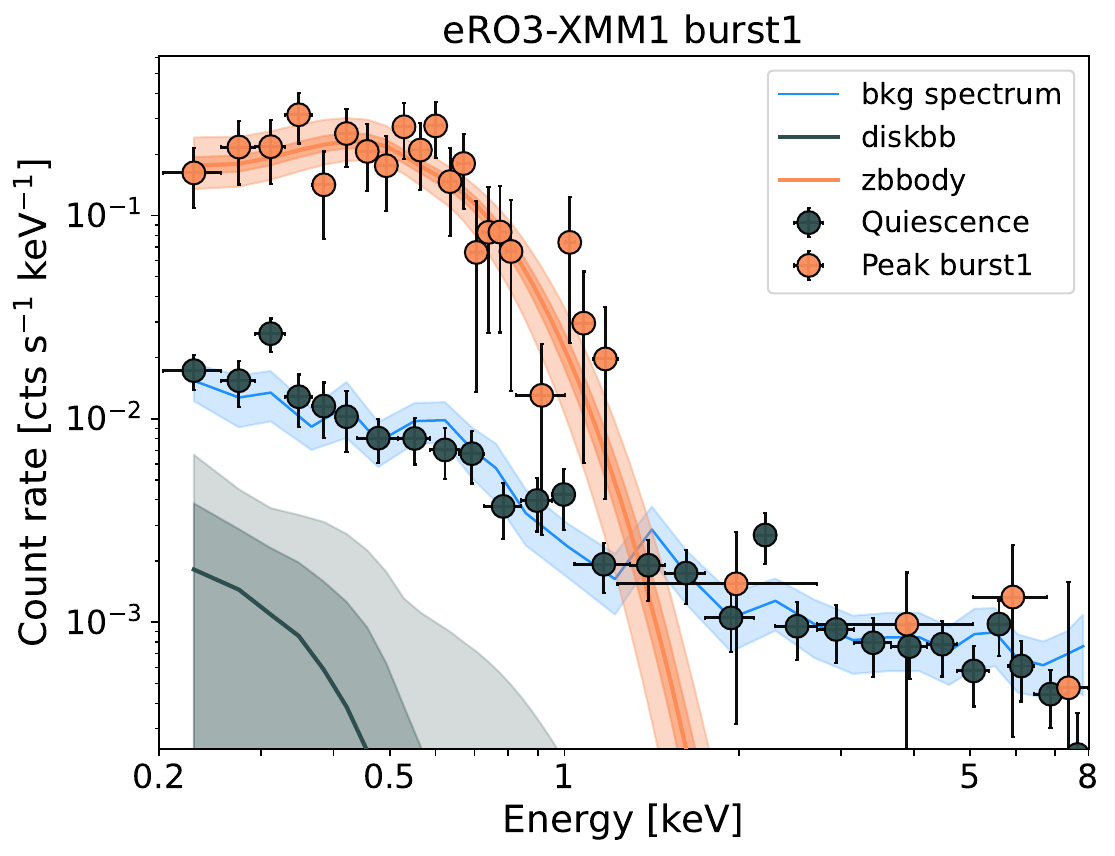}
    	\caption{Source plus background spectrum of eRO-QPE3 in quiescence (dark gray, background dominated) and at the peak of the first burst in the observation eRO3-XMM1 (orange, following the color-coding of Fig.~\ref{fig:eRO3_lcphases}). The light-blue line and contour represent the background spectrum and its uncertainties. Source plus background data points are shown to put the source-only models (lines and contours with the same colors) into context. Darker model contours are $1\sigma$ percentiles, lighter ones are $3\sigma$.}
    	\label{fig:eRO3XMM1_spec_peak}
    \end{figure}
	
	\subsection{XMM-Newton}	
    \label{sec:xmm}
	
	Two \emph{XMM-Newton} observations were performed on eRO-QPE3, namely ObsID 0883770101 starting on 19 July 2022 (hereafter eRO3-XMM1) and 0883770701 starting on 08 August 2022 (hereafter eRO3-XMM2). A single \emph{XMM-Newton} observation was taken on eRO-QPE4, namely ObsID 0883770401 starting on 10 March 2023 (hereafter eRO4-XMM). We reduced \emph{XMM-Newton} data of EPIC-MOS1 and 2 \citep{Turner+2001:mos} and EPIC-PN \citep{Struder+2001:EPIC} cameras and the Optical Monitor \citep[OM;][]{Mason+2001:OM} using standard tools and prescriptions (SAS v. 20.0.0 and HEAsoft v. 6.29). Event files from EPIC cameras were screened for flaring particle background. For eRO-QPE3, source (background) regions were extracted within a circle of 30" centered on the source (in a nearby source-free region). For eRO-QPE4, apertures of 40" were used. The X-ray position obtained from eROSITA during the survey was refined using \emph{XMM-Newton} data and the task \texttt{eposcorr}. We cross-correlated the X-ray sources in the EPIC-PN image with external optical and infrared catalogs (available through the Processing Pipeline Subsystem). The counterparts of eRO-QPE3 and eRO-QPE4 were excluded to obtain a more unbiased estimate of the possible offset from the nucleus. The resulting 1$\sigma$ positional circle from \emph{XMM-Newton} is shown in green in Fig.~\ref{fig:opt_image} for both eRO-QPE3 (accuracy of $1.5$") and eRO-QPE4 (accuracy of $4.9$"). For both sources, the refined X-ray position is consistent with being nuclear within the current uncertainties.
    
    \subsubsection{eRO-QPE3}

    In eRO3-XMM1 two eruptions are observed, with faster rise than decay, separated by $\sim20\,$h, while in eRO3-XMM2 only one burst was detected (Fig.~\ref{fig:xmm_lc}). We use the burst model described in \citet{Arcodia+2022:ero1_timing}, since it has shown itself to be successful in parametrizing asymmetric QPE light curves:
    \begin{equation}
	\begin{cases}
	\label{eq:1}
	A\,\lambda\,e^{\tau_1/(t_{\rm peak} - t_{\rm as} - t)} & \text{if $t<t_{\rm peak}$}\\
	A\,e^{- (t-t_{\rm peak})/\tau_2} & \text{if $t>=t_{\rm peak}$}\\
	\end{cases}       
	\end{equation}
	which is evaluated at zero for times smaller than the asymptote at $t_{\rm peak} - t_{\rm as}$, where $t_{\rm as}=\sqrt{\tau_1\,\tau_2}$. $\tau_1$ and $\tau_2$ are the characteristic timescales, although only the latter is directly related to the decay timescale. $A$ is the amplitude at the peak and $\lambda=e^{t_{\lambda}}$ a normalization, where $t_{\lambda}=\sqrt{\tau_1/\tau_2}$. Rise and decay times can be defined as a function of $1/e^{n}$ factors with respect to the peak flux, with $n$ being an integer. Rise and decay are obtained from the width and asymmetry factors ($w$ and $k$), where $w(n)=\tau_2\,n - \tau_1/(t_{\lambda}\,n) + t_{\rm as}$ and $k(n)= [\tau_2\,n + \tau_1/(t_{\lambda}\,n) - t_{as}]/w$. We define rise and decay timescales as $\tau_{\rm rise}=w\,(1-k)/2$ and $\tau_{\rm decay}=w\,(1+k)/2$ using $n=1$. We quote median values and related 16th and 84th percentiles. Referring to the three eruptions observed by \emph{XMM-Newton} with time, the rise is $2908_{-435}^{+95}$\,s, $2522_{-192}^{+290}$\,s, $2903_{-175}^{+219}$\,s, respectively, and the decay is $5896_{-308}^{+681}$\,s, $5194_{-357}^{+672}$\,s, $5015_{-512}^{+395}$\,s, respectively. The two eruptions in eRO3-XMM1 are separated by a median recurrence time (and related 16th, 84th percentiles) $20.40_{-0.11}^{+0.25}$\,h. We show an example of this model and its application to QPEs in Fig.~\ref{fig:eRO3_lcphases}, for the first burst of observation eRO3-XMM1. Given the relatively low signal-to-noise in eRO-QPE3, we test the QPEs' energy dependence dividing the light curve in two energy bins only, namely $0.2-0.6\,$keV and $0.8-2.0\,$keV. Adopting the first eruption of eRO3-XMM1 as representative, we obtain a rise (decay) of $3661_{-244}^{+202}$\,s ($6020_{-342}^{+445}$\,s) in the low-energy bin and of $2237_{-398}^{+895}$\,s ($4946_{-1234}^{+1288}$\,s) in the high-energy bin. Furthermore, we obtain that the peak time of the low-energy light curve is $(13.3\pm 0.3)$\,ks after the start of eRO3-XMM1, while it is $12.0_{-0.5}^{+0.6}$\,ks for the high-energy light curve. Therefore, the eruptions of eRO-QPE3 follow the known energy-dependence (\citetalias{Miniutti+2019:qpe1,Giustini+2020:qpe2,Arcodia+2021:eroqpes}; \citealp{Chakraborty+2021:qpe5cand,Arcodia+2022:ero1_timing}), in that they are wider and start later at lower energies.
    

    \begin{figure*}[tb]
		\centering
		\includegraphics[width=\textwidth]{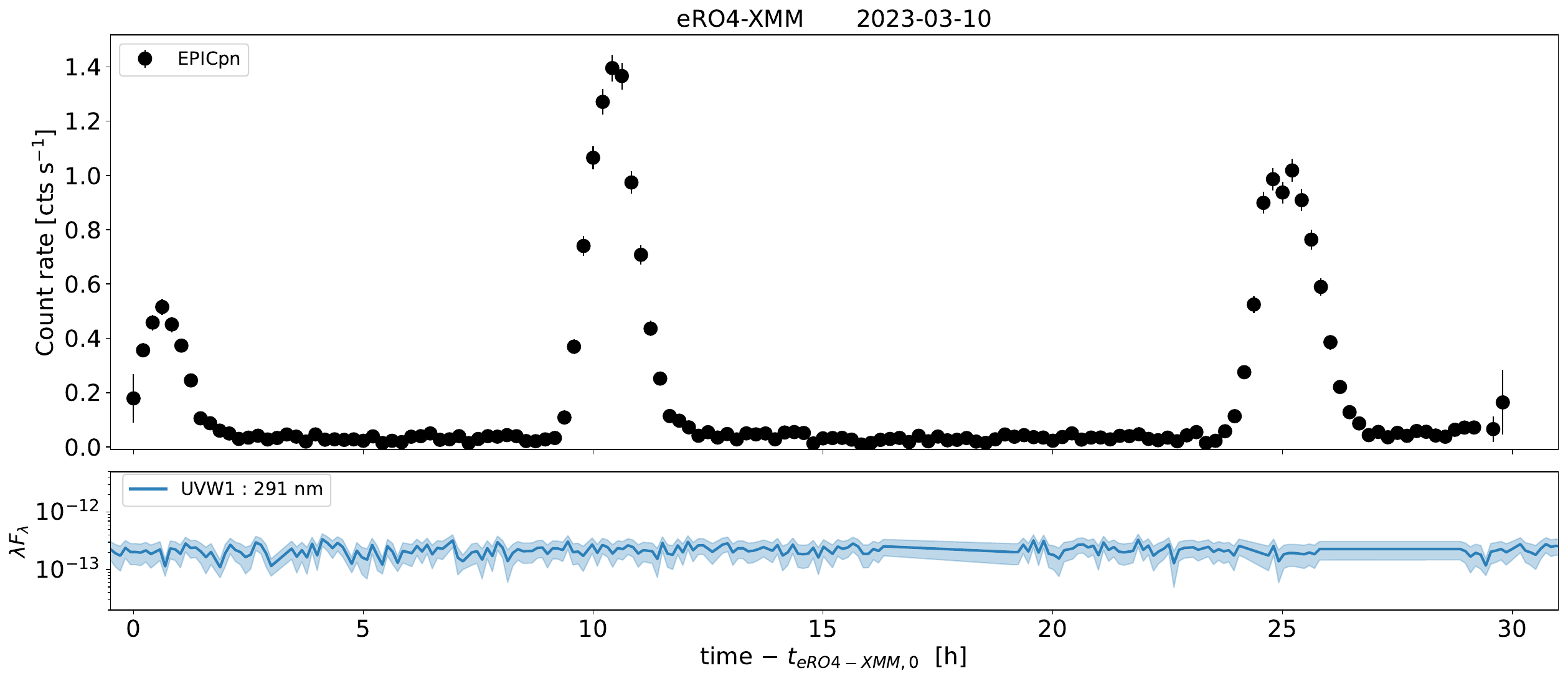}
		\caption{\emph{XMM-Newton} $0.2-10.0\,$keV EPICpn light curve of eRO-QPE4. The three eruptions are separated by $\sim9.82$\,h and $\sim14.70\,$h. Here, $t_{\rm eRO4-XMM,0}$ corresponds to MJD\,$\sim60\,013.127$.}
		\label{fig:xmm_lc_ero4}
	\end{figure*}

    \begin{figure}[tb]
	\centering
    \includegraphics[width=0.99\columnwidth]{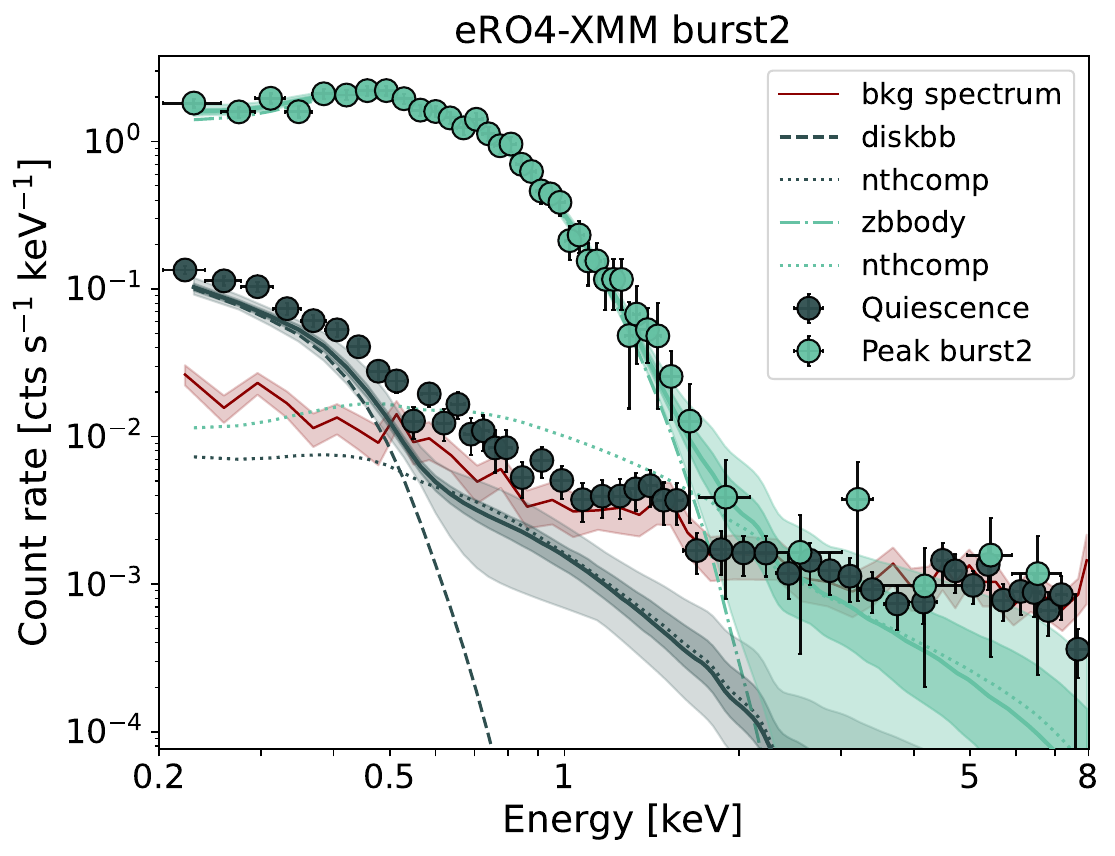}
	\caption{Spectrum of eRO-QPE4 in quiescence (dark gray) and at the peak of the second burst in the observation eRO4-XMM (light green, following the color-coding of Fig.~\ref{fig:eRO3_lcphases}). The dark-red line and contour represent the background spectrum and its uncertainties. Source plus background data points are shown to put the source-only models (lines and contours with the same colors) into context. Darker model contours are $1\sigma$ percentiles, lighter ones are $3\sigma$. Individual model components are labeled.}
	\label{fig:eRO4XMM_spec_peak}
    \end{figure}

    The quiescent state of eRO-QPE3 is too faint to be securely detected above background. We adopt a disk spectrum to provide a flux upper limit of $F_{\rm 0.2-2.0\,keV}<1.1 \times 10^{-14}$\,erg\,s$^{-1}$\,cm$^{-2}$ (Fig.~\ref{fig:eRO3XMM1_spec_peak}). The fit posterior on the flux, or disk normalization, is unconstrained, while the disk temperature is loosely constrained to $kT_{\rm disk}=46_{-18}^{+21}$\,eV. Despite the weak constraint on the temperature, we conservatively consider the quiescence of eRO-QPE3 to be undetected at the \emph{XMM-Newton} epoch with a soft X-ray luminosity upper limit of $L_{0.2-2.0\, \rm keV}<1.6 \times 10^{40}$\,erg\,s$^{-1}$. 
    We divided the first eruption of eRO3-XMM1, as example, in five epochs, namely rise1, rise2, peak, decay1 and decay2 (see Fig.~\ref{fig:eRO3_lcphases}). We fit each epoch with a black body (see more details in Appendix~\ref{sec:xmm_spec}) and report best-fit parameters in Table~\ref{tab:erass1_spec}. We present more details on the spectral evolution during the eruptions in Sect.~\ref{sec:energy_evol}. Here, we quote results of the peak spectrum (see Fig.~\ref{fig:eRO3XMM1_spec_peak}): we obtained $kT_{\rm QPE}=111_{-5}^{+6}$\,eV and $F_{\rm 0.2-2.0\,keV} = 3.1_{-0.2}^{+0.3} \times 10^{-13}$\,erg\,s$^{-1}$\,cm$^{-2}$ ($L_{\rm 0.2-2.0\,keV} = (4.2 \pm 0.4) \times 10^{41}$\,erg\,s$^{-1}$), thus $\gtrsim25-30$ times brighter than the quiescence flux in the soft X-ray band. The bolometric luminosity of the black body component is $L_{\rm QPE,bol} = (4.9 \pm 0.5) \times 10^{41}$\,erg\,s$^{-1}$.
    
     
    The OM UVW1 (U) filter was used throughout eRO3-XMM1 (eRO3-XMM2) with a series of $\sim4400$\,s-long exposures. The subpanels of Fig.~\ref{fig:xmm_lc} show the OM light curves in the respective filters. Similarly to other QPE sources \citepalias{Miniutti+2019:qpe1,Giustini+2020:qpe2,Arcodia+2021:eroqpes}, no simultaneous variability is observed, with the caveat that these data are likely contaminated by the host galaxy and possibly nuclear star cluster. The mean flux is $\lambda F_{\lambda,344\,nm} = (2.2 \pm 0.5) \times 10^{-13}$\,erg\,s$^{-1}$\,cm$^{-2}$ and $\lambda F_{\lambda,291\,nm} = (1.0 \pm 0.5) \times 10^{-13}$\,erg\,s$^{-1}$\,cm$^{-2}$ in the U (344\,nm) and UVW1 (291\,nm) filter, respectively, with magnitudes of $19.44\pm0.09$ and $19.87\pm0.11$, respectively.

    \subsubsection{eRO-QPE4}
    \label{sec:ero4xmm}

    In eRO4-XMM, three eruptions are observed (Fig.~\ref{fig:xmm_lc_ero4}). Their duration appears shorter compared to eRO-QPE3 and they appear more symmetric. We fit each eruption in eRO4-XMM with both a Gaussian model and the asymmetric model of Eq.~\ref{eq:1}. The latter model performs significantly worse (in terms of residuals and goodness of fit) and we adopt the Gaussian model for the eruptions in eRO-QPE4. We show an example in Fig.~\ref{fig:eRO3_lcphases}, for the second burst of observation eRO4-XMM. The median (with related 16th, 84th percentile values) of the Gaussian width is $1678_{-90}^{+83}$\,s, $1765_{-23}^{+24}$\,s and $2081_{-29}^{+30}$\,s, for the three eruptions, respectively. The peak-to-peak separation between the QPEs is $9.82_{-0.02}^{+0.03}$\,h and $14.70_{-0.01}^{+0.02}\,$h. We divided the light curve in energy bins of $0.2-0.4\,$keV, $0.4-0.6\,$keV, $0.6-0.8\,$keV, $0.8-1.0\,$keV and $1.0-2.0\,$keV (hereafter $E_1$ to $E_5$, respectively). Adopting the second eruption in eRO4-XMM as example, given its higher signal-to-noise, we obtain that the median (with related 16th, 84th percentile values) of the Gaussian width is $(1759\pm 22)$\,s, $1630_{-27}^{+25}$\,s, $1417_{-32}^{+33}$\,s, $1292_{-52}^{+51}$\,s and $1299_{-116}^{+120}$\,s, respectively, hence decreasing from $E_1$ to $E_5$. Adopting the $E_5$ Gaussian peak as reference (which occurs $10.28_{-0.03}^{+0.04}$\,h after the start of eRO4-XMM), we obtain delays of $603_{-30}^{+31}$\,s, $464_{-30}^{+35}$\,s, $165_{-40}^{+43}$\,s and $-30_{-70}^{+67}$\,s, for $E_1$, $E_2$, $E_3$ and $E_4$, respectively. Apart from $E_4$ which is consistent with the properties of $E_5$ within uncertainties, the eruptions of eRO-QPE4 follow the known energy-dependence (\citetalias{Miniutti+2019:qpe1,Giustini+2020:qpe2,Arcodia+2021:eroqpes}; \citealp{Chakraborty+2021:qpe5cand,Arcodia+2022:ero1_timing}), in that they are wider and start later at lower energies.

    The quiescent state of eRO-QPE4 is detected with a soft spectrum (Fig.~\ref{fig:eRO4XMM_spec_peak}) similar to that of other QPE sources in quiescence (\citetalias{Miniutti+2019:qpe1,Giustini+2020:qpe2,Arcodia+2021:eroqpes}, \citealp{Chakraborty+2021:qpe5cand}). We fit the quiescence with a disk model and subsequently add a Comptonization component (Fig.~\ref{fig:eRO4XMM_spec_peak}), which yields a disk temperature of $kT_{\rm disk}=(43\pm2)\,$eV with a flux $F_{\rm 0.2-2.0\,keV} = (3.9\pm0.4) \times 10^{-13}$\,erg\,s$^{-1}$\,cm$^{-2}$. This corresponds to $L_{\rm 0.2-2.0\,keV} = (1.7\pm0.3) \times 10^{42}$\,erg\,s$^{-1}$ for the disk component alone, or to a bolometric luminosity of $L_{\rm disk,bol} = 2.0_{-0.3}^{+0.6} \times 10^{43}$\,erg\,s$^{-1}$. The Comptonization component is much fainter, with $F_{\rm 0.2-2.0\,keV} = 1.4_{-0.4}^{+0.8} \times 10^{-14}$\,erg\,s$^{-1}$\,cm$^{-2}$, and its slope is unconstrained. More details are presented in Appendix~\ref{sec:xmm_spec} and Table~\ref{tab:ero4_spec}. As done for eRO-QPE3, the QPE flares of eRO-QPE4 were separated in five epochs, namely rise1, rise2, peak, decay1 and decay2 (bottom panel of Fig.~\ref{fig:eRO3_lcphases} for the second burst of the observation eRO4-XMM). The quiescence model was held fixed by imposing its parameters to vary only within the 10th-90th percentile interval of the posteriors obtained during the quiescence fit alone. The fit results for the different burst phases are reported in Table~\ref{tab:ero4_spec}. For the epochs rise1, rise2, decay1 and decay2 the best-fit model adopted for the burst is that of the quiescence plus a black body component. For the peak spectrum we add a Comptonization component (see Fig.~\ref{fig:eRO4XMM_spec_peak}). More details are presented in Appendix~\ref{sec:xmm_spec} and Table~\ref{tab:ero4_spec}. This model yields $kT_{\rm QPE}=(123\pm2)$\,eV and $F_{\rm 0.2-2.0\,keV} = (2.9\pm0.1) \times 10^{-12}$\,erg\,s$^{-1}$\,cm$^{-2}$ ($L_{\rm 0.2-2.0\,keV} = (1.27\pm0.03) \times 10^{43}$\,erg\,s$^{-1}$). The bolometric luminosity of the black body component is $L_{\rm QPE,bol} = (1.44\pm0.04) \times 10^{43}$\,erg\,s$^{-1}$. Similarly to eRO-QPE3, the OM UVW1 filter was used throughout and no simultaneous variability is observed. The mean flux is $\lambda F_{\lambda,291\,nm} = (2.1 \pm 0.5) \times 10^{-13}$\,erg\,s$^{-1}$\,cm$^{-2}$ and the mean magnitude $18.96\pm0.15$.
		
	\subsection{NICER}
    \label{sec:nicer}
    \emph{NICER} data were processed using \texttt{HEAsoft} v6.32.1, using \texttt{NICERDAS} v11a. We grouped the data into 200-second Good-Time Intervals (GTIs) and adopted custom filtering choices of unrestricted undershoot (\texttt{underonly\_range}=*-*) and overshoot rates (\texttt{overonly\_range}=*-*), with per-FPM and per-MPU autoscreening disabled to prevent aggressive event filtering. We manually discarded focal plane modules (FPMs) with $0-0.2$\,keV rates or $5-18$\,keV count rates $>3\sigma$ higher than average, or above an absolute threshold of 5 counts sec$^{-1}$. As both eRO-QPE3 and eRO-QPE4 are super-soft and faint, the emission above 5\,keV is entirely background-dominated, whereas the $0-0.2$\,keV band is undershoot-dominated, making these bands effective proxies for assessing the background conditions.

\begin{figure}[tb]
	\centering
    \includegraphics[width=0.99\columnwidth]{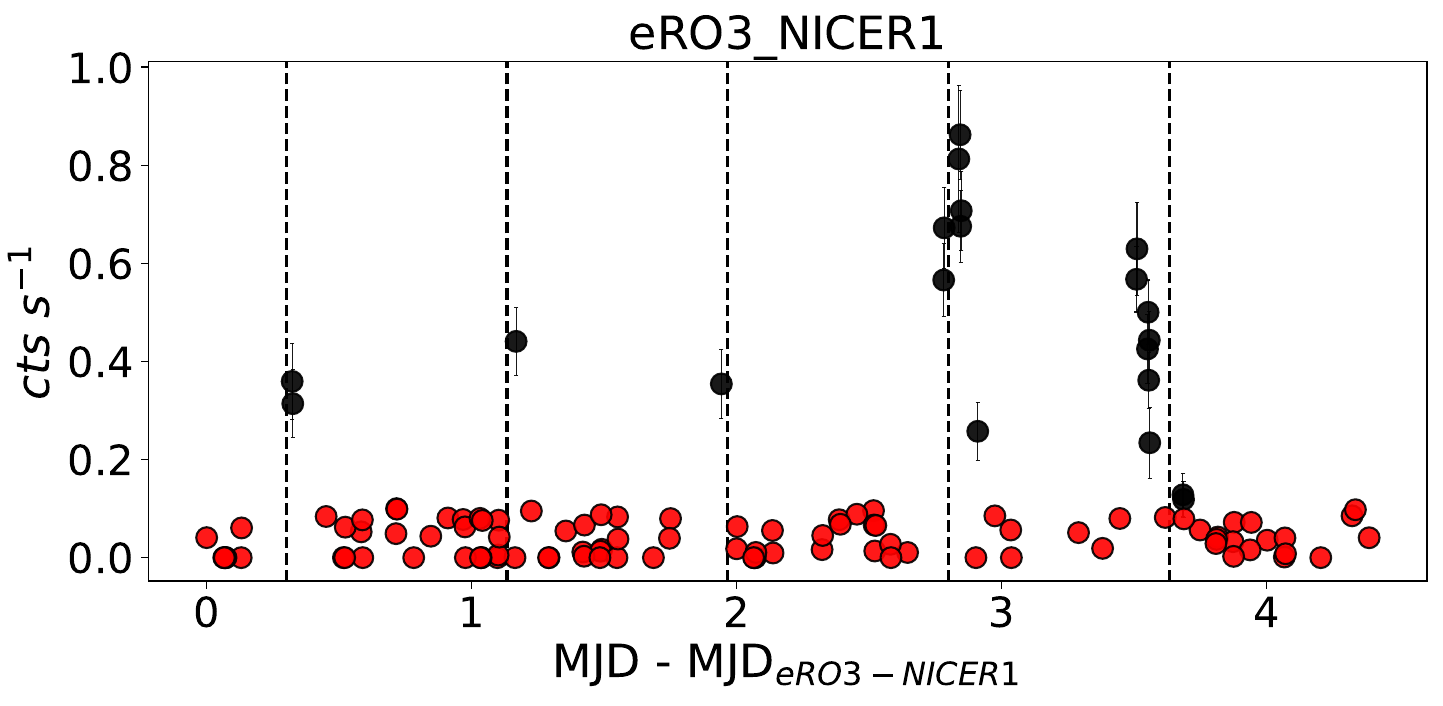}
    \includegraphics[width=0.99\columnwidth]{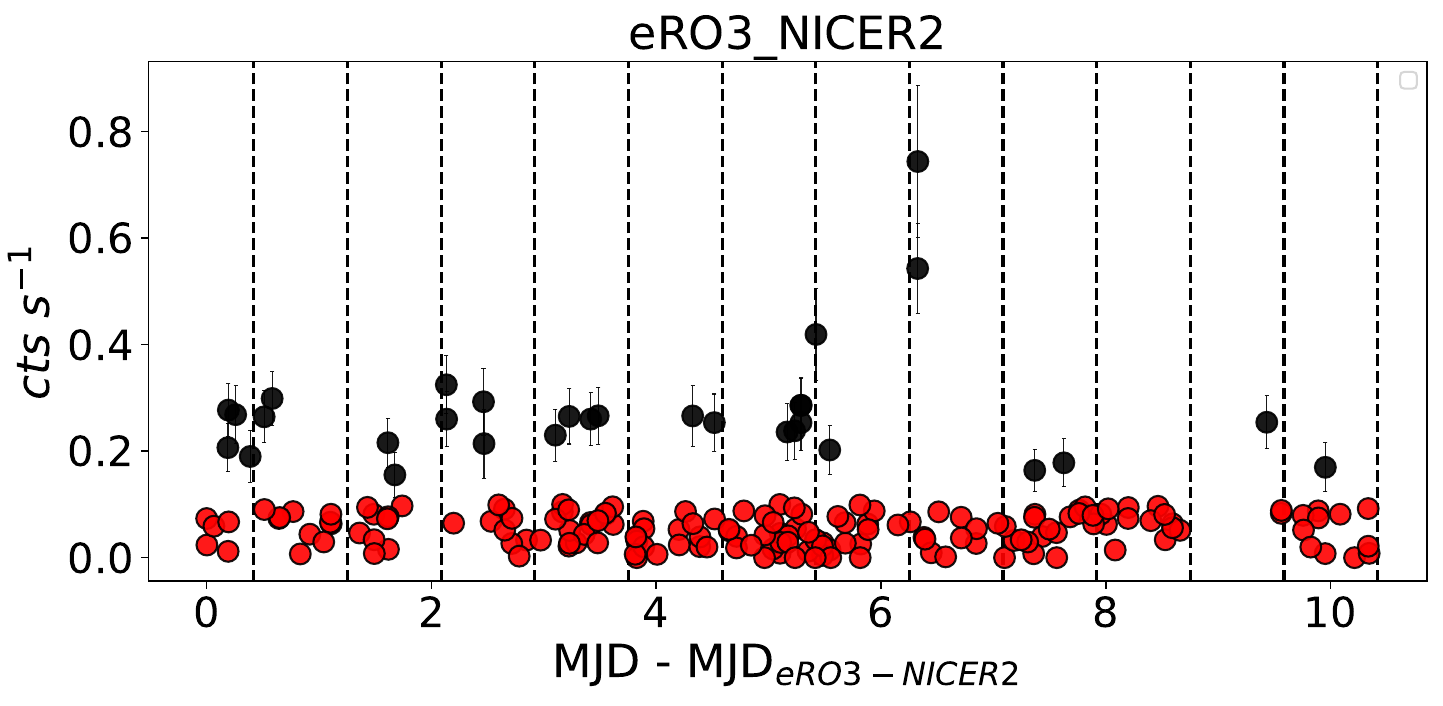}
	\caption{\emph{NICER} background-subtracted $0.2-2.0$\,keV light curve of eRO-QPE3, in the epoch eRO3-NICER1 (\emph{top}, starting at MJD$\sim59697.644$) and eRO3-NICER2 (\emph{bottom}, starting at MJD$\sim59822.782$). Black points represent secure detections of a source component on top of background emission, contrary to the epochs shown in red which are consistent with background-only. Vertical dashed lines represent evenly spaced recurrences of $20\,$h, as a guide for the eye.}
	\label{fig:eRO3_nicer}
    \end{figure}

    Following the methodology of \citet{Chakraborty+2024:ero1} we create the light curves in Figs.~\ref{fig:eRO3_nicer} and ~\ref{fig:eRO4_nicer} with time-resolved spectroscopy across the GTIs. After screening the event lists, we used the \texttt{SCORPEON}\footnote{\href{https://heasarc.gsfc.nasa.gov/lheasoft/ftools/headas/niscorpeon.html}{\texttt{https://heasarc.gsfc.nasa.gov/lheasoft/ftools/\\headas/niscorpeon.html}}} background model to estimate the contribution from the diffuse X-ray background and non X-ray background. We fit the entire broadband ($0.2-18$\,keV) array counts with PyXspec\footnote{\href{https://heasarc.gsfc.nasa.gov/docs/xanadu/xspec/python/html/index.html}{\texttt{https://heasarc.gsfc.nasa.gov/docs/xanadu/xspec/\\python/html/index.html}}}. Along with the \texttt{SCORPEON} background we fit each GTI with a source model represented by \texttt{tbabs}$\times$\texttt{zbbody}. We consider a source detection as any GTI in which the blackbody normalization is $>1\sigma$ inconsistent with zero, namely a non-background component is required by the fit at the $1\sigma$ level. The source count rates are the summed counts contained only in the blackbody component. We consider all other cases as  non detections and we adopt as upper limit the $3\sigma$ upper error uncertainty on the flux of the blackbody model that did not make the significance cut.
    
    \subsection{eRO-QPE3}

    eRO-QPE3 was observed for a total of $\sim26.2$\,ks between 28 April 2022 and 3 May 2022, and again for a total of $\sim40.3$\,ks between 31 August 2022 and 11 September 2022, namely a few months before and about a month after eRO3-XMM1 (Fig.~\ref{fig:eRO3_nicer}). These observation sets are hereafter named eRO3-NICER1 and eRO3-NICER2, respectively. In both sets, the source is significantly detected in a transient fashion, although in most cases the full eruption is not resolved, at times due to nonoptimal background conditions preventing a secure detection of a source component. 
    
    Only two clear bright eruptions are seen in eRO3-NICER1 (Fig.~\ref{fig:eRO3_nicer}, top), separated by roughly $\sim17$\,h, which is significantly shorter than the $\sim20.4\,$h recurrence seen in eRO3-XMM1 (Fig.~\ref{fig:xmm_lc}). eRO-QPE3 is then significantly detected above background for a few short exposures in eRO3-NICER1. As these transient detections are roughly separated by $\sim20\,$h (see vertical dashed lines in Fig.~\ref{fig:eRO3_nicer}, as guide for the eye), we are confident that they all correspond to eruption phases. Although current data do not allow for a more precise constraint on the exact recurrence times, we can conclude that in eRO-QPE3 QPEs have a typical recurrence in the range $\sim17-20\,$h, with significant scatter as seen in other sources (e.g., eRO-QPE1; \citetalias{Arcodia+2021:eroqpes}). The brightest eruptions in eRO3-NICER1 reach a maximum flux of $F_{\rm 0.2-2.0\,keV} = 9.8_{-1.3}^{+1.2} \times 10^{-13}$\,erg\,s$^{-1}$\,cm$^{-2}$ and a temperature $kT=133_{-8}^{+10}\,$eV, whilst the average of all bursts including the other transient detections is $F_{\rm 0.2-2.0\,keV} = 6.5_{-3.1}^{+2.3} \times 10^{-13}$\,erg\,s$^{-1}$\,cm$^{-2}$ and $kT=107_{-20}^{+24}\,$eV. The maximum values are larger than those observed in eRO3-XMM1, perhaps indicating a long-term decay of peak flux and temperature. In eRO3-NICER2, only part of a bright eruption is observed (Fig.~\ref{fig:eRO3_nicer}, bottom), with a peak flux and temperature roughly consistent with those of the maximum values observed in eRO3-NICER1. At several other epochs the source is found with a flux of $F_{\rm 0.2-2.0\,keV} \sim 5 \times 10^{-13}$\,erg\,s$^{-1}$\,cm$^{-2}$, similar to the peak flux observed in eRO3-XMM1. This would suggest that most QPEs peak at a level consistent with eRO3-XMM1, but occasional brighter eruptions reach a brighter flux, as that seen at the maximum of eRO3-NICER1 and eRO3-NICER2. The repetition is unclear and not as close to $\sim20\,$h as in eRO3-NICER1, although in most snapshots we were not able to securely assess the presence of a source component. However, these lower-flux detections close to the background count rate are much harder to be securely disentangled from the latter, compared to brighter eruptions. Hence, we refrain from further interpretations of eRO3-NICER2 at this stage and defer more thorough analysis to future work.
        

    \subsection{eRO-QPE4}

    \begin{figure}[tb]
	\centering
    \includegraphics[width=0.99\columnwidth]{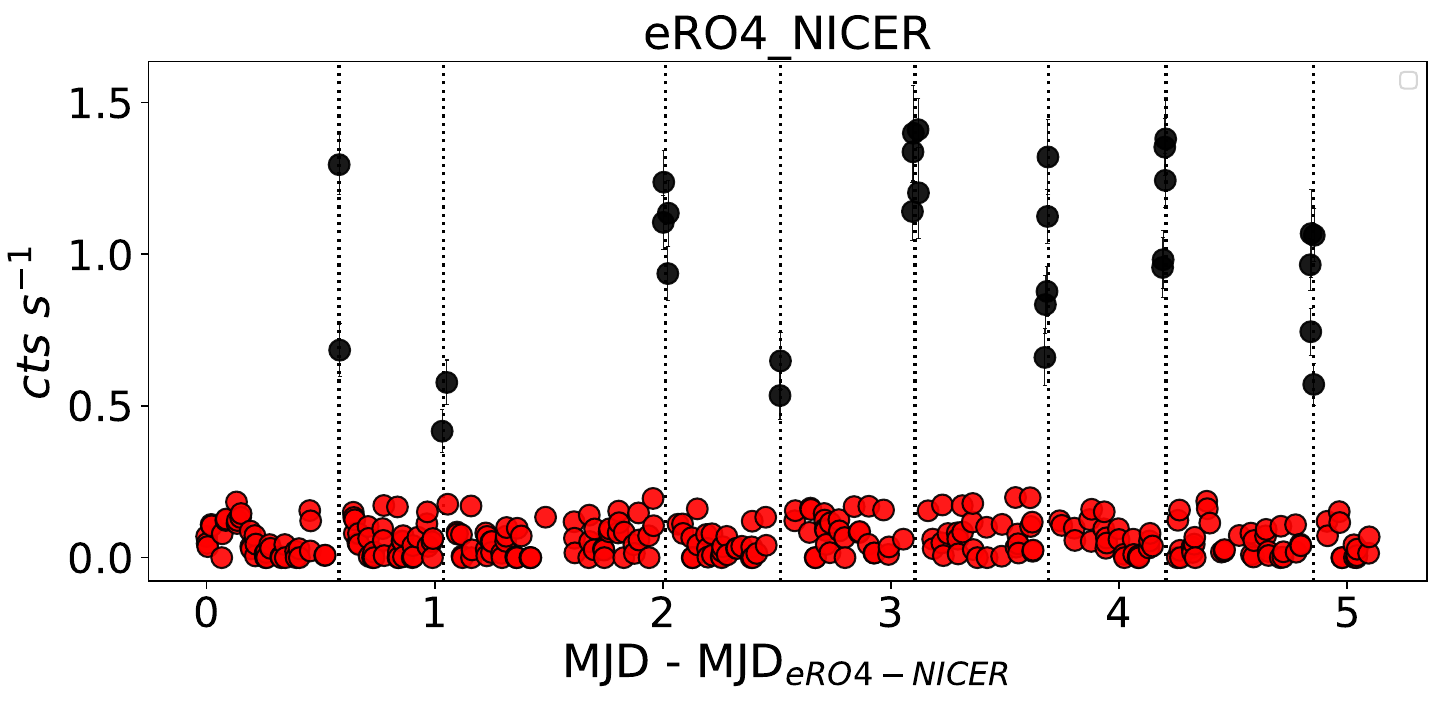}
	\caption{As in Fig.~\ref{fig:eRO3_nicer} but for eRO-QPE4, starting at MJD$\sim59971.637$. Vertical dotted lines represent approximate locations for the eruptions' peak (thus they are not evenly spaced), as a guide for the eye.}
	\label{fig:eRO4_nicer}
    \end{figure}

    eRO-QPE4 was observed for a total of $\sim66$\,ks between 27 January 2023 and 2 February 2023 (Fig.~\ref{fig:eRO4_nicer}), hereafter named eRO4-NICER. The source is significantly detected in a transient fashion, although in most cases the full eruption is not resolved due to the short duration of eruptions ($\approx30$\,min of FWHM, based on \emph{XMM-Newton}, Fig.~\ref{fig:xmm_lc_ero4}) compared to the typical gap in \emph{NICER}'s monitoring. In Fig.~\ref{fig:eRO4_nicer}, vertical dotted lines represent approximate locations for the eruptions, as a guide for the eye. Inferred separations are in the range $\sim(11-15.5)$\,h and appear somewhat regular (considering that an eruption was likely lost in the gap around MJD$_{\rm eRO4-NICER}$ + 1.5 in Fig.~\ref{fig:eRO4_nicer}). However, they do not come in a clear alternating pattern such as in GSN\,069 and eRO-QPE2 (\citetalias{Miniutti+2019:qpe1,Arcodia+2021:eroqpes}; \citealp{Arcodia+2022:ero1_timing}), nor do they show very large scatter such as the eruptions in eRO-QPE1 (\citetalias{Arcodia+2021:eroqpes}; \citealp{Chakraborty+2024:ero1}) or RXJ1301.9+2747 (\citetalias{Giustini+2020:qpe2}; Giustini et al., in prep.), but rather show an intermediate behavior. The peak flux and temperature range from $F_{\rm 0.2-2.0\,keV} = 1.6_{-0.1}^{+0.2} \times 10^{-12}$\,erg\,s$^{-1}$\,cm$^{-2}$ and $kT=121_{-7}^{+6}\,$eV for the lowest peaks, to $F_{\rm 0.2-2.0\,keV} = (2.4 \pm 0.2) \times 10^{-12}$\,erg\,s$^{-1}$\,cm$^{-2}$ and $kT=(118 \pm 5)\,$eV. This is reasonably consistent with \emph{XMM-Newton} (Fig.~\ref{fig:xmm_lc_ero4} and Table~\ref{tab:ero4_spec}), given that not all NICER eruptions are resolved and some peaks have been missed.
        

    \subsection{\emph{Swift}}	
    \label{sec:swift}

    Two serendipitous \emph{Swift-XRT} observations were taken with eRO-QPE3 in the field of view, on 2020-08-25 and 2020-12-26, for an exposure of $\sim584$\,s and $\sim767$\,s, respectively. Aperture photometry was performed at the location of eRO-QPE3 with the Python package \texttt{photutils}. The source aperture adopted was $40$", while background counts were extracted in an annulus with inner and outer radius of $120$" and $360$", respectively. Count rates were converted to fluxes using \texttt{WebPIMMS} and adopting the best-fit models of the closest eROSITA observation. eRO-QPE3 was detected during the first observation, close to eRASS2, at about the same flux level of the eRASS2 bright state, perhaps indicative of a serendipitous QPE observation (see Sec.~\ref{sec:ero3_longterm}). The second observation did not detect the source, at the same depth of the eRASS3 upper limit of the faint phase. eRO-QPE3 was in the UVOT field of view only in the second observations, for a total exposure of $\sim390\,$s with the UVW1 filter at 268\,nm. Performing aperture photometry with the task \texttt{uvotsource}, eRO-QPE3 is only detected at 2.6$\sigma$, hence it is formally undetected. The corresponding flux upper limit is $<1.3 \times 10^{-13}$\,erg\,s$^{-1}$\,cm$^{-2}$, so consistent with the OM-UVW1 detection of $\sim1.0 \times 10^{-13}$\,erg\,s$^{-1}$\,cm$^{-2}$ at 291\,nm.

    We triggered \emph{Swift} ToO follow up observations of eRO-QPE4 after the eRASS4 detection. eRO-QPE4 was also serendipitously observed with \emph{Swift} before the eRASS4 discovery. We used the XRT online data analysis tool\footnote{\url{http://www.swift.ac.uk/user_objects}} \citep{evans_etal2009} to check whether eRO-QPE4 was detected for each observation. The tool was also used to extract the X-ray spectra for observations in which it was detected, and to estimate the $3\sigma$ count rate upper limits for non-detections. The X-ray spectra were rebinned to have at least one count in each bin. An absorbed accretion disk model (\texttt{diskbb}) model was used to calculate the flux and upper limits. Most observations are non detections, shallower than the eROSITA and \emph{XMM-Newton} fluxes, hence uninformative. In between the eRASS4 detection and eRO4-XMM, there are a few \emph{Swift-XRT} detections in the range $F_{\rm 0.2-2.0\,keV}\sim (1-4) \times 10^{-12}$\,erg\,s$^{-1}$\,cm$^{-2}$, which is compatible with the eruptions fluxes observed in eRASS4 and with \emph{XMM-Newton}, thus supporting a serendipitous detection during an eruption. eRO-QPE4 was not in the UVOT FoV of the archival observations. We thus only analyzed the UVOT data for the ToO observations taken after eRASS4. We used the UVOT analysis pipeline provided in \textsc{HEASoft} software (version \texttt{6.31}) with UVOT calibration version \texttt{20201215} to analyze the UVOT data. Source counts were extracted from a circular region with a radius of $5''$ centered at the source position. A $20''$ radius circle from a source-free region close to the position of eRO-QPE4 was chosen as the background region. The task \texttt{uvotsource} was used extract the photometry. The UVOT/UVW1 (UVOT/UVW2) flux is fairly consistent among the several snapshots, with an average flux at $\sim2.8 \times 10^{-13}$\,erg\,s$^{-1}$\,cm$^{-2}$ ($\sim2.4 \times 10^{-13}$\,erg\,s$^{-1}$\,cm$^{-2}$).
    
	\subsection{Radio observations}
    We observed the coordinates of eRO-QPE3 and eRO-QPE4 on two occasions each with the Australia Telescope Compact Array (ATCA) between October 2022 and August 2023 (proposal ID C3513/C3527). All radio data were reduced using standard procedures in the Common Astronomy Software Application \citep[CASA v5.6.3,][]{CASA2022} including flux and bandpass calibration with PKS 1934-638 and phase calibration with PKS 1406-267 (eRO-QPE3), PKS 0458-020 (eRO-QPE4 4\,cm), and PKS 0420-014 (eRO-QPE4 16\,cm). For eRO-QPE3 we observed the target on two occasions separated by approximately 8\,months with the ATCA dual 4\,cm receiver and the CABB correlator  producing 2x2\,GHz of bandwidth split into 2048x1\,MHz channels centered on 5.5\,GHz and 9\,GHz. For eRO-QPE4 we initially observed at 5.5 and 9\,GHz with the dual 4\,cm receiver and then followed up 3\,months later with an ATCA observation with the 16\,cm receiver and the CABB correlator producing 2\,GHz of bandwidth split into 2048x1\,MHz channels centered on 2.1\,GHz. Images of the target field were made with the CASA task \texttt{tclean}. No radio emission was detected at the coordinates of either QPE source in any of the observations. A summary of the ATCA observations including the 3$\sigma$ flux density upper limits is given in Table \ref{tab:ATCAobs}. The radio non-detections of eRO-QPE3 and eRO-QPE4 indicate it is unlikely the host galaxies have strong AGN activity in the nuclei, in agreement with the optical proxies.

    \begin{table}[]
        \centering
        \caption{Summary of ATCA radio observations of eRO-QPE3 and eRO-QPE4.}
        \begin{tabular}{clll}
        \hline
             Date & Array config. & Frequency & Flux Density \\
              & &  (GHz) & ($\rm{\mu}$Jy) \\
             \hline
             eRO-QPE3 \\
             \hline
             2022-Oct-01 & 6D & 5.5 & $<54$ \\
             2022-Oct-01 & 6D & 9 & $<49$ \\
             2023-Jun-13 & 6D & 5.5 & $<51$ \\
             2023-Jun-13 & 6D & 9 & $<49$ \\
             \hline
             eRO-QPE4 \\
             \hline
             2023-May-05 & 1.5A & 5.5 & $<48$ \\
             2023-May-05 & 1.5A & 9 & $<39$ \\
             2023-Aug-20 & 6D & 2.1 & $<298$ \\
             \hline
             
        \end{tabular}
        \label{tab:ATCAobs}
    \end{table}

 
    \section{Spectral evolution of QPEs}
    \label{sec:energy_evol}

    \begin{figure*}[tb]
		\centering
		\includegraphics[width=\columnwidth]{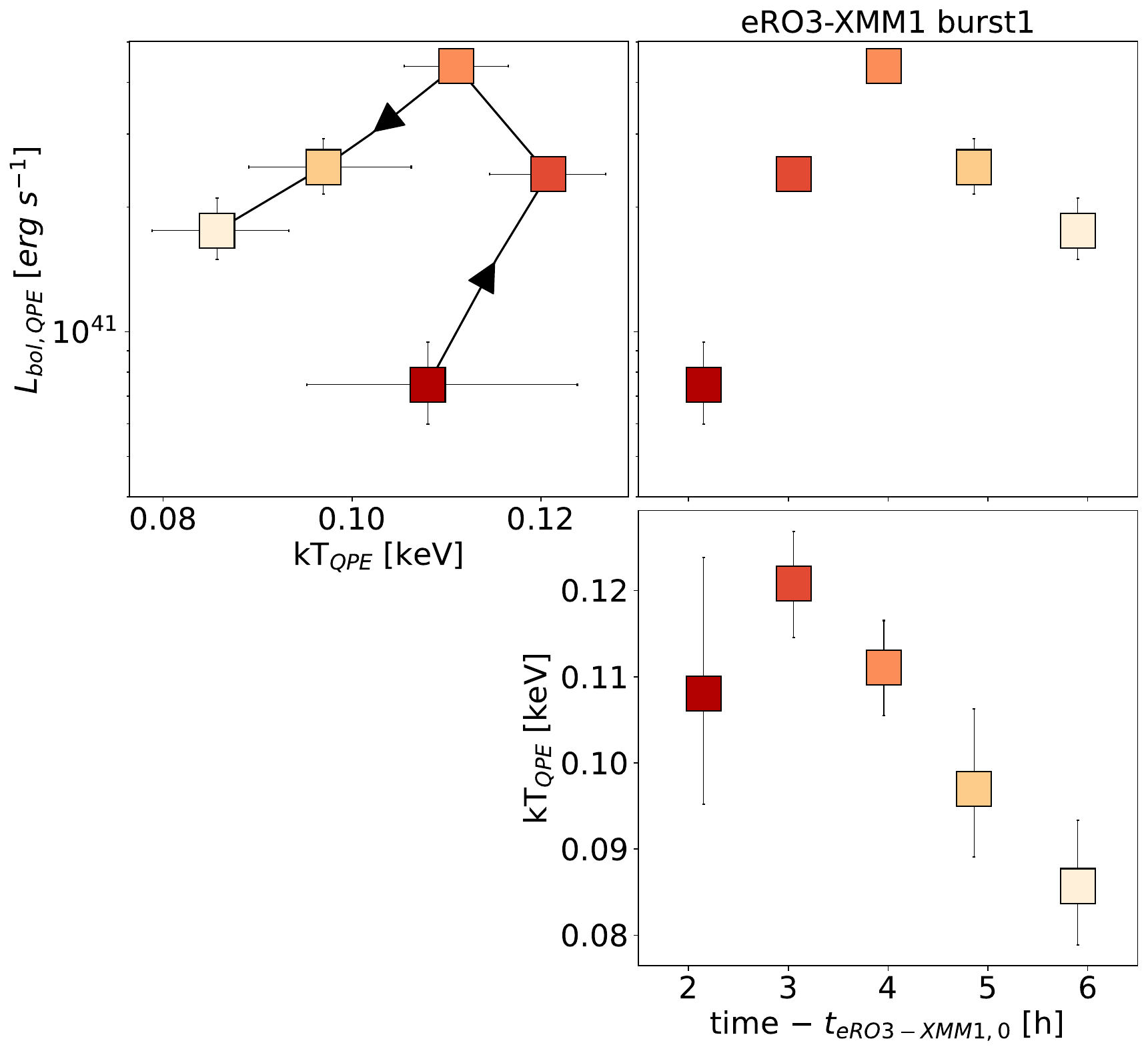}
        \includegraphics[width=\columnwidth]{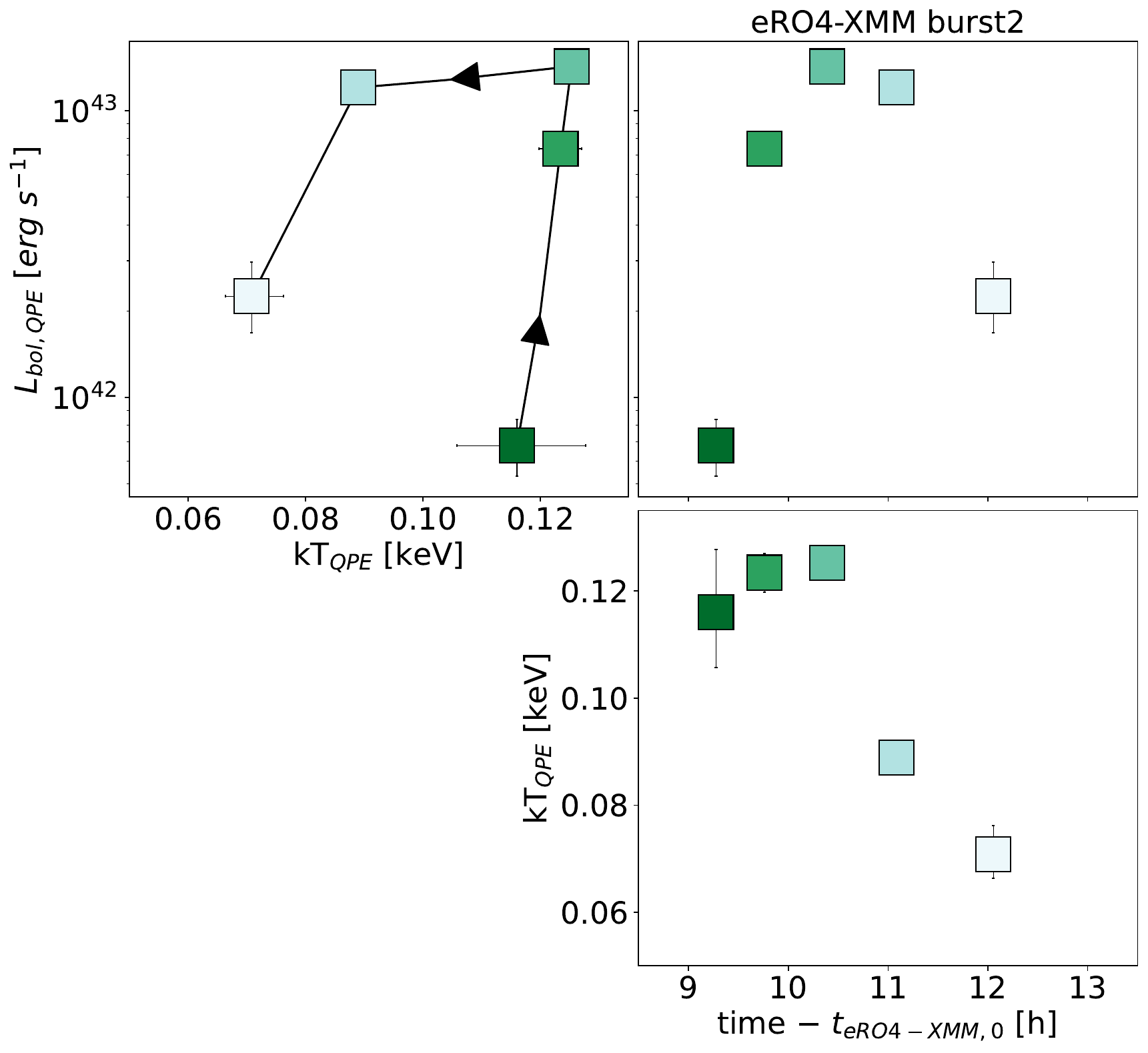}
		\caption{Spectral evolution of QPEs in eRO-QPE3 (\emph{left}) and eRO-QPE4 (\emph{right}). The top-right (bottom) subpanels show the luminosity (temperature) evolution with time of the QPE component. Darker to lighter colors represent the evolution from start to end of the eruptions. The first eruptions of the eRO3-XMM1 observation and the second of the eRO4-XMM observation are taken as reference for the QPEs in eRO-QPE3 and eRO-QPE4, respectively. The top-left subpanels show the luminosity-temperature coevolution, analogous to the rate-hardness evolution in \citet{Arcodia+2022:ero1_timing} which first reported the hysteresis cycle.
  }
		\label{fig:energyevol}
	\end{figure*}

    Given the growing number of repeating X-ray transients, it is useful to identify common properties that indicate a common emission process, or suggest otherwise. For instance, \citet{Arcodia+2022:ero1_timing} first reported the specific spectral evolution of QPEs in eRO-QPE1 as a soft X-ray spectrum which, for the same count rate or flux, is harder during the rise than during the decay. Moreover, the peak hardness of the spectrum is reached before the peak flux of the eruption. The same behavior was identified in GSN\,069 \citep{Miniutti+2023:gsnrebr} and RX J1301.9+2747 (Giustini et al., in prep.). 
    Identifying this hysteresis cycle in a repeating X-ray transient is the smoking gun of the same emission process. This is particularly important since the timing properties of the known QPEs are quite diverse (\citetalias{Miniutti+2019:qpe1,Giustini+2020:qpe2,Arcodia+2021:eroqpes}; \citealp{Arcodia+2022:ero1_timing}), and can even significantly change in the same source \citep{Miniutti+2023:gsnreapp,Miniutti+2023:gsnrebr}.    

    We show the spectral evolution of eRO-QPE3 and eRO-QPE4 in Fig.~\ref{fig:energyevol}. We choose model-dependent quantities such as luminosity and temperature, compared to pure data-driven quantities such as hardness-ratio and count rate. This is because, at least in eRO-QPE4, the quiescence component needs to be decomposed \citep[e.g., as done for GSN\,069,][]{Miniutti+2023:gsnrebr}. This was not required for eRO-QPE1 \citep{Arcodia+2022:ero1_timing} as the source is not detected between the eruptions. Therefore, Fig.~\ref{fig:energyevol} assumes that QPEs can be modeled by a black body. Results would be identical with any model with an exponential decay due to a characteristic temperature, for instance Bremsstrahlung or an accretion disk. In Fig.~\ref{fig:energyevol}, the first eruption of eRO3-XMM1 and the second of eRO4-XMM are used as representative, although other eruptions are shown in Fig.~\ref{fig:energyevol_backup}. The eruption evolves from darker to lighter colors of the respective color maps, as shown in Fig.~\ref{fig:eRO3_lcphases}. The eruptions were divided in 5 epochs with the goal to compare rise and decay at similar count rate or luminosity. As there is not a one-to-one relation between count rate and luminosity, this is not necessarily achieved. As a matter of fact, as much as the $L_{\rm bol}$ and $kT$ values would change slightly depending on the definition of the epochs, the hysteresis behavior in the $L_{\rm bol}-kT$ plane (or count rate versus hardness ratio) would remain evident. This is confirmed by the same pattern in other eruptions of both eRO-QPE3 and eRO-QPE4 (Fig.~\ref{fig:energyevol_backup}) and other QPE sources \citep{Arcodia+2022:ero1_timing,Miniutti+2023:gsnrebr}. We also note that, similarly to eRO-QPE1 \citep{Arcodia+2022:ero1_timing} and GSN\,069 \citep{Miniutti+2023:gsnrebr}, the peak temperature is reached before the peak luminosity.

    As done in \citet{Miniutti+2023:gsnrebr}, we estimate the size of this emitting region, under the assumption that it is indeed black body emission. We assume that the observer sees half of a radiating spherical surface. For eRO-QPE3 (Fig.~\ref{fig:energyevol}, left) we obtain radii of $\sim1.2 \times 10^{10}\,$cm$^{2}$, $\sim1.2 \times 10^{10}\,$cm$^{2}$, $\sim1.7 \times 10^{10}\,$cm$^{2}$, $\sim1.6 \times 10^{10}\,$cm$^{2}$ and $\sim1.8 \times 10^{10}\,$cm$^{2}$ for rise1, rise2, peak, decay1 and decay2, respectively. Therefore, we infer an increase in size of a factor $\sim1.5$ from start to end of the eruptions, lower than what was inferred for GSN\,069 \citep{Miniutti+2023:gsnrebr}. For eRO-QPE4 (Fig.~\ref{fig:energyevol}, right), we obtain radii of $\sim2.4 \times 10^{10}\,$cm$^{2}$, $\sim7.0 \times 10^{10}\,$cm$^{2}$, $\sim9.5 \times 10^{10}\,$cm$^{2}$, $\sim17.3 \times 10^{10}\,$cm$^{2}$ and $\sim11.6 \times 10^{10}\,$cm$^{2}$ for rise1, rise2, peak, decay1 and decay2, respectively. Therefore, we infer an increase in size of a factor $\sim7$ from rise1 to decay1 and of a factor $\sim5$ from rise1 to decay2, which is larger than what was inferred for GSN\,069 \citep{Miniutti+2023:gsnrebr}. Regardless of the fine details which are deferred to future homogeneous work extended to all QPE sources, the common thread is that during the $L_{\rm bol}-kT$ hysteresis the size of the emitting region increases, which is qualitatively consistent with the latest models of disk-orbiter collisions \citep{Franchini+2023:qpemodel,Linial+2023:qpemodel2,Tagawa+2023:qpemodel}. The absolute numbers are, of course, model-dependent and any departure from this spectral model would likely change the absolute values of the size, but not the qualitative evolution during the eruptions, provided the X-ray spectrum is indeed the exponential tail of a thermal spectrum.
	
	\section{Long-term evolution}

    \subsection{eRO-QPE3: Quiescent flux decays until disappearance}
    \label{sec:ero3_longterm}

    \begin{figure}[tb]
		\centering
		\includegraphics[width=0.99\columnwidth]{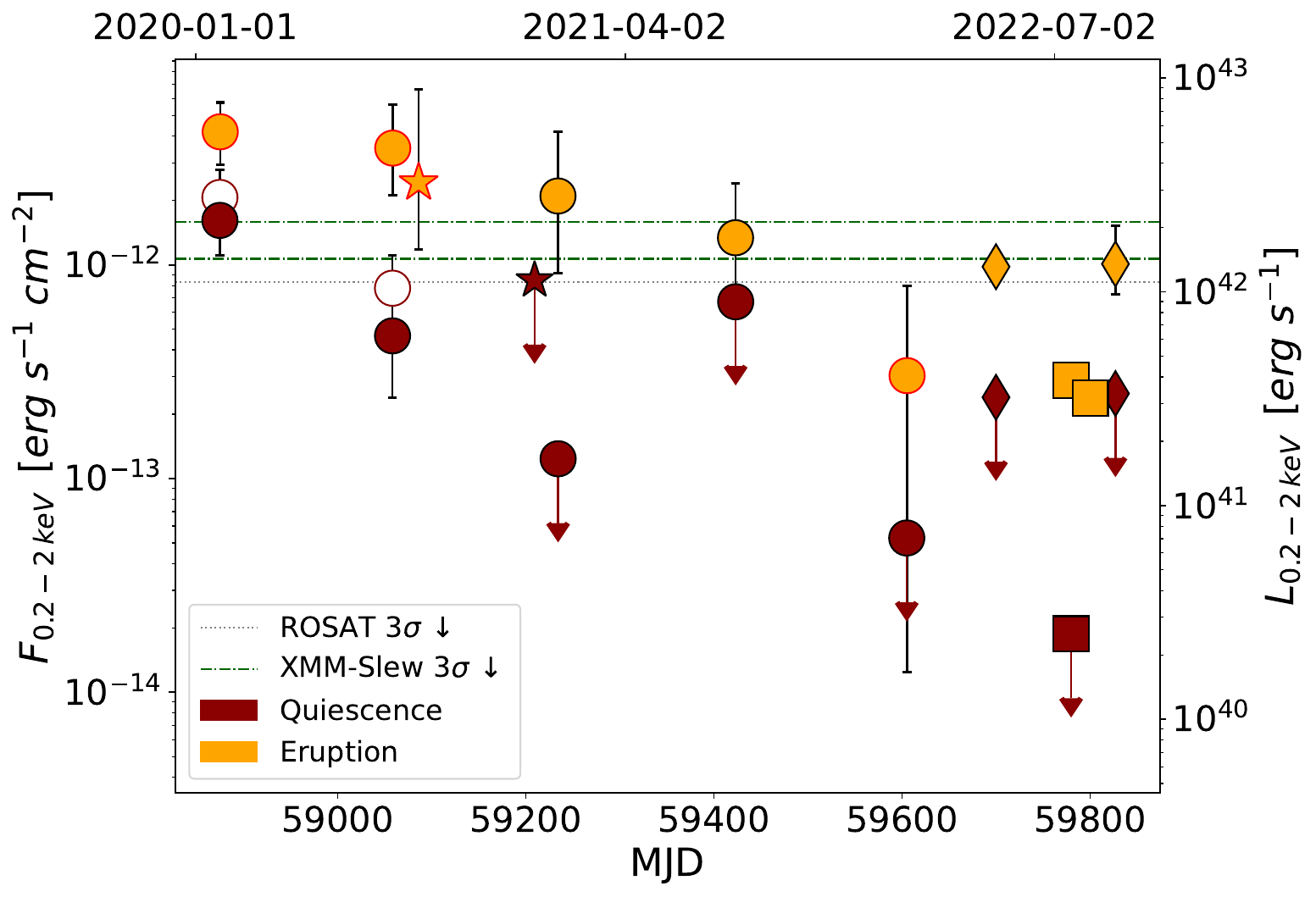}
		\caption{Long-term evolution of eRO-QPE3, separating flux states between quiescence and eruption (which includes the former). eROSITA points are shown as circles, squares for \emph{XMM-Netwon}, diamonds for \emph{NICER} (the brightest eruptions in the light curves of Fig.~\ref{fig:eRO3_nicer}) and stars for \emph{Swift-XRT}. All uncertainties are $3\sigma$. Red contours on orange points highlight observations in which the identification of an eruption is less secure. White symbols for eRASS1 and eRASS2 correspond to the flux of the full observation without separating different states. Horizontal lines highlight \emph{ROSAT} and \emph{XMM-Newton} archival upper limits, as stated in the legend.}
		\label{fig:ero3_longterm}
	\end{figure}

    In Fig.~\ref{fig:ero3_longterm}, we show the long-term evolution of eRO-QPE3 inferred with eROSITA, \emph{XMM-Newton}, \emph{NICER} and \emph{Swift-XRT}. We separate flux states between quiescence (dark red data points) and eruption (orange). This separation is straight-forward in \emph{XMM-Newton} and \emph{NICER} data and at least in eRASS3 and eRASS4 data (see Fig.~\ref{fig:eRO3_erass_lc}). In eRASS1 and eRASS2, since the quiescence is detected and given eROSITA's sampling, the identification of the eruption state is informed by the presence of QPEs at later times. However, formally one cannot exclude high-amplitude variability of the quiescent state with no eruptions. This potential no-QPE flux state is shown in white with dark red contours and the related possible eruption states are shown with red contours on orange points. For eRASS5, we consider the eruption state as ambiguous since the variability is not as significant and the flux chain has a tail extending to faint fluxes, therefore it is formally not well constrained at $3\sigma$ (hence the large uncertainty in Fig.~\ref{fig:ero3_longterm}). The \emph{Swift-XRT} detection is also ambiguous: its flux is as bright as the eruption state in eRASS2, but no variability can be inferred from the few hundred seconds of exposure to unambiguously identify it as an eruption. Despite these caveats, throughout the rest of the discussion we assume that all the orange points caught eRO-QPE3 in eruption, and the dark red points represent the flux of the intra-QPEs' quiescence.
    
    Quite interestingly, eRO-QPE3 showed a disappearing quiescence component (Fig.~\ref{fig:eRO3_erass_lc} and Table~\ref{tab:erass1_spec}), being detected in eRASS1 and eRASS2 and never since, including \emph{Swift-XRT} and eRASS3 observations taken within 6 months. This is consistent with what has been observed for GSN\,069 \citep{Shu+2018:gsn,Miniutti+2023:gsnrebr}, albeit over a decade and with a much better data coverage before the QPEs' discovery. Furthermore, the two QPE candidates presented in \citet{Chakraborty+2021:qpe5cand} and \citet{Quintin+2023:tormund} also showed evidence of eruptions on top of a decaying baseline flux. In general, a possible interpretation of the decaying quiescence in GSN\,069 and the two candidates is that the accretion flow was induced by a TDE. The presence of a precursor TDE is secure in AT2019vcb \citep[``Tormund'';][]{Quintin+2023:tormund}, but since only the possible start of an X-ray eruption was observed, the association between QPEs and TDEs is still open. Were this connection to be unambiguously proven, one can interpret the long-term evolution of eRO-QPE3 in a similar way. However, the available data (only two epochs are detected in eRO-QPE3) do not allow us to achieve a precise constraint or prediction on this long-term evolution. 
    We can test whether the spectral evolution between the disk component in eRASS1 and eRASS2 is consistent with a cooling thin disk. The disk temperature decreased from $\sim100\,$eV to $\sim89\,$eV from eRASS1 to eRASS2 (Table~\ref{tab:erass1_spec}). Approximating the disk emission as a black body with constant area with the temperature of the inner radius, the eRASS2 bolometric emission would be $\sim(89/100)^4\sim0.6$ times that of eRASS1. The disk bolometric emission of eRASS1 and eRASS2, estimated from the spectral fits, is $L_{\rm bol} = 4.1_{-2.3}^{+1.6}\times10^{42}\,$erg\,s$^{-1}$ and $L_{\rm bol}\sim8.4_{-8.3}^{+4.7}\times10^{41}\,$erg\,s$^{-1}$, respectively. This corresponds to a decrease to $\approx20\%$ of the eRASS1 luminosity, although given the large $1\sigma$ uncertainties of the eRASS2 value, a 60$\%$ decrease is still compatible. Regardless of the exact type and nature of the observed decay, eRO-QPE3 is the first QPE source in which the intra-QPE quiescence is detected at some earlier phases and never again. Other secure QPE sources have either always shown it, or never (\citetalias{Miniutti+2019:qpe1,Giustini+2020:qpe2,Arcodia+2021:eroqpes}). Hence, eRO-QPE3 would bridge the gap between QPE sources that always showed a (possibly time-evolving) quiescence spectrum in between QPEs \citep[e.g., GSN\,069,][]{Miniutti+2023:gsnrebr} and QPE sources which never showed a quiescence spectrum (e.g., eRO-QPE1, \citetalias{Arcodia+2021:eroqpes}). This suggests that the absence of intra-QPE quiescence in some QPE sources is merely observational, as it might have faded compared to an earlier epoch. 

    Furthermore, we also show in Fig.~\ref{fig:ero3_longterm} $3\sigma$ archival upper limits from ROSAT (from 1991) and the \emph{XMM-Newton} Slew Survey (from 2000 and 2013; \citealp{Saxton+2008:slew}). Comparing these with the eRASS1 flux (both if the quiescence includes or not the putative eruption), we notice that ROSAT would have detected the source if it were as bright as eRASS1. Therefore, the accretion flow of eRO-QPE3 likely became brighter (e.g., perhaps radiatively efficient) in the soft X-rays some time between 1991 and 2020.	

    \begin{figure}[tb]
		\centering
		\includegraphics[width=0.99\columnwidth]{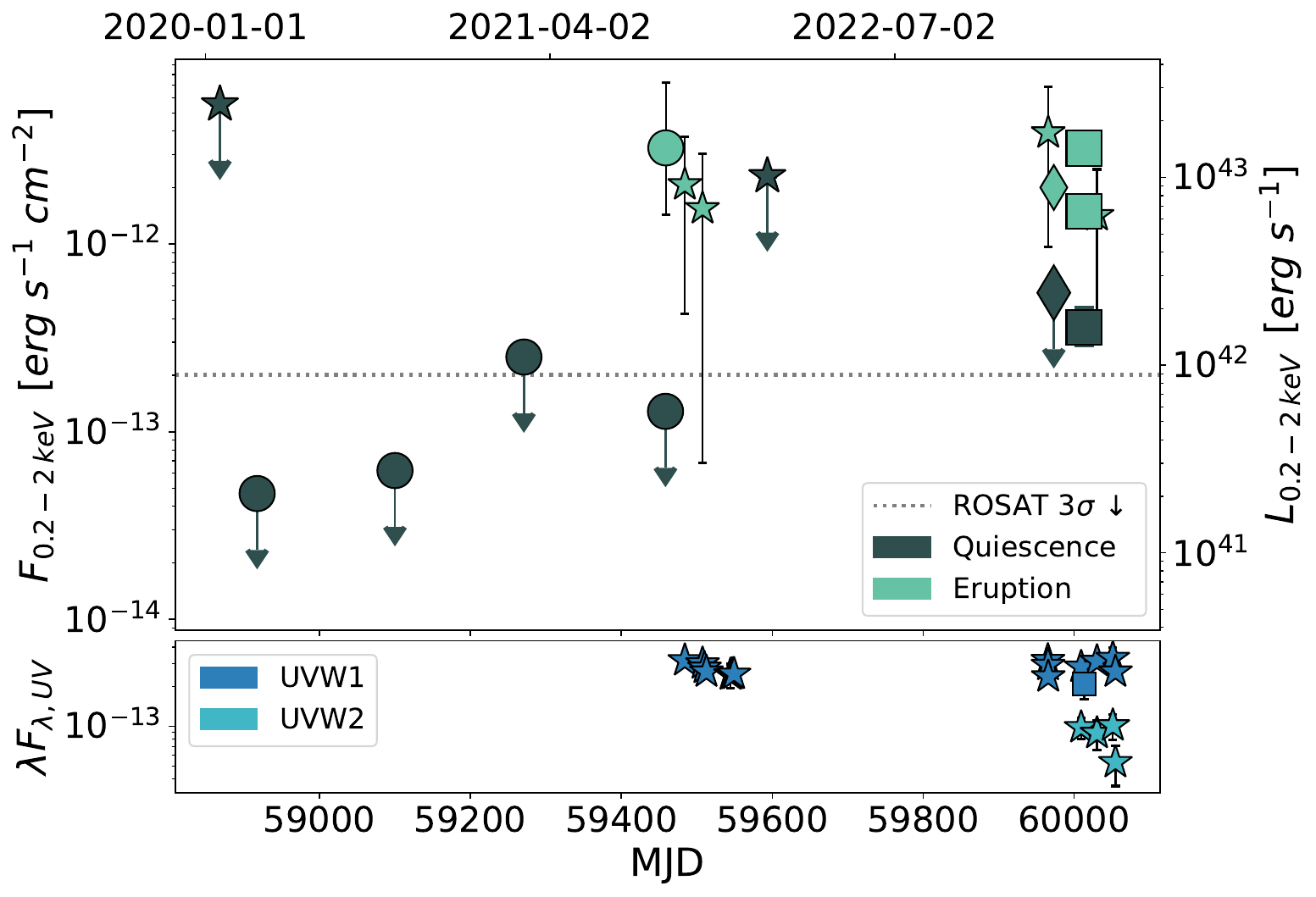}
		\caption{Long-term evolution of eRO-QPE4, separating flux states between quiescence and eruption. eROSITA points are shown as circles, squares for \emph{XMM-Netwon}, diamonds for \emph{NICER} and stars for \emph{Swift-XRT}. \emph{Swift-XRT} and \emph{NICER} eruptions points are considered such based on their flux level similar to the range shown by \emph{XMM-Netwon}. All uncertainties are $3\sigma$. The lower-panel shows \emph{Swift-UVOT} data with the UVW1 (light blue) and UVW2 filter (azure).}
		\label{fig:ero4_longterm}
	\end{figure}

    Despite \emph{XMM-Newton} data and most \emph{NICER} data (unless there are large gaps in the monitoring) allow the identification of the peak, the eROSITA and \emph{Swift-XRT} data points would catch the source in an unknown part of the eruptions. Therefore, all data points except \emph{XMM-Newton} and \emph{NICER} are lower limits to the peak. Without \emph{NICER} monitoring covering several burst cycles, one might have concluded that QPEs must have significantly faded over only a few months. However, \emph{NICER} data taken both before and after eRO3-XMM1 and eRO3-XMM2 (which showed $F_{\rm 0.2-2.0\,keV}\sim3.1 \times 10^{-13}$\,erg\,s$^{-1}$\,cm$^{-2}$) unveil the presence of some eruptions significantly brighter ($F_{\rm 0.2-2.0\,keV}\sim1.0 \times 10^{-12}$\,erg\,s$^{-1}$\,cm$^{-2}$) than the others, which are instead compatible with the eRO3-XMM1 peak flux. This indicates that whilst on average the peak flux might still be overall decreasing, current data also unveil the presence of significant diversity in amplitudes, as seen in other QPE sources \citepalias{Giustini+2020:qpe2,Arcodia+2021:eroqpes}. We show the flux of these brigthest eruptions in Fig.~\ref{fig:eRO3_nicer}. Finally, we note that no significant optical variability is apparent within the available public datasets from optical sky monitors \citep[e.g., ASAS-SN,][]{Shappee+2014:ASASSN} covering the epochs shown in Fig.~\ref{fig:ero3_longterm}.
    
    \subsection{eRO-QPE4: quiescent flux appears or brightens after the eruptions}
    \label{sec:ero4_longterm}

    Fig.~\ref{fig:ero4_longterm} shows the long-term evolution of the X-ray and UV flux of eRO-QPE4. The UV flux is overall constant within the available exposuress. On the contrary, the X-ray quiescence flux (dark gray points) is only constrained by \emph{XMM-Newton} with a detection. However, it is interesting to note that both the archival ROSAT (dotted line) and eROSITA 3$\sigma$ upper limits suggest that the quiescent accretion disk must have been much fainter or absent at the ROSAT/eROSITA epochs. Current data surely rule out the presence of a radiatively efficient accretion flow, as bright as that seen by \emph{XMM-Newton}, prior to the eRASS4 QPE discovery. This would be in agreement with many other QPE sources with no evidence of bright X-ray sources much earlier than the onset of QPEs. However, we are not able to constrain the start of the QPE behavior. Regarding the peaks of the eruptions, the eROSITA, \emph{Swift-XRT} and \emph{NICER} fluxes are compatible with the range spanned by the faintest and brightest eruption in eRO4-XMM (Fig.~\ref{fig:xmm_lc_ero4}). Finally, we note that similarly to eRO-QPE3 no significant optical variability is observed \citep[e.g., ASAS-SN and ZTF;][]{Shappee+2014:ASASSN,Bellm+2019:ztf} during the epochs shown in Fig.~\ref{fig:ero4_longterm}.

	\section{Discussion}
	\label{sec:disc}
	
	\subsection{QPEs in preexisting AGN?}


    The X-ray properties of all QPE sources suggest that the nucleus is currently active, as the soft X-ray spectrum observed in quiescence is indicative of radiatively efficient accretion (\citetalias{Miniutti+2019:qpe1,Giustini+2020:qpe2,Arcodia+2021:eroqpes}, \citealp{Chakraborty+2021:qpe5cand}). In eRO-QPE4 the quiescence emission was a few times fainter, or nonexistent, prior to the QPE discovery (Fig.~\ref{fig:ero4_longterm}). This would support a transient nature for the quiescent accretion disk and disfavor a long-lived radiatively efficient accretion flow. However, in other QPE sources it is much harder to say whether or not this was the case. In fact, most QPE sources do not have constraining archival X-ray observations prior to the current activity phase, and results from the radio and optical bands are currently ambiguous. For instance, eRO-QPE2 is detected as a faint radio source, although its radio emission is consistent with that predicted from star-formation alone (Arcodia et al., in prep.). All other eROSITA QPE sources are undetected in radio (e.g., this work), which also rules out a preexisting powerful AGN in these nuclei and agrees with the lack of a canonical hard and bright X-ray power-law component in QPE sources. From the optical band, given the lack of broad lines detected, we do not have information on the recent nuclear activity. Narrow-line diagnostics place the select number of QPE sources in different places of the typical diagnostic diagrams including regions indicating the need of a nuclear ionizing source \citep[e.g., an AGN,][]{Wevers+2022:hosts}. However, these diagnostics cannot securely determine whether these lines were excited by the current activity epoch, while spatially resolved observations will help solving this ambiguity. As a matter of fact, GSN\,069 was found to be evidently powered by an AGN using narrow-line diagnostics \citep{Wevers+2022:hosts}. However, recent work by \citet{Patra+2023:oiii} making use of archival HST data of GSN\,069 found evidence of a compact nuclear [OIII] region, which indicates that the accretion activity is only $\sim10-100\,$y old. More radially extended [OIII] regions, which are not indicative of the current accretion activity, would instead contaminate typical spectra which populate the narrow-line diagnostic diagrams. In absence of a comparable systematic study with spatially resolved observations on all QPE sources, simplistic narrow-line diagnostics have to be interpreted with caution. Significant progress can be made in the near future using integral-field unit data, and high angular resolution spectroscopic and photometric observations, to understand the spatial distribution of the gas in these galaxies and how it relates to the current QPE activity.

	\subsection{What is new on QPEs based on these discoveries}

    \begin{figure}[tb]
	\centering
    \includegraphics[width=0.99\columnwidth]{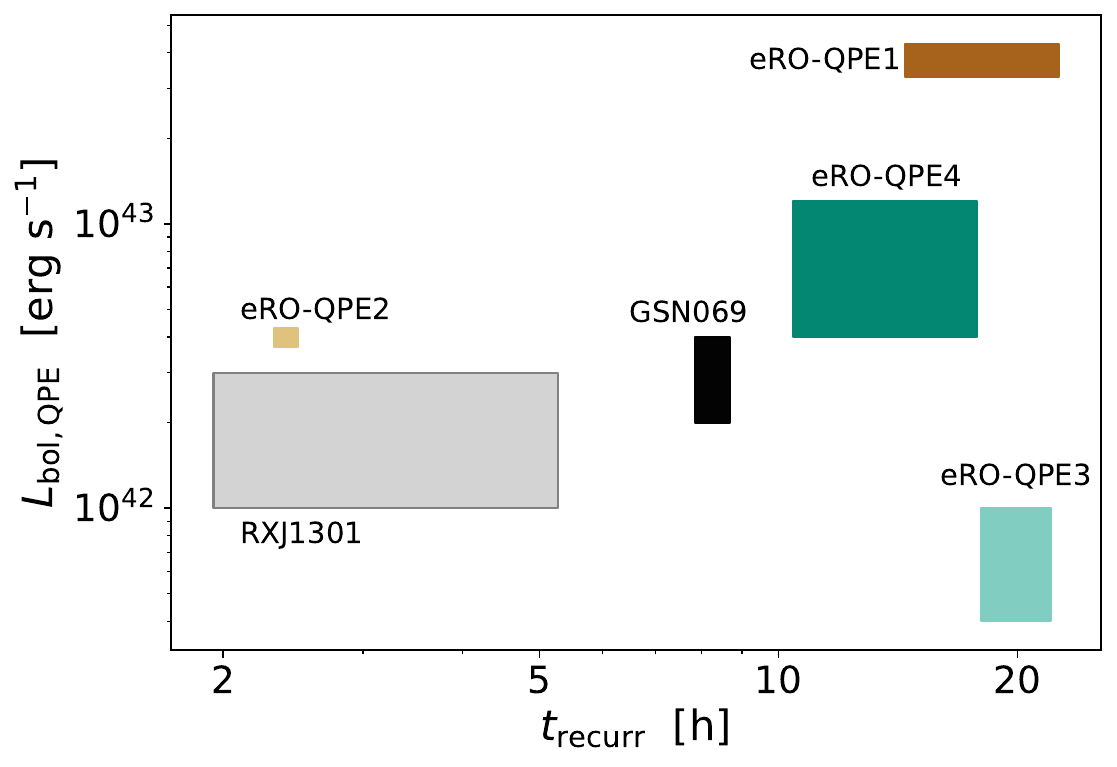}
    \includegraphics[width=0.95\columnwidth]{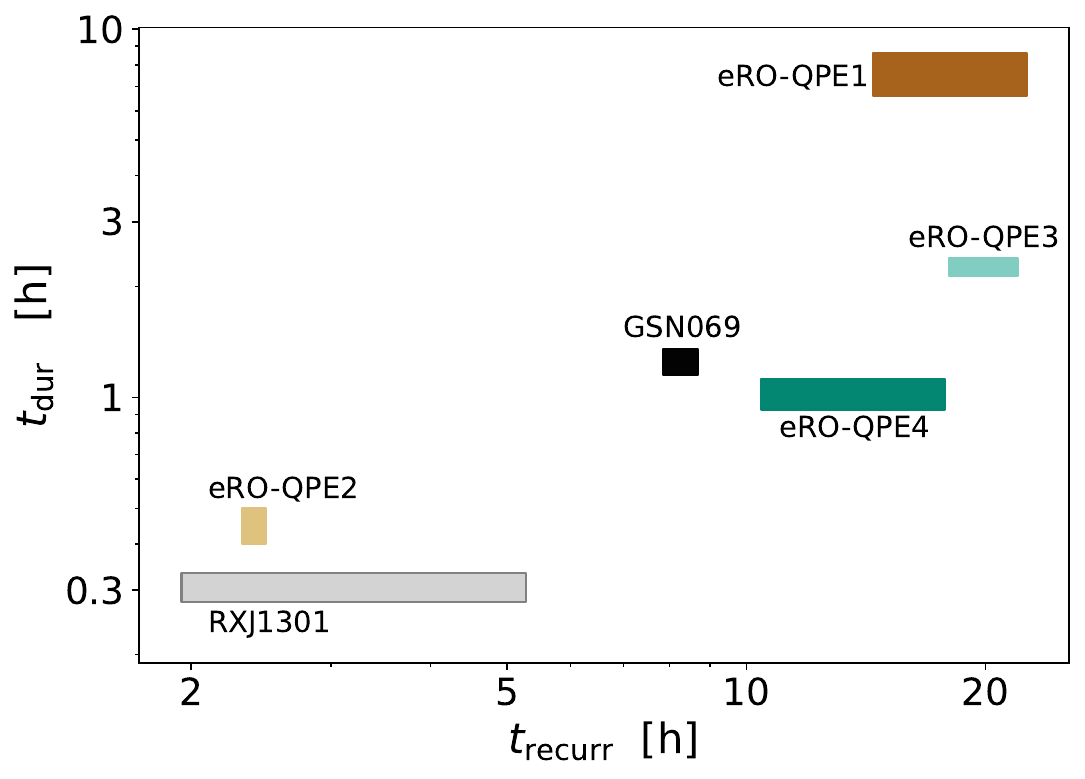}
	\caption{Relation between recurrence and peak luminosity (\emph{top}) and recurrence  and duration timescales (\emph{bottom}) in secure QPE sources, shown as labeled. The extent of shaded areas represents the standard deviation, when available, or the span within observed values in a given source. eRO-QPE3 is a clear outlier of the top panel, ruling out the presence of a predictive correlation between timescales and peak luminosity.}
	\label{fig:correlations}
    \end{figure}

    The discovery of QPEs on top of a decaying X-ray continuum in eRO-QPE3 (Fig.~\ref{fig:ero3_longterm}) offers a new piece of evidence that strengthens the connection between QPEs and previous TDEs in the same nucleus \citep{Shu+2018:gsn,Chakraborty+2021:qpe5cand,Sheng+2021:tdegsn,Miniutti+2023:gsnrebr,Quintin+2023:tormund}. Moreover, eRO-QPE3 showed the faintest peak luminosity (average of $L_{\rm 0.5-2.0\,keV}^{\rm peak}\sim9.7\times10^{41}\,$erg\,s$^{-1}$ in the eRASS1-4 epoch, down to $\sim4.2\times10^{41}\,$erg\,s$^{-1}$ in the eRO3-XMM1 epoch) and the highest recurrence times ($\approx20\,h$) among secure QPEs, therefore invalidating any strict relation between peak luminosity and recurrence time across the QPE population. In fact, eRO-QPE1 shows similar recurrence times but a luminosity up to 10-100 times higher \citepalias{Arcodia+2021:eroqpes}. We show this in the top panel of Fig.~\ref{fig:correlations}, where the extent of shaded areas represents the standard deviation, when available, or the span within observed values in a given source (\citetalias{Miniutti+2019:qpe1,Giustini+2020:qpe2,Arcodia+2021:eroqpes}; \citealp{Arcodia+2022:ero1_timing,Miniutti+2023:gsnreapp,Miniutti+2023:gsnrebr}; \citealp{Chakraborty+2024:ero1}; Giustini et al., in prep.). This lack of correlation disfavor accretion-based QPE emission models, in which both peak luminosity and accretion timescales would depend on the black hole mass. Instead, for EMRI models, the somewhat more diverse orbital parameters (semi-major axis, eccentricity) and the orbiter's mass and nature (i.e., stars of different kind, degenerate stars and/or black holes) would more easily accommodate scenarios in which the peak luminosity and average recurrence are decoupled. Conversely, as shown in previous work \citep{Chakraborty+2021:qpe5cand,Guolo+2023:swift} recurrence and duration timescales do appear to be somewhat correlated (bottom panel of Fig.~\ref{fig:correlations}), although we refrain from attempting to fit any scaling due to the low number of sources and the lesson learned from the top panel of Fig.~\ref{fig:correlations}. Nonetheless, for the sake of designing a somewhat informed follow-up monitoring based on a single observed eruption \citep[e.g.,][]{Quintin+2023:tormund}, this tentative relation may be used based on the current QPE sources.

    eRO-QPE4 shows evidence of a harder component appearing close to the QPE peak (Fig.~\ref{fig:eRO4XMM_spec_peak}), although at this stage it is unclear what its origin might be (Appendix~\ref{sec:xmm_spec}). Modeling it as Comptonization of the QPE thermal continuum, we infer a flux ratio of $\sim 1:94$ in the $0.2-2.0\,$keV band compared to the latter. However, this is merely a possible interpretation, as after all the inferred slope is rather unconstrained and its signal very weak. Nonetheless, we note that until now, a harder component was only seen in quiescence (and not ubiquitously, \citetalias{Miniutti+2019:qpe1,Giustini+2020:qpe2,Arcodia+2021:eroqpes}; \citealp{Chakraborty+2021:qpe5cand}), but never during the eruptions, let alone in a transient fashion.

    In eRO-QPE4 we see clearly, for the first time, that the quiescence emission as detected during the QPE epoch must have been either absent or much fainter previous to the QPE behavior (Fig.~\ref{fig:ero4_longterm}). What is particularly intriguing is that in eRASS4, when the first QPE was observed, the quiescence emission was also much fainter than the eRO4-XMM epoch. Given the lower sensitivity with \emph{Swift-XRT} and \emph{NICER}, their upper limits are shallower than the eRO4-XMM level therefore we are not able to constrain how fast the brightening was.

    The long-term emission of both eRO-QPE3 and eRO-QPE4 can only be constrained qualitatively. In the former case the fainter peak flux, close to \emph{NICER}'s background level, prevents us from securely resolving the fainter eruptions, which appear to be more frequent than the bright ones around the \emph{NICER} and \emph{XMM-Newton} epochs (Fig.~\ref{fig:xmm_lc} and~\ref{fig:eRO3_nicer}). In eRO-QPE4 eruptions have a FWHM of $\approx30\,$min, which is on the order of the typical separation in a \emph{NICER} monitoring given its $\sim90\,$-min orbit. Nonetheless, based on \emph{NICER} data we are able to identify the timing properties of eRO-QPE4 as somewhat intermediate compared to the other known QPEs. The eruptions are not strictly regular or in alternation, as from Fig.~\ref{fig:eRO4_nicer} we roughly constrain the last five uninterrupted recurrences to be $\sim12.1$\,h, $\sim14.2$\,h, $\sim14.0$\,h, $\sim12.4$\,h and $\sim15.5$\,h, combined with the $\sim9.8\,$h and $\sim14.7\,$h separations seen by \emph{XMM-Newton} (Fig.~\ref{fig:xmm_lc_ero4}). They are somewhat similar to the irregular recurrences in eRO-QPE1 (\citetalias{Arcodia+2021:eroqpes}; \citealp{Chakraborty+2024:ero1}) or RXJ1301.9+2747 (\citetalias{Giustini+2020:qpe2}; Giustini et al., in prep.), although in eRO-QPE4 they do not show burst superposition such as in eRO-QPE1 \citep{Arcodia+2022:ero1_timing}, nor closely spaced eruptions such as RXJ1301.9+2747 (Giustini et al., in prep.). Hence, eRO-QPE4 shows intermediate properties (between regular and irregular arrival times) confirming what the latest orbital models suggest \citep[e.g.,][]{Linial+2023:qpemodel2,Franchini+2023:qpemodel}, namely that precession would induce different timing behavior in a continuum from regular (no disk precession, lower eccentricity) and irregular (disk and orbital precession, higher eccentricity).
    
    We also note that in eRO-QPE3, in which the bursts recur every $\approx20$\,h, the asymmetry (with a fast rise and slow decay) appears evident (Fig.~\ref{fig:eRO3_lcphases}, top), whereas in eRO-QPE4 bursts are consistent with being symmetric within the current statistics (Fig.~\ref{fig:eRO3_lcphases}, bottom). Interestingly, the QPE sources with the clearest asymmetry (eRO-QPE1 and eRO-QPE3) are also the ones with undetected quiescence. 

    The two discoveries in this work confirm the trend of finding QPEs from low-mass galaxies ($\log M_*\approx 9-10$) and low-mass black holes ($\log M_{\rm BH}\approx 5-7$), where the high-mass end has been now pushed to larger values by eRO-QPE4. We remind readers that the eROSITA search applied to discover the first four eROSITA QPEs is blind in terms of the host galaxy nature, as we only select extragalactic sources with {\ Gaia} without selecting specific galaxy counterparts. This might be a selection effect due to observational and model-dependent reasons. An observational bias could be that for more massive galaxies and black holes, the Wien tail of the accretion flow would be shifted out of band. In this case, if the temperature of the QPE component is somewhat dependent on the temperature of the accretion flow, the QPE emission would also be softer and currently unobservable. So far, the observed disk and QPE temperatures have spanned a fairly consistent range within all the known QPE sources, although the absence of harder eruptions is remarkable, despite being easily observable. Larger sample statistics on the QPE population is required for more conclusive statements. In addition, for partial TDEs and stellar EMRI models the detectability of QPEs is limited to smaller galaxies and MBHs for which the tidal radius is not within the innermost stable circular orbit \citep[e.g.,][]{Hills1975:tde,vanvelzen2018:ratesuppr,King+2023:lowmass,Linial+2023:qpemodel2}.
						
\section{Summary and prospects}

We report on the discovery of two more galaxies showing QPEs in the eROSITA all-sky survey data, eRO-QPE3 and eRO-QPE4. eRO-QPE3 was found to flare once in multiple eROSITA surveys (Fig.~\ref{fig:eRO3_erass_lc}) and its nature as a QPE emitter was later confirmed by \emph{XMM-Newton} (Fig.~\ref{fig:xmm_lc}) and \emph{NICER} (Fig.~\ref{fig:eRO3_nicer}). Eruptions are observed to last $\sim2.1-2.4\,$h and to recur every $\sim17-20\,$h, although the exact recurrence pattern is currently poorly constrained. eRO-QPE4 was instead detected only once in a transient fashion across all eROSITA surveys (Fig.~\ref{fig:eRO4_erass_lc}) and both \emph{XMM-Newton} (Fig.~\ref{fig:xmm_lc_ero4}) and \emph{NICER} (Fig.~\ref{fig:eRO4_nicer}) confirmed the repeating nature of the source. Eruptions in eRO-QPE4 are observed to last $\approx0.5\,$h and recur every $\sim9.8-15.5\,$h.

Many of their properties are in agreement with those of known secure QPEs (\citetalias{Miniutti+2019:qpe1,Giustini+2020:qpe2,Arcodia+2021:eroqpes}), for instance:
\begin{itemize}
\item Their X-ray spectra during the eruptions are soft and thermal in shape, with typical peak temperatures of $kT=111_{-5}^{+6}$\,eV and $kT=(123\pm 4)$\,eV for eRO-QPE3 and eRO-QPE4, respectively, as observed by \emph{XMM-Newton}
\item The emission in quiescence, when detected, is consistent with the Wien tail of a radiatively efficient accretion disk, with inner temperatures $kT_{\rm in}\sim(90-100)$\,eV and $kT_{\rm in}\sim50$\,eV for eRO-QPE3 and eRO-QPE4, respectively (all spectral parameters are reported in Table~\ref{tab:erass1_spec} and~\ref{tab:ero4_spec})
\item Both sources show a clear energy dependence during the eruptions, with a harder rise than decay at the same flux level (Fig.~\ref{fig:energyevol} and~\ref{fig:energyevol_backup}), which was first reported for eRO-QPE1 in \citet{Arcodia+2022:ero1_timing} and then recovered in both GSN 069 \citep{Miniutti+2023:gsnrebr} and RXJ1301.9+2747 (Giustini et al., in prep.). At this stage, we consider it the only consistent property across all secure QPE sources, and therefore it is a key requirement to name an X-ray repeater as such. 
\item Their optical counterparts are local galaxies with a seemingly inactive spectrum (Fig.~\ref{fig:opt_image}) and even if a secure classification is compromised by the current data quality \citep[e.g., see][]{Wevers+2022:hosts}, a preexisting powerful AGN is ruled out. This is in agreement with the lack of a bright hard power-law spectrum indicative of a corona, the lack of broad lines, infrared photometry suggestive of an AGN-like torus obscurer and with the deep radio non-detections (Table~\ref{tab:ATCAobs}).
\item Both eRO-QPE3 and eRO-QPE4 are found in low-mass galaxies. In eRO-QPE3 $M_*$ is in the range $(0.7-2.6)\times10^9\,M_{\astrosun}$, which in turn implies $M_{\rm BH}$ in the range $(0.9-5.3)\times10^6\,M_{\astrosun}$ \citep{Reines+2015:mstar}. Using the stellar velocity dispersion, the inferred $M_{\rm BH}$ \citep{Gultekin+2009:msigma} is however $\sim4\times10^7\,M_{\astrosun}$, which only confirms how these scaling relations are poorly calibrated at low masses. In eRO-QPE4, all estimates are more consistent and point toward a slightly more massive galaxy, with $M_*$ being in the range $(0.6-1.6)\times10^{10}\,M_{\astrosun}$ and black hole, with $M_{\rm BH}$ in the range $(1.7-6.8)\times10^7\,M_{\astrosun}$, with the velocity dispersion also yielding $\sim2.7\times10^7\,M_{\astrosun}$. eRO-QPE4 therefore extends the known QPE population to slightly higher masses.
\end{itemize}

\noindent Furthermore, there are many novelties brought by the new discoveries, which only further highlight the importance to find more QPE sources to build meaningful statistics for a population study. For instance, we report the following:
\begin{itemize}
\item eRO-QPE3 showed QPEs on top of a decaying quiescence emission (Fig.~\ref{fig:ero3_longterm}), in support of other evidence that connects QPEs to a previous TDE \citep{Chakraborty+2021:qpe5cand,Sheng+2021:tdegsn,Miniutti+2023:gsnrebr,Quintin+2023:tormund}.
\item eRO-QPE3 shows the longest recurrence time ($\approx20$\,h) and the lowest peak luminosity ($L_{\rm 0.5-2.0\,keV}^{\rm peak}\sim\rm{few}\times10^{41}\,$erg\,s$^{-1}$) among known QPEs (Fig.~\ref{fig:correlations}, top), thus further strengthening that there is no correlation between these two quantities. Such timescale-luminosity correlations are instead predicted in many accretion-instability models.
\item eRO-QPE4 shows evidence, for the first time, of a harder component appearing close to the QPE peak (Fig.~\ref{fig:eRO4XMM_spec_peak}), although it is very faint and of an unclear origin (Appendix~\ref{sec:xmm_spec}).
\item In eRO-QPE4 we see clearly, for the first time, that the quiescence emission detected during the QPE epoch must have been either absent or much fainter previous to the QPE behavior (Fig.~\ref{fig:ero4_longterm}), supporting a short-lived nature of the radiatively efficient accretion flow seen in-between QPEs.
\item The long-term recurrence pattern of both eRO-QPE3 and eRO-QPE4 cannot be accurately constrained based on current data, although in eRO-QPE4 (Fig.~\ref{fig:eRO4_nicer}) it appears somewhat intermediate between regular alternating and irregular, perhaps bridging the gap between the apparent dichotomy that would have been inferred from the only four sources available \citep[e.g.,][]{Arcodia+2022:ero1_timing}; a continuum of variability patterns would be expected in the case of orbital phenomena with orbital and/or disk precession in play.
\end{itemize}

The detection of eRO-QPE4 as a non-repeating bright and soft extragalactic flare, compared to the repeated nature of eRO-QPE1, eRO-QPE2 \citepalias{Arcodia+2021:eroqpes} and eRO-QPE3 (this work), opens up a new way to select more candidates within the available eROSITA surveys. Furthermore, the wealth of eROSITA data can still be used for searches for QPEs performed in a more informed way, by targeting galaxies strikingly similar to those of the known QPE sources \citep[e.g.,][]{Wevers+2022:hosts}. Looking at the future, wide-area X-ray telescopes that are sensitive in the soft X-rays ($<2\,$keV) are needed to efficiently detect QPEs. In the meantime, thorough archival searches are still very useful \citep{Webbe+2023:ML} and have already proved successful in providing QPE candidates \citep{Chakraborty+2021:qpe5cand,Quintin+2023:tormund}. Finally, we note that the homogeneous nature of our search with the eROSITA telescope allows us to estimate the intrinsic abundance rate of QPEs, which will be presented in a companion paper \citep{Arcodia+2024:rates}.
	
\begin{acknowledgements}
We thank the teams of both \emph{XMM-Newton} and \emph{NICER} for their prompt efforts in the scheduling of our observations. R.A. thanks G. Miniutti and M. Giustini for the many lively discussions about QPEs.

R.A. received support for this work by NASA through the NASA Einstein Fellowship grant No HF2-51499 awarded by the Space Telescope Science Institute, which is operated by the Association of Universities for Research in Astronomy, Inc., for NASA, under contract NAS5-26555. R.A. thanks T. Nowa and B. Banks and their sets for inspiring the writing.

GP acknowledges funding from the European Research Council (ERC) under the European Union’s Horizon 2020 research and innovation programme (grant agreement No 865637), support from Bando per il Finanziamento della Ricerca Fondamentale 2022 dell’Istituto Nazionale di Astrofisica (INAF): GO Large program and from the Framework per l’Attrazione e il Rafforzamento delle Eccellenze (FARE) per la ricerca in Italia (R20L5S39T9).

This work was supported by the Australian government through the Australian Research Council’s Discovery Projects funding scheme (DP200102471).

This work is based on data from eROSITA, the soft X-ray instrument aboard SRG, a joint Russian-German science mission supported by the Russian Space Agency (Roskosmos), in the interests of the Russian Academy of Sciences represented by its Space Research Institute (IKI), and the Deutsches Zentrum für Luft- und Raumfahrt (DLR). The SRG spacecraft was built by Lavochkin Association (NPOL) and its subcontractors, and is operated by NPOL with support from the Max Planck Institute for Extraterrestrial Physics (MPE). The development and construction of the eROSITA X-ray instrument was led by MPE, with contributions from the Dr. Karl Remeis Observatory Bamberg \& ECAP (FAU Erlangen-Nuernberg), the University of Hamburg Observatory, the Leibniz Institute for Astrophysics Potsdam (AIP), and the Institute for Astronomy and Astrophysics of the University of Tuebingen, with the support of DLR and the Max Planck Society. The Argelander Institute for Astronomy of the University of Bonn and the Ludwig Maximilians Universitaet Munich also participated in the science preparation for eROSITA. The eROSITA data shown here were processed using the eSASS software system developed by the German eROSITA consortium.

This work has made use of data from the European Space Agency (ESA) mission
{\it Gaia} (\url{https://www.cosmos.esa.int/gaia}), processed by the {\it Gaia}
Data Processing and Analysis Consortium (DPAC,
\url{https://www.cosmos.esa.int/web/gaia/dpac/consortium}). Funding for the DPAC
has been provided by national institutions, in particular the institutions participating in the {\it Gaia} Multilateral Agreement. This research has made use of the VizieR catalogue access tool, CDS, Strasbourg, France (DOI : 10.26093/cds/vizier). The original description of the VizieR service was published in 2000, A\&AS 143, 23.

Part of the funding for GROND (both hardware as well as personnel) was generously granted from the Leibniz-Prize to Prof. G. Hasinger (DFG grant HA 1850/28-1).

MK acknowledges support from DLR grant FKZ 50 OR 2307.

We acknowledge the use of the matplotlib package \citep{Hunter2007:matplotlib}.

\end{acknowledgements}

	%
	%
\bibliographystyle{aa} 
\bibliography{bibliography} 

\begin{appendix}        
     
\section{X-ray spectral analysis}
\subsection{eROSITA data of eRO-QPE3}
\label{sec:erass_spec}

\begin{figure}[h]
	\centering
    \includegraphics[width=0.85\columnwidth]{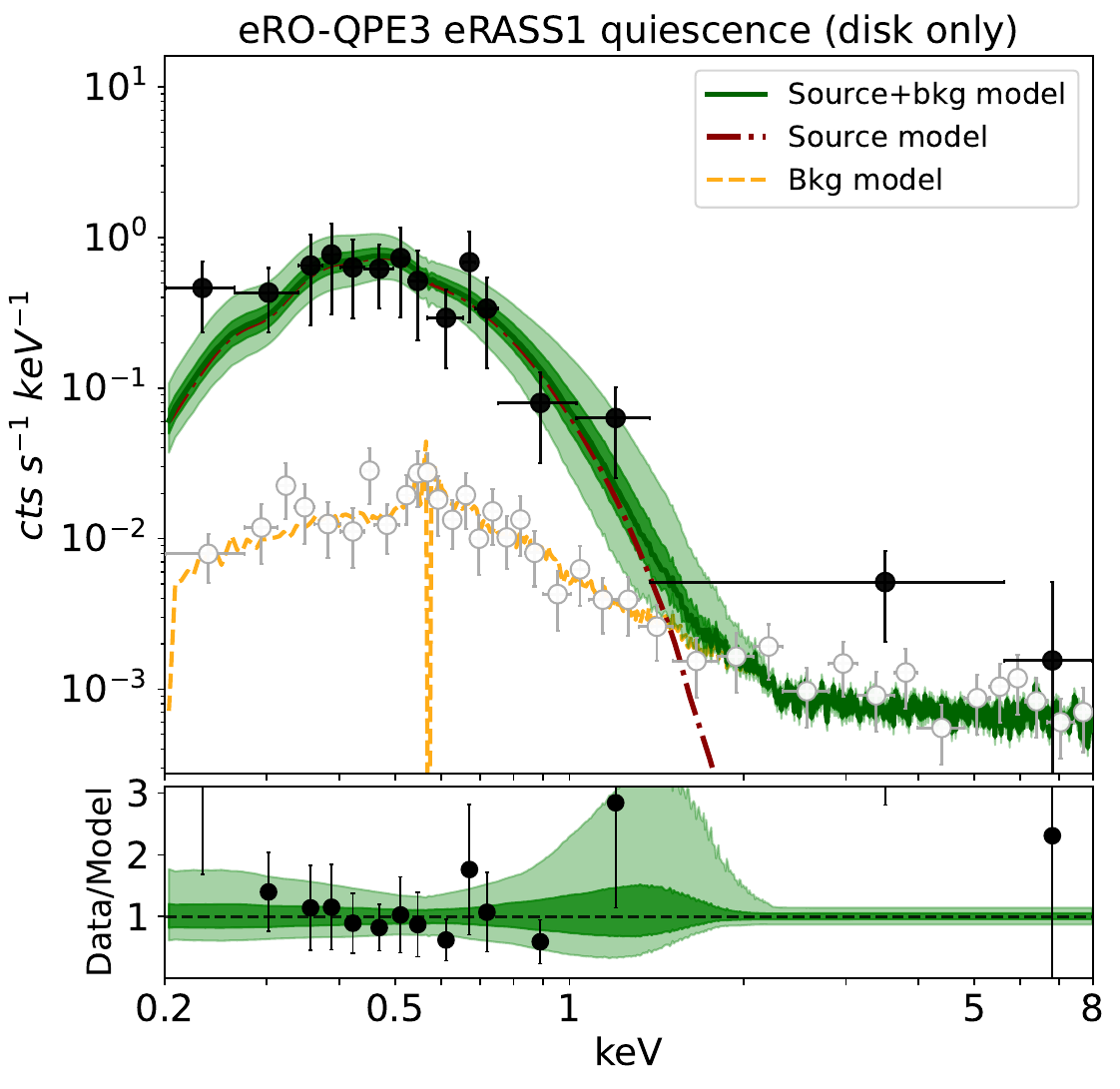}
    \includegraphics[width=0.85\columnwidth]{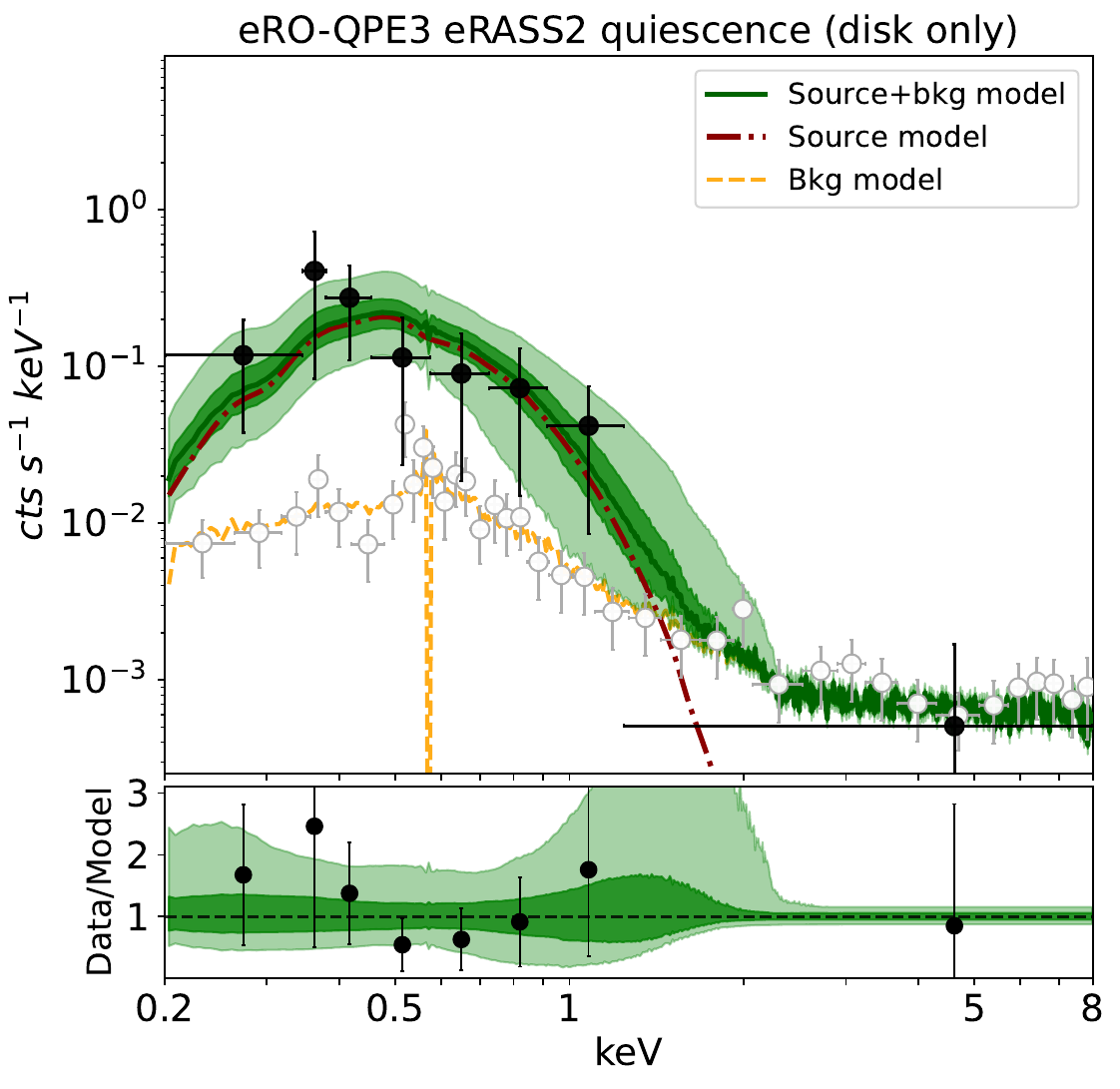}
	\caption{Spectral fit for the eRASS1 (\emph{top}) and eRASS2 (\emph{bottom}) spectra in quiescence modeled with a \texttt{diskbb}. Black points are source plus background data, empty gray points show the background alone. The green line and related light green (dark green) shaded regions are the source plus background model median and 1st-99th (16th-84th) percentiles, respectively. The orange dashed lines shows the background model alone. In the lower panel, the data-model ratio is shown. The individual absorption-corrected source model component (here a \texttt{diskbb}) is shown with a red line, as indicated by the legend.}
	\label{fig:eRO3_erass1_spec_diskbbonly}
\end{figure}

\begin{figure}[h]
	\centering
    \includegraphics[width=0.85\columnwidth]{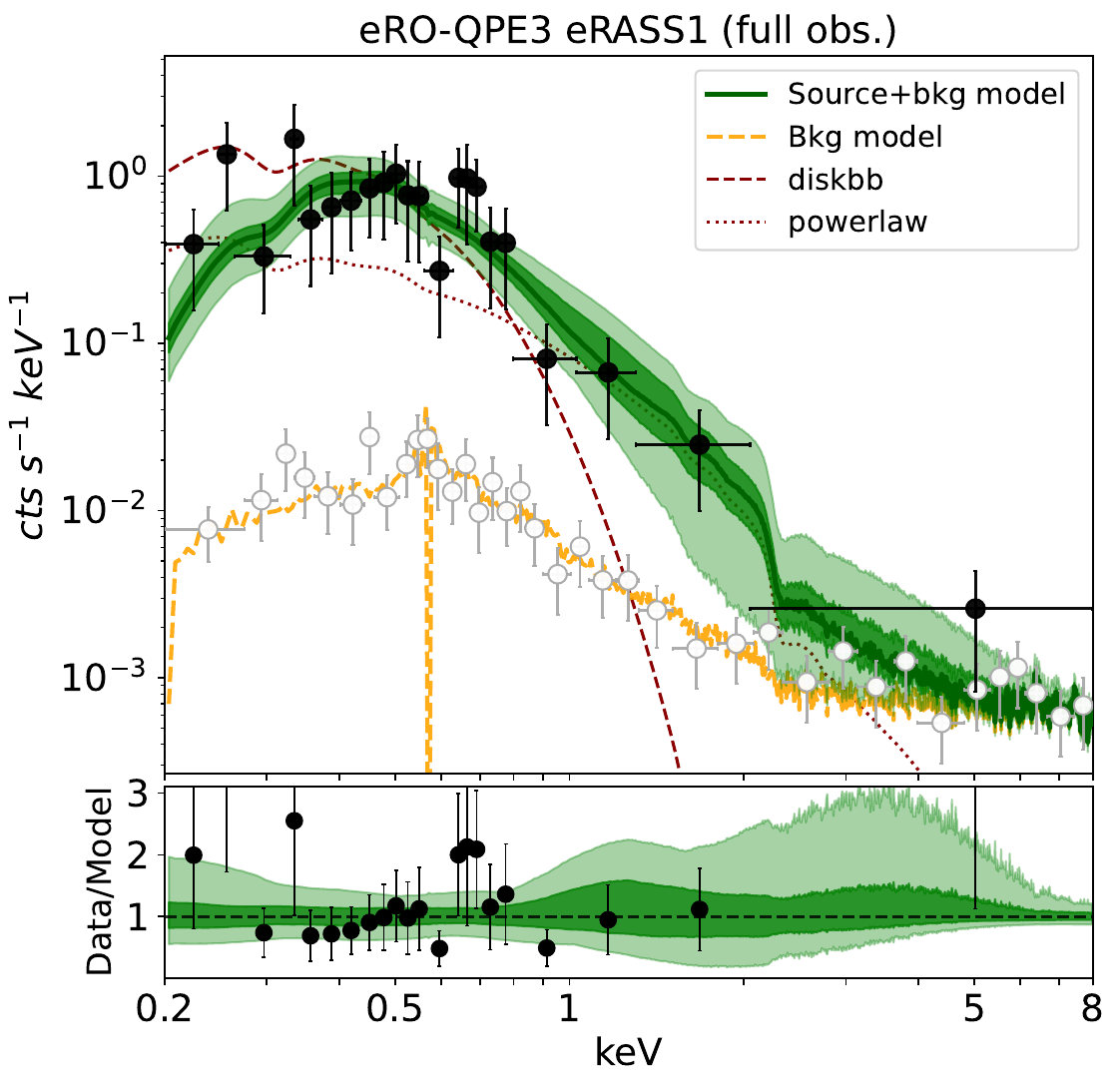}
	\caption{As in Fig.~\ref{fig:eRO3_erass1_spec_diskbbonly}, but for the quiescence model (\texttt{diskbb} and \texttt{zpowerlw}) applied to the eRASS1 spectrum of the full observations, without separating faint eROdays from the bright (top-left panel of Fig.~\ref{fig:eRO3_erass_lc}).}
	\label{fig:eRO3_erass1_spec_noqpe}
\end{figure}

\begin{figure}[tb]
	\centering
    \includegraphics[width=0.85\columnwidth]{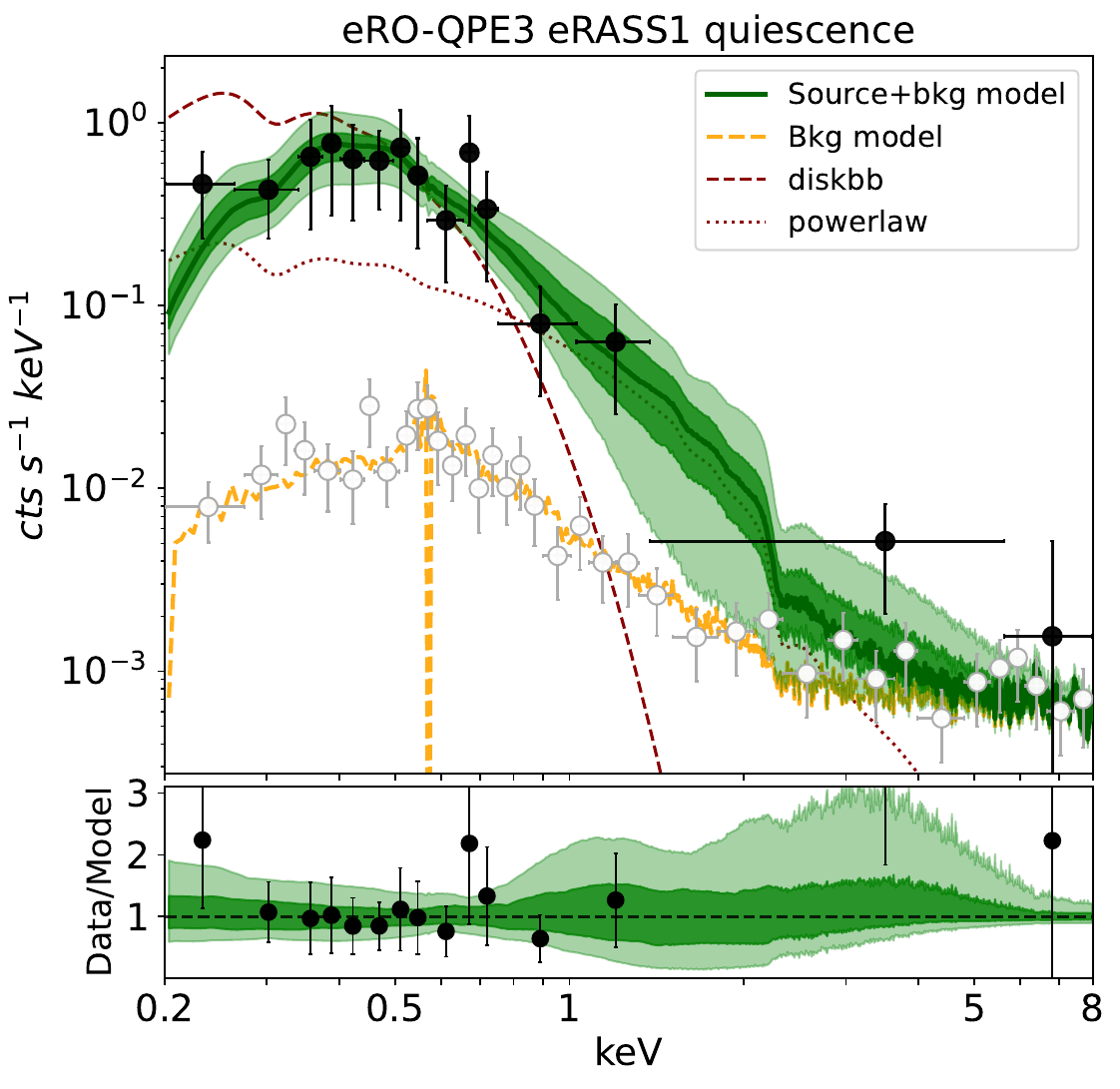}
    \includegraphics[width=0.85\columnwidth]{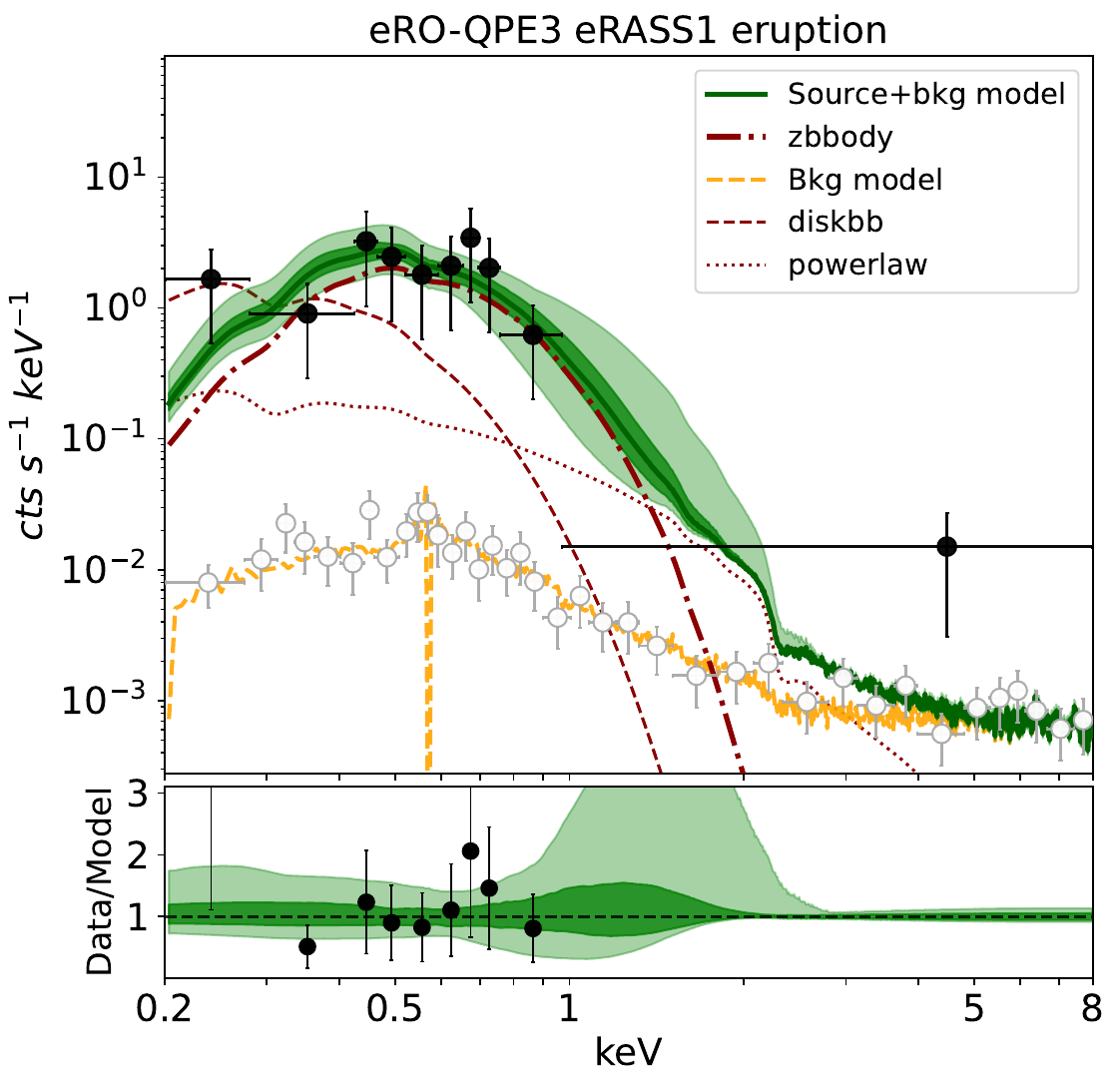}
	\caption{As in Fig.~\ref{fig:eRO3_erass1_spec_diskbbonly}, but for the chosen best-fit models for both eRASS1 in quiescence (\emph{top}, \texttt{diskbb} and \texttt{zpowerlw}) and in eruption (\emph{bottom}, adding a \texttt{zbbody}).}
	\label{fig:eRO3_erass1_spec}
\end{figure}

The eRASS1 spectrum in quiescence, if modeled with a simple accretion disk, shows marginal residual at high-energies (top panel of Fig.~\ref{fig:eRO3_erass1_spec_diskbbonly}) and a median peak temperature of $kT_{disk}\sim133\,$eV. This temperature would be unusually high compared to other QPE-sources in quiescence, in which the presence of a fainter harder component is often observed (\citetalias{Miniutti+2019:qpe1,Giustini+2020:qpe2,Arcodia+2021:eroqpes}; \citealp{Chakraborty+2021:qpe5cand}). Therefore, we add a power-law component to account for the residuals, mimicking thermal Comptonization of the accretion disk photons (top panel of Fig.~\ref{fig:eRO3_erass1_spec}). Compared to a disk model alone, $\log Z$ improves by 2.4, however we note that spectral simulations are needed to calibrate the $\delta \log Z$ for this model comparison. This more complex model is mostly chosen as best-fit model due to the comparison with other quiescence spectra of QPE sources, obtained with higher signal-to-noise spectra (\citetalias{Miniutti+2019:qpe1,Giustini+2020:qpe2,Arcodia+2021:eroqpes}; \citealp{Chakraborty+2021:qpe5cand}). With the addition of the power-law component at higher energies, the median (and 16th, 84th percentiles) peak temperature of the disk model becomes $kT_{\rm disk}=100_{-20}^{+18}\,$eV. The slope of the power-law component $\Gamma_X$ is unconstrained, with a slight preference toward softer values\footnote{A maximum value of $\Gamma_X=3.5$ is allowed for the fit.}.  The median (and related 16th, 84th percentiles) flux of the disk component is $F_{\rm 0.2-2.0\,keV} = 1.4_{-0.6}^{+0.4} \times 10^{-12}\,$erg\,s$^{-1}$\,cm$^{-2}$, while that of the power-law component is $F_{\rm 0.2-2.0\,keV} = 2.7_{-1.7}^{+8.4} \times 10^{-13}\,$erg\,s$^{-1}$\,cm$^{-2}$. The total flux in the observed X-ray band is $F_{\rm 0.2-8.0\,keV} = 1.9_{-0.6}^{+0.7} \times 10^{-12}\,$erg\,s$^{-1}$\,cm$^{-2}$. This model is held fixed during the eruptions and a black body component is added (bottom panel of Fig.~\ref{fig:eRO3_erass1_spec}). Since the eRASS1 light curve (top-left panel of Fig.~\ref{fig:eRO3_erass_lc}) does not show variability of the same amplitude compared to eRASS2-3-4, we explore the possibility that no QPEs were present during eRASS1 and that the quiescence continuum was simply highly variable. For instance, in GSN\,069, the only other secure QPE source with a clearly decaying quiescence emission over months-years as in eRO-QPE3, QPEs were not detected when the quiescence emission was the brigthest \citep{Miniutti+2023:gsnrebr}. Therefore we also fit the quiescence model (\texttt{diskbb} and \texttt{zpowerlw}) to the full eRASS1 observation, without separating faint eROdays from the bright one (top-left panel of Fig.~\ref{fig:eRO3_erass_lc}). In this work, we do not aim to reach a conclusive statement on the presence of QPEs in eRASS1 or not. Model comparison cannot be performed due to the use of different good-time intervals (GTIs) for the spectra in the two scenarios. We show this fit in Fig.~\ref{fig:eRO3_erass1_spec_noqpe} and we report the fit parameters in Table~\ref{tab:erass1_spec} together with other eRASS1 fit results.

\begin{table*}[t]
	\small
	\setlength{\tabcolsep}{4pt}
	\caption{Spectral fit results for eRO-QPE3.}
	\label{tab:erass1_spec}
	\centering
	\begin{threeparttable}
		\begin{tabular}{ccccccccc}
			\toprule
            \multicolumn{1}{c}{Epoch} &
			\multicolumn{1}{c}{Spectrum} &
			\multicolumn{1}{c}{Model} &
			\multicolumn{1}{c}{$kT_{\rm disk}$} &
			\multicolumn{1}{c}{$F^{\rm disk}_{\rm 0.2-2.0\,keV}$} &
			\multicolumn{1}{c}{$F^{\rm pl}_{\rm 0.2-2.0\,keV}$} &
			\multicolumn{1}{c}{$kT_{\rm QPE}$} &
			\multicolumn{1}{c}{$F^{\rm QPE}_{\rm 0.2-2.0\,keV}$} &
            \multicolumn{1}{c}{$\Delta \log Z$}
            \\
              &       &          & [eV]   &  [erg\,s$^{-1}$\,cm$^{-2}$]  & [erg\,s$^{-1}$\,cm$^{-2}$]  &  [eV]   &  [erg\,s$^{-1}$\,cm$^{-2}$]  &  
            \\
			\midrule
			eRASS1 & Quiescence      &     \texttt{disk}       &     $133_{-12}^{+13}$   &  $(1.4\pm0.2) \times 10^{-12}$  &    --   &  --   &  --  & 2 \\
             &       &     \textbf{\texttt{disk+pl}}       &     $100_{-20}^{+18}$   &  $1.4_{-0.6}^{+0.3} \times 10^{-12}$  &    $2.9_{-1.9}^{+6.7} \times 10^{-13}$   &  --   &  --  & 0 \\
             & QPE      &     \textbf{\texttt{disk+pl+bb}}       &     $100$   &  $1.4 \times 10^{-12}$  &    $2.9 \times 10^{-13}$   &  $122_{-15}^{+19}$   &  $(3.0\pm0.7) \times 10^{-12}$  & -- \\
             & Full      &     \texttt{disk+pl}       &     $110_{-27}^{+15}$   &  $1.7_{-1.3}^{+0.4} \times 10^{-12}$  &   $5.3_{-3.8}^{+14.3} \times 10^{-13}$   &  --   &  --  & -- \\
             \midrule
             eRASS2 & Quiescence      &     \texttt{disk}       &     $151_{-24}^{+29}$   &  $3.9_{-0.8}^{+1.1} \times 10^{-13}$  &    --   &  --   &  --  & 1.5 \\
              &       &     \texttt{disk+pl} (free)       &     $80_{-59}^{+85}$   &  $<4.2 \times 10^{-13}$  &   $4.2_{-4.0}^{+1.7} \times 10^{-13}$   &  --   &  --  & 0.5 \\
               &       &     \textbf{\texttt{disk+pl}}       &     $89_{-60}^{+42}$   &  $3.12_{-3.11}^{+1.51} \times 10^{-13}$  &   $2.4_{-1.1}^{+0.9} \times 10^{-13}$   &  --   &  --  & 0 \\
               &   QPE    &     \textbf{\texttt{disk+pl+bb}}       &     $89$   &  $3.1 \times 10^{-13}$  &   $2.4 \times 10^{-13}$   &   $82_{-9}^{+10}$   &  $3.2_{-0.7}^{+1.0} \times 10^{-12}$  & -- \\
             \midrule
             eRASS3 & Quiescence      &     \texttt{disk}       &     --   &  $<2.7\times 10^{-14}$  &    --   &  --   &  --  &  -- \\
              & QPE      &     \texttt{bb}       &     --   &  --  &    --   &  $72_{-9}^{+12}$   &  $3.2_{-0.7}^{+1.0} \times 10^{-12}$  &  -- \\
             \midrule
             eRASS4 & Quiescence      &     \texttt{disk}       &     --   &  $<3.5\times 10^{-13}$  &    --   &  --   &  --  &  -- \\
              & QPE      &     \texttt{bb}       &     --   &  --  &    --   &  $112_{-16}^{+18}$   &  $1.4_{-0.3}^{+0.5} \times 10^{-12}$  &  -- \\
             \midrule
             eRASS5 & Quiescence      &     \texttt{disk}       &     --   &  $<7.2\times 10^{-15}$  &    --   &  --   &  --  &  -- \\
              & QPE      &     \texttt{bb}       &     --   &  --  &    --   &  $\approx294_{-103}^{+240}$   &  $3.2_{-1.6}^{+2.4} \times 10^{-13}$  &  -- \\
              \midrule
              XMM1 burst1 & Quiescence      &     \texttt{disk}       &     --   &  $<1.2\times 10^{-14}$  &    --   &  --   &  --  & -- \\
                  & QPE rise1    &     \texttt{bb}       &     --   &  --  &    --   &  $108_{-13}^{+16}$   &  $5.3_{-1.1}^{+1.3} \times 10^{-14}$  &  -- \\
                 & QPE rise2    &     \texttt{bb}       &     --   &  --  &    --   &  $(121\pm6)$   &  $1.7_{-0.1}^{+0.2} \times 10^{-13}$  &  -- \\
               & QPE peak    &     \texttt{bb}       &     --   &  --  &    --   &  $111_{-5}^{+6}$   &  $3.1_{-0.2}^{+0.3} \times 10^{-13}$  &  -- \\
               & QPE decay1    &     \texttt{bb}       &     --   &  --  &    --   &  $97_{-8}^{+9}$   &  $1.7_{-0.2}^{+0.3} \times 10^{-13}$  &  -- \\
               & QPE decay2    &     \texttt{bb}       &     --   &  --  &    --   &  $(86\pm7)$   &  $1.1_{-0.1}^{+0.3} \times 10^{-13}$  &  -- \\
             \bottomrule
		\end{tabular}
	    \begin{tablenotes}
        \small
        \item Fit values show the median and related 16th-84th percentiles of the fit posteriors. We mark in boldface the adopted best-fit model, if more than one is reported, which has the highest Bayesian evidence. The difference between the logarithmic Bayesian evidence values is shown in the last column for these cases. For eRASS2, the \texttt{disk+pl(free)} model has no constraint on the power-law normalization, whilst in the \texttt{disk+pl} model the power-law is tied to that of the disk with the same ratio observed in the eRASS1 spectrum. During the QPE fit, the quiescence model is held fixed, if detected. For eRASS5, the QPE fit is compatible with background within $3\sigma$ and the fit parameters should be interpreted with caution. For \emph{XMM-Newton}, the different phases are shown in Fig.~\ref{fig:energyevol} and~\ref{fig:eRO3_lcphases}. Given the spectroscopic redshift and the cosmology adopted \citep{Hinshaw+2013:wmap9}, the conversion for related luminosity values for eRO-QPE3 is $1.34\times10^{54}$\,cm$^2$ in this paper.
        \end{tablenotes}
   \end{threeparttable}
\end{table*}

\begin{figure}[tb]
	\centering
    \includegraphics[width=0.85\columnwidth]{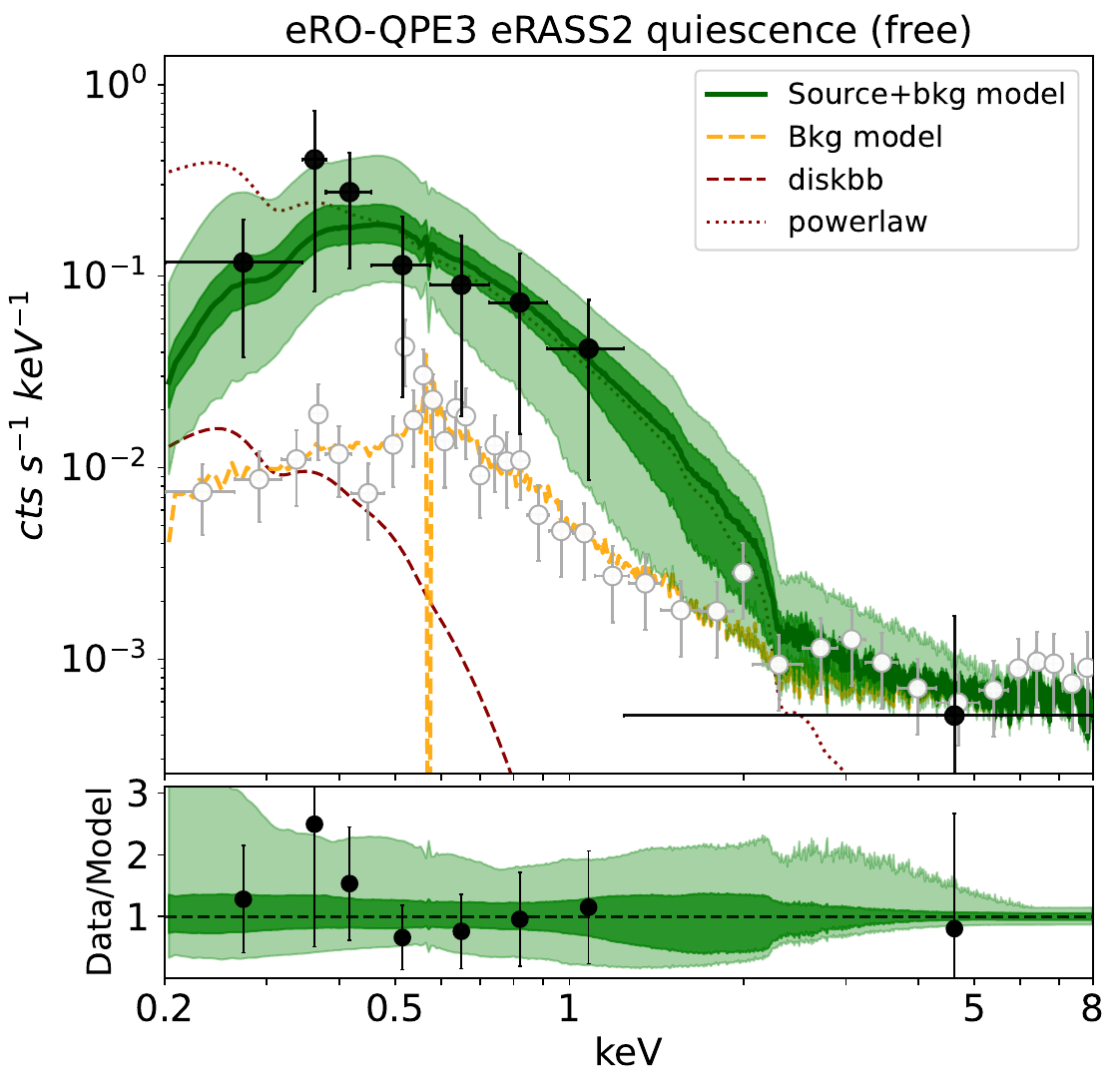}
	\caption{As in Fig.~\ref{fig:eRO3_erass1_spec_diskbbonly}, but for the quiescence spectrum of eRASS2, where both disk and power-law components are free to vary.}
	\label{fig:eRO3_erass2_spec_quntied}
    \end{figure}

\begin{figure}[tb]
	\centering
    \includegraphics[width=0.85\columnwidth]{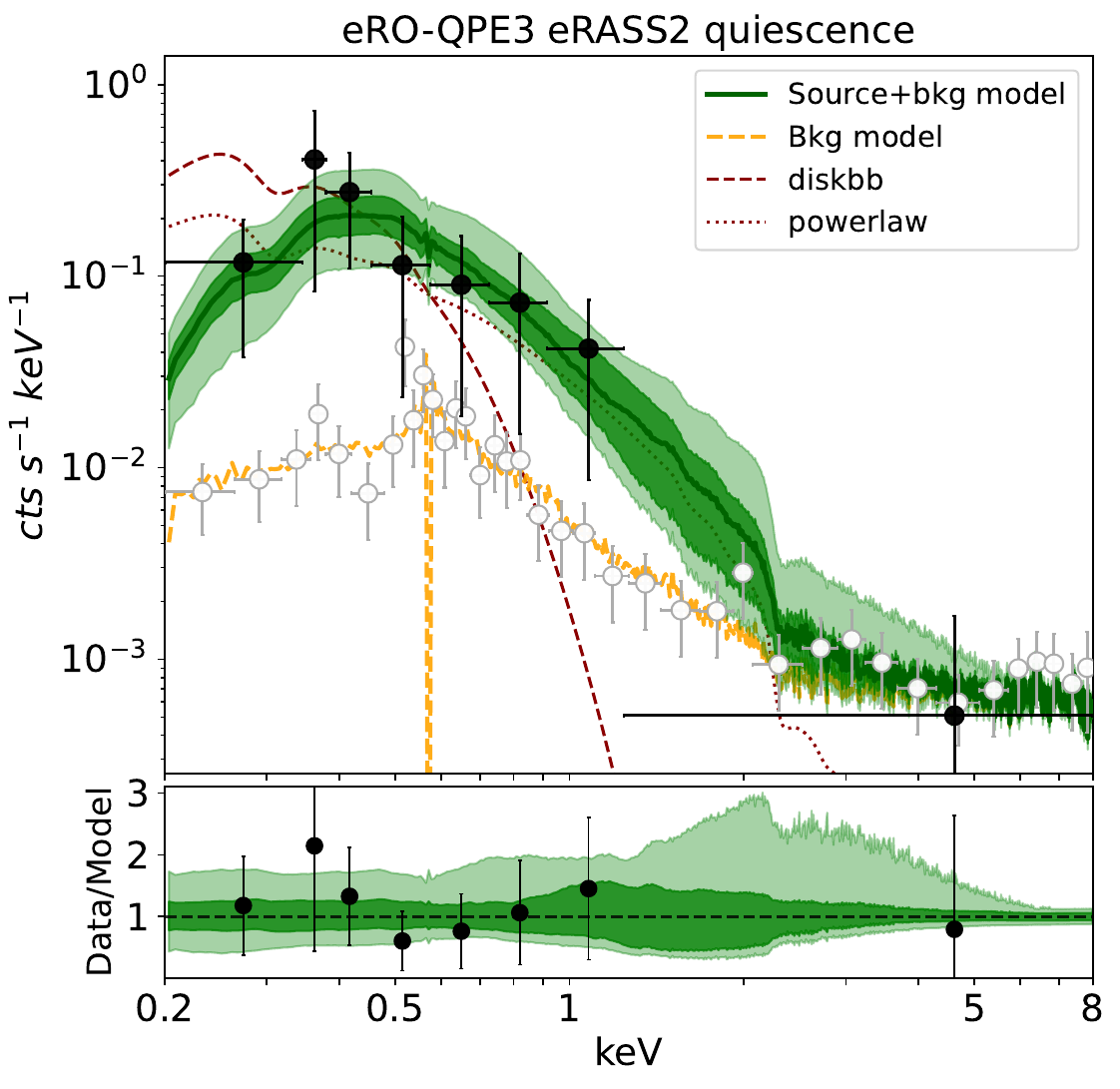}
    \includegraphics[width=0.85\columnwidth]{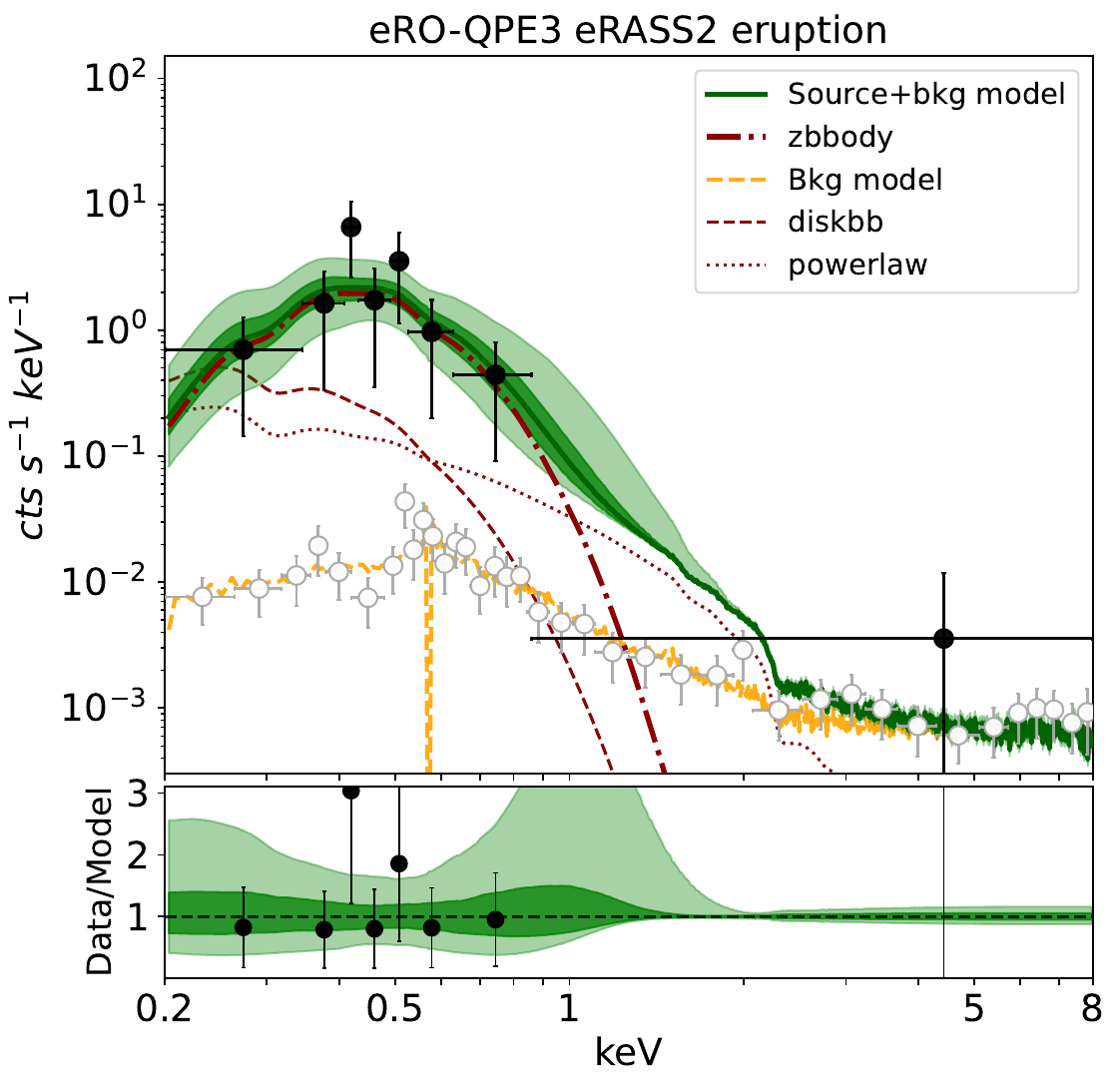}
	\caption{As in Fig.~\ref{fig:eRO3_erass1_spec}, but showing the best-fit models for eRASS2, in quiescence (\emph{top}) and eruption (\emph{bottom}).}
	\label{fig:eRO3_erass2_spec}
    \end{figure}

    The eRASS2 spectrum in quiescence appears fainter than in eRASS1 (see Fig.~\ref{fig:eRO3_erass1_spec_diskbbonly} and Fig.~\ref{fig:eRO3_erass_lc}) and can be adequately fit by either a \texttt{diskbb} or a power-law continuum. Models provide equally good fits due to the low counts statistics (32 counts within $0.2-8.0\,$keV versus, e.g., the 95 in eRASS1). Adopting the disk model, based on results of other QPE sources, we obtain $kT_{\rm disk}\sim151\,$eV, which is higher than the disk temperature in eRASS1 (Table~\ref{tab:erass1_spec}). We consider unlikely that the quiescence flux is lower and the temperature higher \citep[e.g.,][for the long-term evolution in GSN\,069]{Miniutti+2023:gsnreapp,Miniutti+2023:gsnrebr}. Therefore, we attribute this result on the eRASS2 quiescence spectrum to the use of an incomplete source model. Based on eRASS1 and results of other QPEs in quiescence, we adopt the \texttt{diskbb} plus \texttt{zpowerlw} model as reference, despite not being statistically required compared to a single component. If left free to vary, the power-law component dominates the fit and the disk component is fit at background level (Fig.~\ref{fig:eRO3_erass2_spec_quntied}), with a highly-uncertain temperature ($kT_{\rm disk}=80_{-59}^{+85}\,$eV). The median (and related 16th, 84th percentiles) flux of the power-law component is $F_{\rm 0.2-2.0\,keV} = 4.2_{-4.0}^{+1.7} \times 10^{-13}\,$erg\,s$^{-1}$\,cm$^{-2}$ while the disk component is an upper limit at $<4.2\times 10^{-13}\,$erg\,s$^{-1}$\,cm$^{-2}$, with a median value of the flux posterior at $\sim1\times 10^{-14}\,$erg\,s$^{-1}$\,cm$^{-2}$. Based on eRASS1 data of eRO-QPE3 and results obtained for other QPE sources with higher signal-to-noise spectra (\citetalias{Miniutti+2019:qpe1,Giustini+2020:qpe2,Arcodia+2021:eroqpes}; \citealp{Chakraborty+2021:qpe5cand}), we assume that this fit result is likely due to the greater flexibility of the power-law model and model degeneracies. A more realistic result with a \texttt{diskbb} plus \texttt{zpowerlw} would find the former brighter than the latter. We therefore impose that the normalization of the power-law is tied to that of the disk with the same ratio observed in the eRASS1 spectrum. The assumption is that both components have faded. We note that, for the scope of this work, the main aim is to obtain a reliable quiescence model to be held fixed in the QPE spectrum, to isolate this component. We verified a posteriori that the use of either free or tied power-law components does not affect the fit of the QPE spectrum in the bright state. We therefore adopt this tied \texttt{diskbb} plus \texttt{zpowerlw} model as reference for the quiescence (top panel of Fig.~\ref{fig:eRO3_erass2_spec}). The median (and related 16th, 84th percentiles) temperature of the disk becomes $kT_{\rm disk}=89_{-60}^{+42}\,$eV, thus slightly more constrained and still compatible with the former value. Its flux becomes $F_{\rm 0.2-2.0\,keV} = 3.12_{-3.11}^{+1.52} \times 10^{-13}\,$erg\,s$^{-1}$\,cm$^{-2}$, therefore still compatible with background. This confirms that we did not artificially imposed the presence of a disk, but merely allowed the posterior of its flux to be more reasonably brighter. The flux of the power-law component is $F_{\rm 0.2-2.0\,keV} = 2.4_{-1.1}^{+0.9} \times 10^{-13}\,$erg\,s$^{-1}$\,cm$^{-2}$. The total flux in the observed X-ray band is $F_{\rm 0.2-8.0\,keV} = 5.8_{-2.9}^{+1.7} \times 10^{-13}\,$erg\,s$^{-1}$\,cm$^{-2}$. As done for eRASS1, this quiescence model is held fixed during the brighter eROday (the orange point in the top-medium panel of Fig.~\ref{fig:eRO3_erass_lc}) and a black body component is added (see bottom panel of Fig.~\ref{fig:eRO3_erass2_spec}). The median (and 16th, 84th percentiles) peak temperature of the QPE component is $kT_{\rm QPE}=82_{-9}^{+10}\,$eV and its flux is $F_{\rm 0.2-2.0\,keV} = 3.2_{-0.7}^{+1.0} \times 10^{-12}\,$erg\,s$^{-1}$\,cm$^{-2}$. The median temperature is colder compared to the bright eROday in eRASS1, despite the flux being compatible. However, given the short $\sim40$s exposure of an eROday compared to the typical QPE duration ($0.5-7\,$h, \citetalias{Arcodia+2021:eroqpes}), eRASS data catch the eruption at different phases and they are lower limits of the QPE peaks. The total flux during the bright eROday, quiescence included, in the observed X-ray band is $F_{\rm 0.2-8.0\,keV} = 3.9_{-0.8}^{+1.0} \times 10^{-12}\,$erg\,s$^{-1}$\,cm$^{-2}$. Therefore, despite the decreasing quiescence level, the flux in the bright eROday is compatible between eRASS1 and eRASS2.

    \begin{figure}[tb]
	\centering
    \includegraphics[width=0.85\columnwidth]{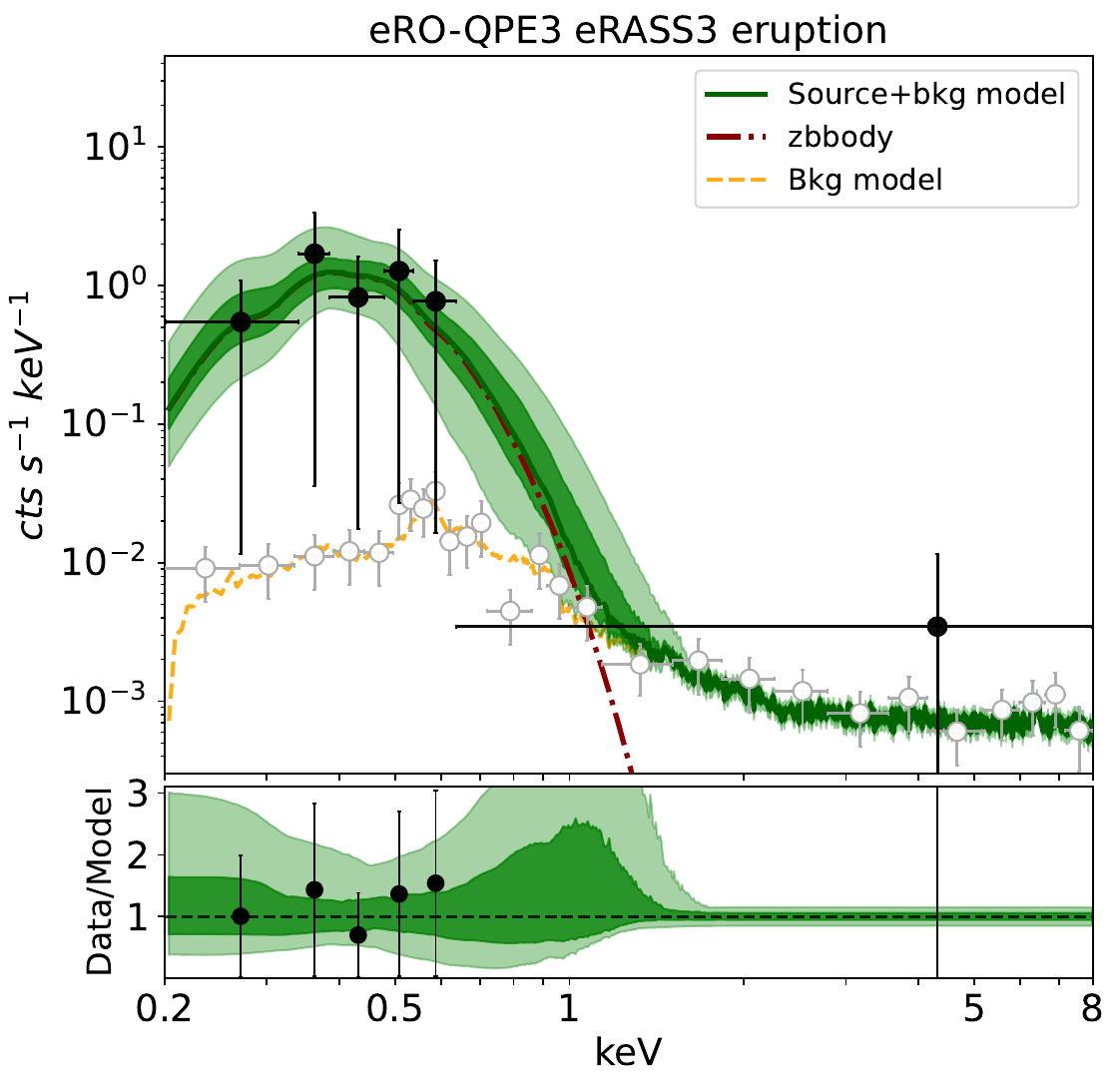}
	\caption{As in Fig.~\ref{fig:eRO3_erass1_spec_diskbbonly}, but for the QPE spectrum of eRASS3.}
	\label{fig:eRO3_erass3_spec_burst}
    \end{figure}

    \begin{figure}[tb]
	\centering
    \includegraphics[width=0.85\columnwidth]{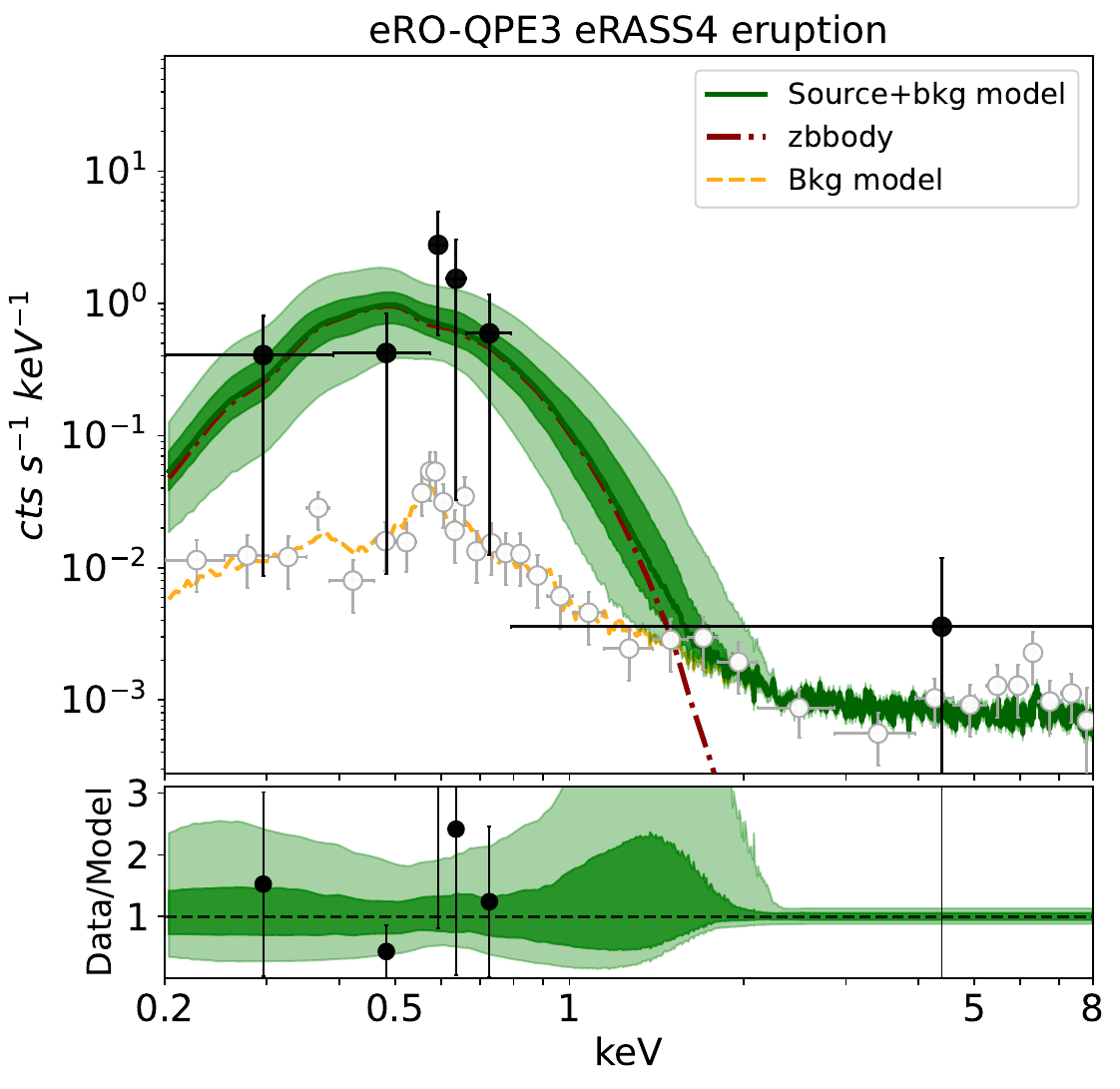}
	\caption{As in Fig.~\ref{fig:eRO3_erass1_spec_diskbbonly}, but for the QPE spectrum of eRASS4.}
	\label{fig:eRO3_erass4_spec_burst}
    \end{figure}

    \begin{figure}[tb]
	\centering
    \includegraphics[width=0.85\columnwidth]{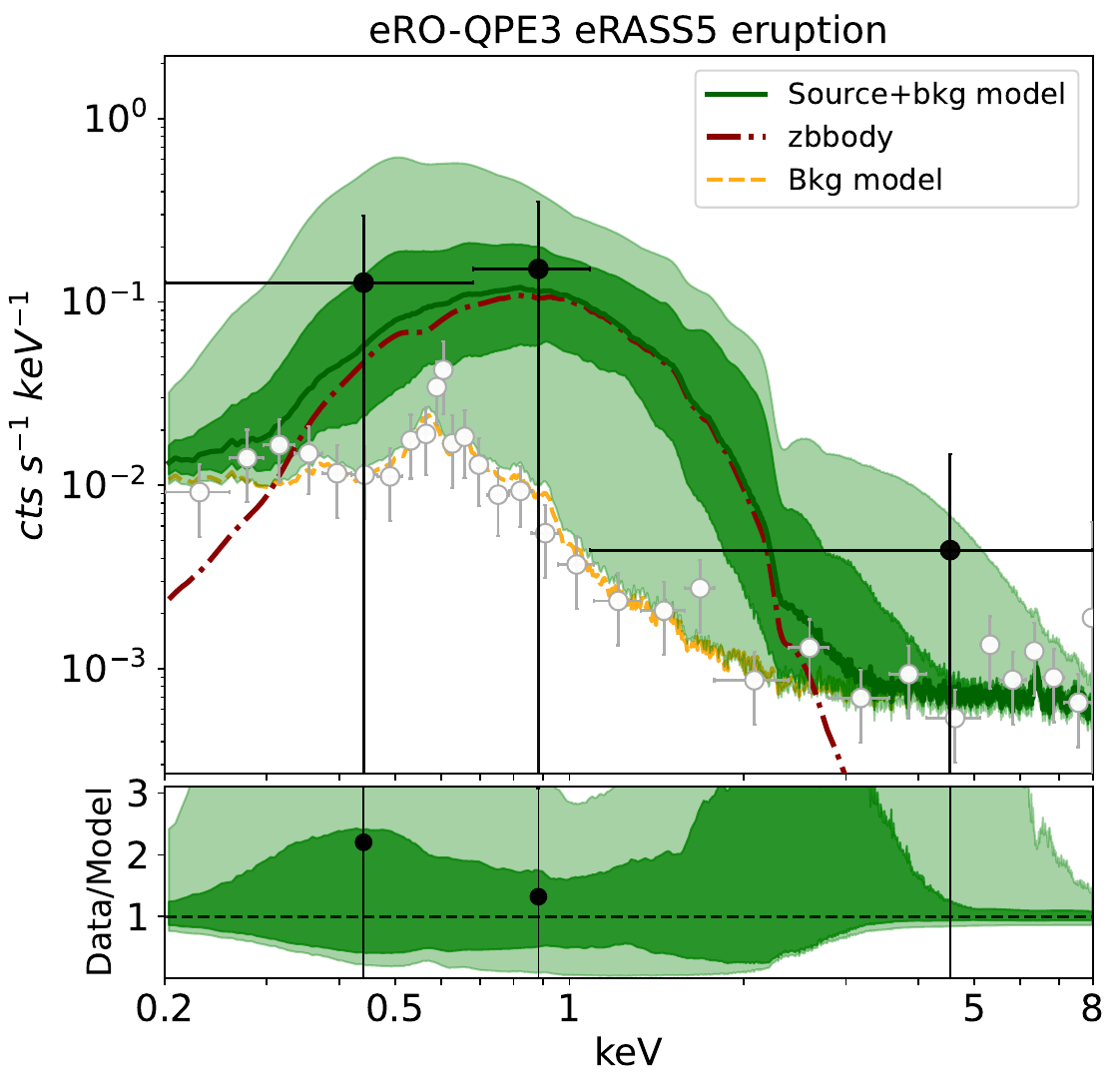}
	\caption{As in Fig.~\ref{fig:eRO3_erass1_spec_diskbbonly}, but for the QPE spectrum of eRASS5. As the fit model is compatible with background within the $3\sigma$ contours, eRASS5 fit results are to be interpreted with caution.}
	\label{fig:eRO3_erass5_spec_burst}
    \end{figure}

    \subsection{\emph{XMM-Newton} spectral analysis}
    \label{sec:xmm_spec}

    \subsubsection{eRO-QPE3}

    \begin{figure}[tb]
	\centering
    \includegraphics[width=0.9\columnwidth]{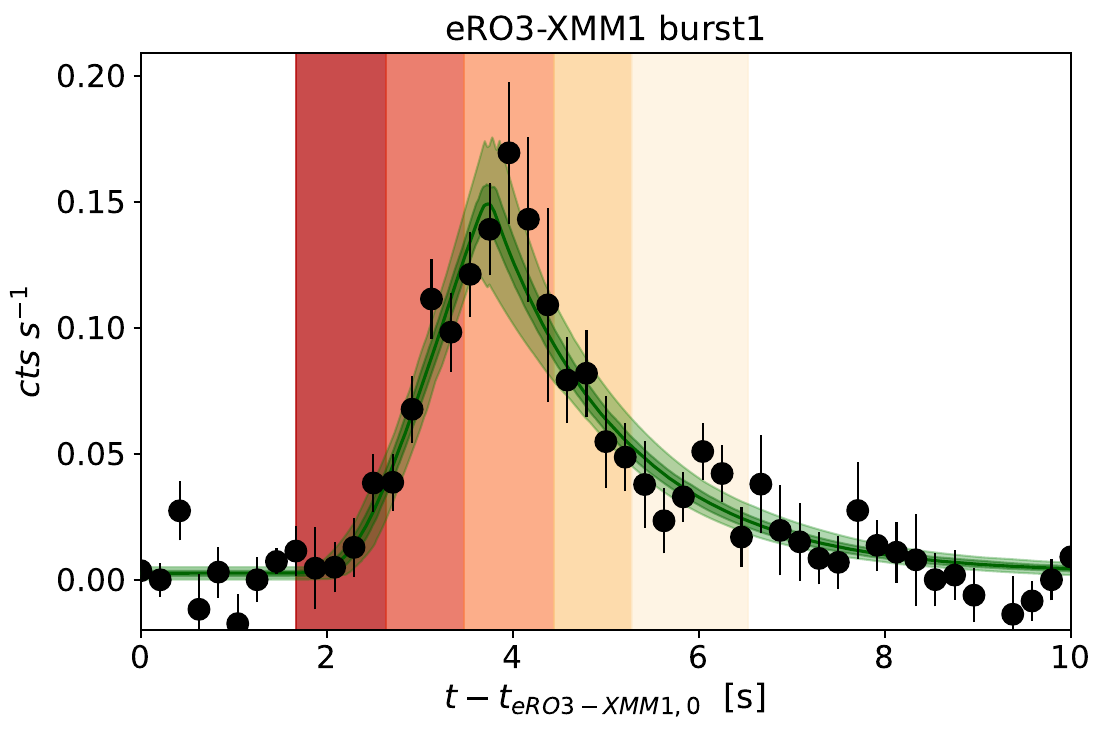}
	\includegraphics[width=0.9\columnwidth]{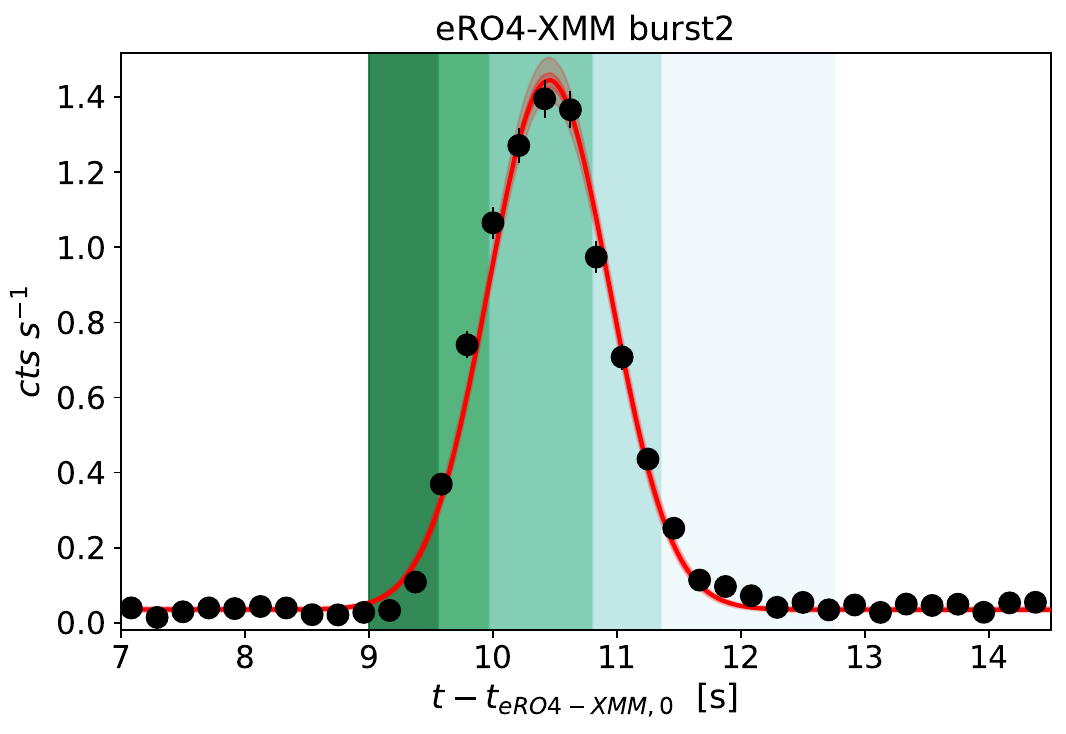}
    \caption{\emph{Top:} a chunk of the eRO3-XMM1 observation (see Fig.~\ref{fig:xmm_lc}). The best-fitting model (Eq.~\ref{eq:1}) is superimposed as a green line (median) and related percentile contours (equivalent to $1\sigma$ and $3\sigma$). The vertical shaded areas represent the phases of the burst (namely rise1, rise2, peak, decay1, decay2, with the same color-coding as Fig.~\ref{fig:energyevol} for eRO-QPE3). \emph{Bottom:} same as the top panel, but for the eRO4-XMM observation. The best-fitting model is here shown in red, and the phases of the burst follow the color-coding of eRO-QPE4 in Fig.~\ref{fig:energyevol}.}
	\label{fig:eRO3_lcphases}
    \end{figure}

    The quiescent state of eRO-QPE3 was selected from GTIs excluding the bursts from both eRO3-XMM1/2 observations: namely taking GTIs between times $774645000 - 774651000$\,s and $774675000 - 774724000$\,s of eRO3-XMM1 and GTIs $<776410000$\,s and $>776440000\,$s of eRO3-XMM2. The resulting quiescence spectrum is too faint to be securely detected above background (dark-gray points in Fig.~\ref{fig:eRO3XMM1_spec_peak}). We adopt a disk spectrum to provide a flux upper limit of $F_{\rm 0.2-2.0\,keV}<1.1 \times 10^{-14}$\,erg\,s$^{-1}$\,cm$^{-2}$. The fit posterior on the flux, or disk normalization, is unconstrained (dark-gray line and contours in Fig.~\ref{fig:eRO3XMM1_spec_peak}), while the disk temperature is loosely constrained to $kT_{\rm disk}=46_{-18}^{+21}$\,eV. We conservatively consider the quiescence of eRO-QPE3 to be undetected at the \emph{XMM-Newton} epoch, although the weak constraints of temperature and luminosity ($L_{\rm disk,bol}<2.3 \times 10^{41}$\,erg\,s$^{-1}$, with a median value of the unconstrained posterior at $L_{\rm disk,bol}\sim3.5 \times 10^{40}$\,erg\,s$^{-1}$) are consistent with quiescent states of other QPEs (\citetalias{Miniutti+2019:qpe1,Giustini+2020:qpe2,Arcodia+2021:eroqpes}). Therefore, we suggest that the quiescent state of eRO-QPE3 is compatible with background (hence not formally detected), but not significantly fainter.

    The QPE flares are separated in two rise states, one peak and two decays. We show an example of this in Fig.~\ref{fig:eRO3_lcphases} for the first burst of the observation eRO3-XMM1. The color-coding, darker to lighter going from rise1 to decay2, follows that of Fig.~\ref{fig:energyevol}. Separations are defined in terms of total count rate, so that, for instance, the separation between rise1 and rise2 occurs roughly at the same count rate as that between decay1 and decay2. Naturally, they are somewhat arbitrary and, as much as temperature or flux values would slightly change with different definitions, no major conclusion of this work would, not the hysteresis shown in Fig.~\ref{fig:energyevol}. For the example in Fig.~\ref{fig:eRO3_lcphases}, the separating times are 774651000, 774654500, 774657500, 774661000, 774664000 and 774668500\,s, respectively. Since the quiescent state is formally undetected, we fit all phases of all bursts with a black body model, representative of the QPE. The individual temperature and luminosities are shown in Fig.~\ref{fig:energyevol} and Table~\ref{tab:erass1_spec}. For the peak spectrum (orange in Fig.~\ref{fig:eRO3XMM1_spec_peak}), we obtain $kT_{\rm QPE}=111_{-5}^{+6}$\,eV and $F_{\rm 0.2-2.0\,keV} = 3.1_{-0.2}^{+0.3} \times 10^{-13}$\,erg\,s$^{-1}$\,cm$^{-2}$ ($L_{\rm 0.2-2.0\,keV} = (4.2\pm 0.4) \times 10^{41}$\,erg\,s$^{-1}$), thus $\gtrsim25-30$ times brighter than the quiescence flux upper limit in the soft X-ray band. The bolometric luminosity of the black body component is $L_{\rm QPE,bol} = (4.9\pm0.5) \times 10^{41}$\,erg\,s$^{-1}$. Based on the eRASS1 and eRASS2 results (and \emph{XMM-Newton} analysis of eRO-QPE4; see Sect.~\ref{sec:ero4xmm} and see below), we test the addition a Comptonization component to the QPE phases. However, during all phases the flux of this component is compatible with background (thus, with zero flux) and the $log Z$ of the fit is orders of magnitude worse. Therefore, this component is not statistically required given the current data quality. We adopt the simple black body as reference model for the eruptions of eRO-QPE3 observed by \emph{XMM-Newton}.  

    \subsubsection{eRO-QPE4}

    \begin{table*}[t]
	\footnotesize
	\setlength{\tabcolsep}{3.5pt}
	\caption{Spectral fit results for eRO-QPE4.}
	\label{tab:ero4_spec}
	\centering
	\begin{threeparttable}
		\begin{tabular}{cccccccccc}
			\toprule
            \multicolumn{1}{c}{Epoch} &
			\multicolumn{1}{c}{Spectrum} &
			\multicolumn{1}{c}{Model} &
			\multicolumn{1}{c}{$kT_{\rm disk}$} &
			\multicolumn{1}{c}{$F^{\rm disk}_{\rm 0.2-2.0\,keV}$} &
			\multicolumn{1}{c}{$F^{\rm disk,comp}_{\rm 0.2-2.0\,keV}$} &
			\multicolumn{1}{c}{$kT_{\rm QPE}$} &
			\multicolumn{1}{c}{$F^{\rm QPE}_{\rm 0.2-2.0\,keV}$} &
            \multicolumn{1}{c}{$F^{\rm QPE,comp}_{\rm 0.2-2.0\,keV}$} &
            \multicolumn{1}{c}{$\Delta \log Z$}
            \\
              &       &          & [eV]   &  [erg\,s$^{-1}$\,cm$^{-2}$]  & [erg\,s$^{-1}$\,cm$^{-2}$]  &  [eV]   &  [erg\,s$^{-1}$\,cm$^{-2}$]  &  [erg\,s$^{-1}$\,cm$^{-2}$] &
            \\
			\midrule
             eRASS:3 & Full      &     \texttt{disk}       &     --   &  $<3.8\times 10^{-15}$  &    --   &  --&  --   &  --  &  -- \\
             \midrule
             eRASS4 & Quiescence      &     \texttt{disk}       &     --   &  $<6.9\times 10^{-14}$  &    --  &  -- &  --   &  --  &  -- \\
              & QPE      &     \texttt{bb}       &     --   &  --  &    --   &  $93_{-14}^{+18}$   &  $3.4_{-1.1}^{+1.7} \times 10^{-12}$  &  -- &  -- \\
              \midrule
             XMM burst2 & Quiescence      &     \texttt{disk}       &     $51_{-2}^{+3}$   &  $(2.8\pm0.2) \times 10^{-13}$  &    --  &  --  &  --   &  --  & 91 \\
              &    &     \textbf{\texttt{disk+comp}}       &    $43\pm2$    &  $(3.9\pm0.4) \times 10^{-13}$  &    $1.4_{-0.4}^{+0.8} \times 10^{-14}$   &  --  &  --  &  --  & 0 \\
                  & QPE rise1    &     \textbf{\texttt{Q+bb}}       &     $\sim43$   &  $\sim3.9 \times 10^{-13}$  &    $\sim1.4\times10^{-14}$   &  $116_{-10}^{+14}$   &  $1.3_{-0.2}^{+0.3} \times 10^{-13}$  &  --  &  -- \\
                  & QPE rise2    &     \textbf{\texttt{Q+bb}}       &     $\sim43$   &  $\sim3.9 \times 10^{-13}$  &    $\sim1.4\times10^{-14}$   &  $123_{-3}^{+4}$   &  $(1.5\pm0.1) \times 10^{-12}$  &  --  &  0 \\
                  &     &     \texttt{Q+bb+comp}     &     $\sim43$   &  $\sim3.9 \times 10^{-13}$  &    $\sim1.4\times10^{-14}$   &  $120_{-6}^{+5}$   &  $(1.5\pm0.1) \times 10^{-12}$  &  $<1.3\times 10^{-13}$ &  2 \\
               & QPE peak    &     \texttt{Q+bb}       &     $\sim43$   &  $\sim3.9 \times 10^{-13}$  &    $\sim1.4\times10^{-14}$   &  $125_{-1}^{+2}$   &  $(2.9\pm0.1) \times 10^{-12}$   &  -- &  0 \\
               &    &     \textbf{\texttt{Q+bb+comp}}       &     $\sim43$   &  $\sim3.9 \times 10^{-13}$  &    $\sim1.4\times10^{-14}$   &  $123\pm2$   &  $(2.9\pm0.1) \times 10^{-12}$   &  $3.1_{-2.1}^{+4.6} \times 10^{-14}$ &  0.5 \\
               & QPE decay1    &     \textbf{\texttt{Q+bb}}       &     $\sim43$   &  $\sim3.9 \times 10^{-13}$  &    $\sim1.4\times10^{-14}$   &  $89\pm2$   &  $(2.1\pm0.1) \times 10^{-12}$  &  -- &  -- \\
               & QPE decay2    &     \textbf{\texttt{Q+bb}}       &     $\sim43$   &  $\sim3.9 \times 10^{-13}$  &    $\sim1.4\times10^{-14}$   &  $71\pm5$   &  $3.2_{-0.6}^{+0.7} \times 10^{-13}$  &  -- &  -- \\
             \bottomrule
		\end{tabular}
	    \begin{tablenotes}
        \item See notes in Table~\ref{tab:erass1_spec}. For \emph{XMM-Newton}, the different phases are shown in Fig.~\ref{fig:energyevol} and the bottom panel of~\ref{fig:eRO3_lcphases}. Here, the quiescence spectrum ("Q") is held fixed during the QPE epochs by letting its parameters free to vary only within the 10th-90th percentiles of the posteriors of the quiescence fit alone (hence the "$\sim$"). Given the spectroscopic redshift and the cosmology adopted \citep{Hinshaw+2013:wmap9}, the conversion for related luminosity values for eRO-QPE4 is $4.42\times10^{54}$\,cm$^2$ in this paper.
        \end{tablenotes}
   \end{threeparttable}
\end{table*}

\begin{figure}[tb]
	\centering
    \includegraphics[width=0.9\columnwidth]{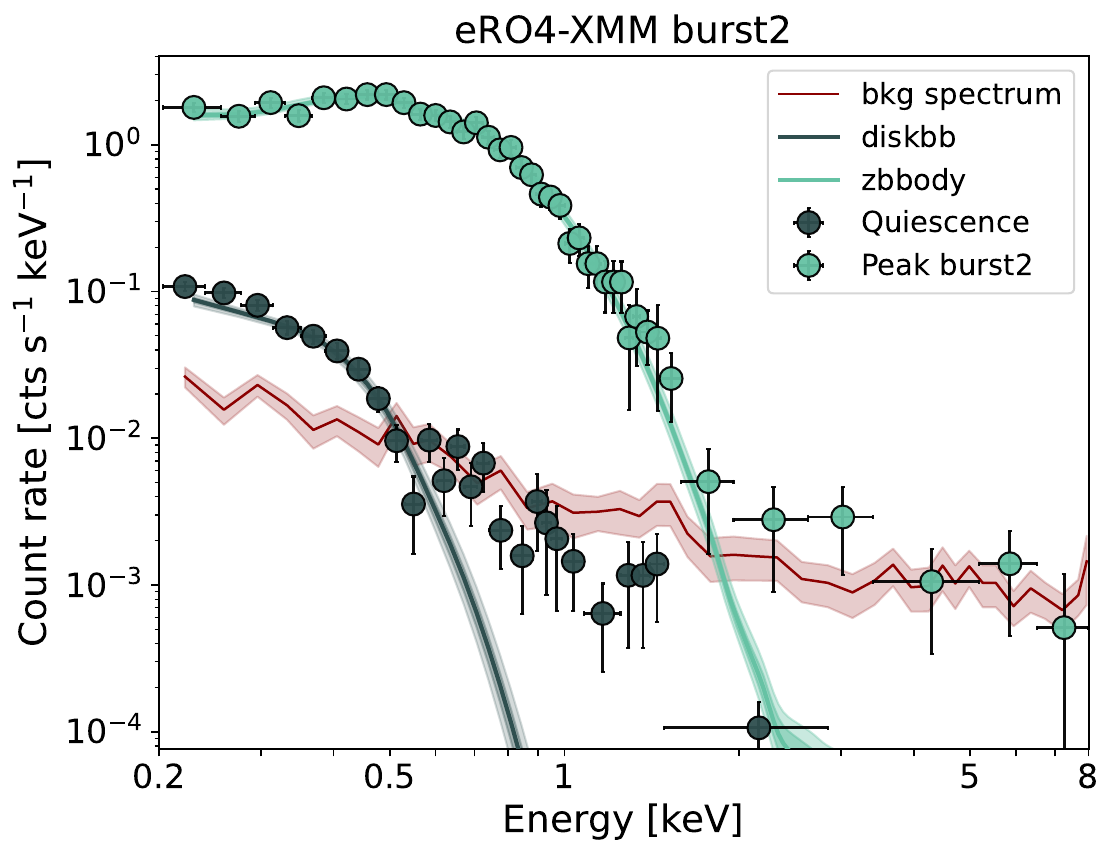}
	\caption{Same as Fig.~\ref{fig:eRO4XMM_spec_peak}, but showing background-subtracted spectra. The models shown are a simple disk spectrum for quiescence and a black body in addition for the peak, both showing significant residuals at higher energies. These residuals are comparable with background or even brighter. The background spectrum (subtracted to yield the data shown) is represented in in red for visualization.}
	\label{fig:eRO4XMM_spec_peak_others}
    \end{figure}

    The quiescent state of eRO-QPE4 was selected from GTIs excluding the bursts from the eRO4-XMM observation: namely taking GTIs between times $794814000 - 794837000$\,s and $794850500 - 794889800$\,s. The source is detected, with a soft spectrum (Fig.~\ref{fig:eRO4XMM_spec_peak}) reminiscent of that of other QPE sources in quiescence. We initially fit the quiescence with a disk model (Fig.~\ref{fig:eRO4XMM_spec_peak_others}) and subsequently add a Comptonization component (Fig.~\ref{fig:eRO4XMM_spec_peak}). As it can be noted in Fig.~\ref{fig:eRO4XMM_spec_peak_others}, a thermal disk model alone fails to account residual signal between $\sim0.6-1.5\,$keV. This signal is significantly above background (Fig.~\ref{fig:eRO4XMM_spec_peak}), as it is comparable to it after background subtraction (Fig.~\ref{fig:eRO4XMM_spec_peak_others}). Adding a Comptonization component vastly improves the $\log Z$ of the spectral fit by 92. Therefore, we adopt the disk plus Comptonization model as reference for the quiescent state of eRO-QPE4, which yields a disk temperature of $kT_{\rm disk}=(43\pm2)$\,eV with a flux $F_{\rm 0.2-2.0\,keV} = (3.9\pm0.4) \times 10^{-13}$\,erg\,s$^{-1}$\,cm$^{-2}$. This corresponds to $L_{\rm 0.2-2.0\,keV} = 1.7_{-0.1}^{+0.2} \times 10^{42}$\,erg\,s$^{-1}$ for the disk component alone, or to a bolometric luminosity of $L_{disk,bol} = 2.0_{-0.3}^{+0.6} \times 10^{43}$\,erg\,s$^{-1}$. The Comptonization component is much fainter, with $F_{\rm 0.2-2.0\,keV} = 1.4_{-0.4}^{+0.8} \times 10^{-14}$\,erg\,s$^{-1}$\,cm$^{-2}$, and its slope is unconstrained with preference for the softest values allowed (i.e., $\Gamma_X = 3.5$). As discussed for eRASS data of eRO-QPE3, Comptonization is merely one possible interpretation for the excess signal observed. In this work, the scope of the quiescence fit is to obtain a good-enough model in order to better isolate the QPE component. We report both models in Table~\ref{tab:ero4_spec}. We have also tested a Comptonization component alone to fit the observed X-ray photons, with the assumption that a colder disk is present. Residuals are much poorer and $\Delta \log Z\sim156$ compared to the disk plus Comptonization quiescence model, hence the latter is preferred.
    
    The QPE flares of eRO-QPE4 were separated in five epochs, namely rise1, rise2, peak, decay1 and decay2 (bottom panel of Fig.~\ref{fig:eRO3_lcphases} for the second burst of the observation eRO4-XMM). The color-coding, darker to lighter going from rise1 to decay2, follows that of Fig.~\ref{fig:energyevol}. For the example in Fig.~\ref{fig:eRO3_lcphases}, the separating times are 794837000, 794839000, 794840500, 794843500, 794845500 and 794850500\,s, respectively. The quiescence model was held fixed by imposing its parameters to vary only within the 10th-90th percentile interval of the posteriors obtained during the quiescence fit alone. Therefore, the bursts parameters are marginalized over the uncertainties of the quiescence model. The fit results for the different burst phases are reported in Table~\ref{tab:ero4_spec}. We fit each phase with a black body alone and also with a Comptonization component of the QPE spectrum in addition. For all phases excluding the peak, the additional Comptonization component is not required by the fit, as established by comparing the $\log Z$ and visualizing the normalization or flux posteriors of the component. Therefore, for these epochs (rise1, rise2, decay1 and decay2) the best-fit model adopted for the burst is that of the quiescence plus a black body component. For the peak spectrum the $\log Z$ of the more complex (black body plus Comptonization) is compatible with that of the simpler model (black body) within typical uncertainties of 0.3 on $\log Z$, thus the former is not formally required. However, the flux and normalization posteriors of the Comptonization component are constrained and the marginal higher-energy residuals accounted for (comparing Fig.~\ref{fig:eRO4XMM_spec_peak} and~\ref{fig:eRO4XMM_spec_peak_others}), all at the expense of a marginal change in the black body properties: as it can be noted in Table~\ref{tab:ero4_spec}, both temperature and flux of the much brighter black body QPE component are unaffected by the much fainter harder component. Therefore, we adopt the more complex model as reference for the QPE peak of eRO-QPE4 at the \emph{XMM-Newton} epoch (see Fig.~\ref{fig:eRO4XMM_spec_peak}). This model yields $kT_{\rm QPE}=(123\pm2)$\,eV and $F_{\rm 0.2-2.0\,keV} = (2.9\pm0.1) \times 10^{-12}$\,erg\,s$^{-1}$\,cm$^{-2}$ ($L_{\rm 0.2-2.0\,keV} = (1.27\pm0.03) \times 10^{43}$\,erg\,s$^{-1}$), thus $\sim7-8$ times brighter than the quiescence luminosity in the same soft X-ray band. The bolometric luminosity is $L_{\rm QPE,bol} = (1.44\pm0.04) \times 10^{43}$\,erg\,s$^{-1}$. In this case, the additional Comptonizaton component requires further discussion, since it is not ubiquitous to QPE sources, not throughout the burst of eRO-QPE4 itself. Again, for the scope of this paper this is irrelevant since the QPE black body properties do not change. Future work should establish whether this signal is significant and physical, or whether, for instance, it could be pile-up effect due to the bright and soft QPE spectrum, since it is only observed at the peak.


    \begin{figure}[tb]
		\centering
		\includegraphics[width=\columnwidth]{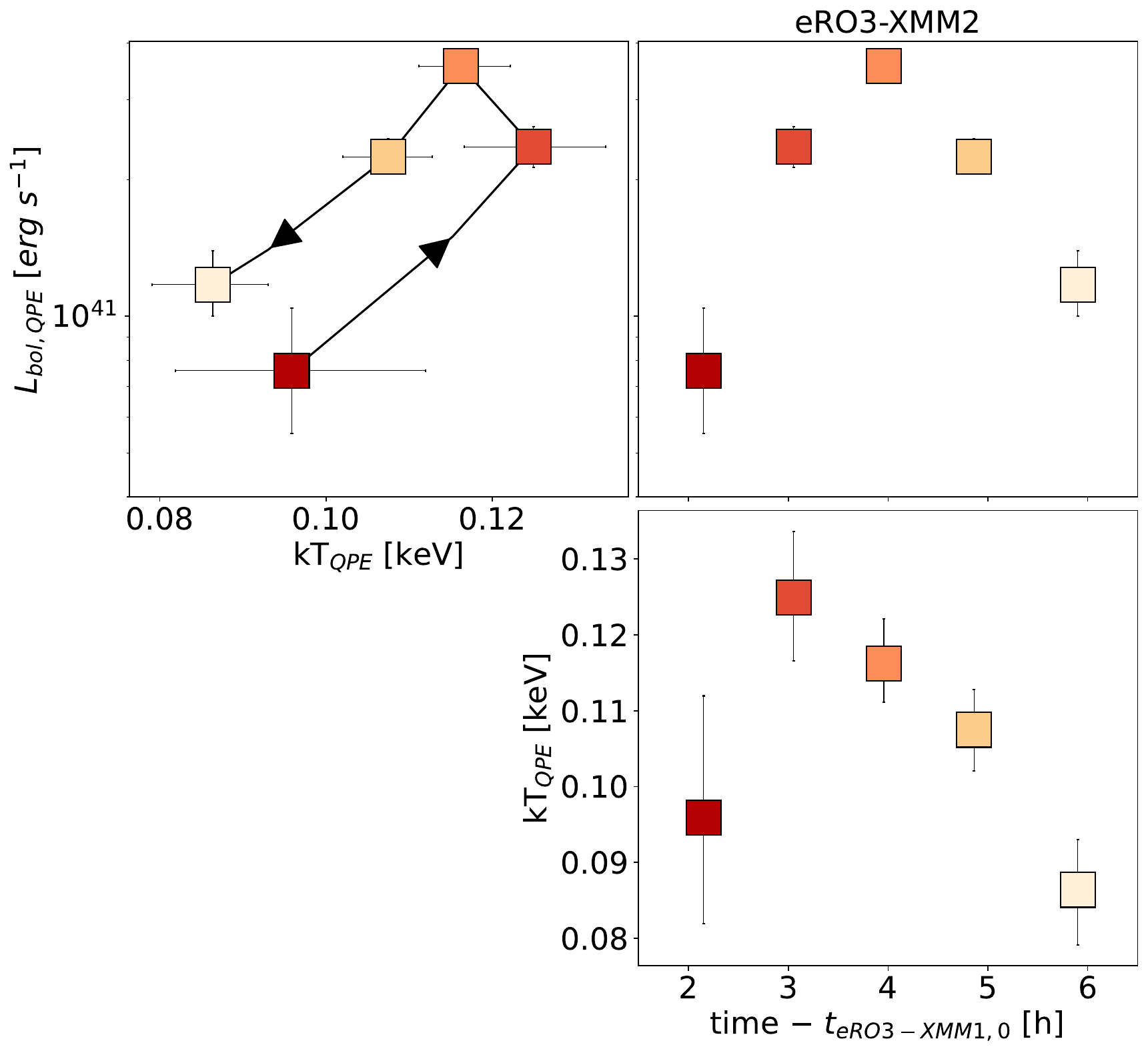}
        \includegraphics[width=\columnwidth]{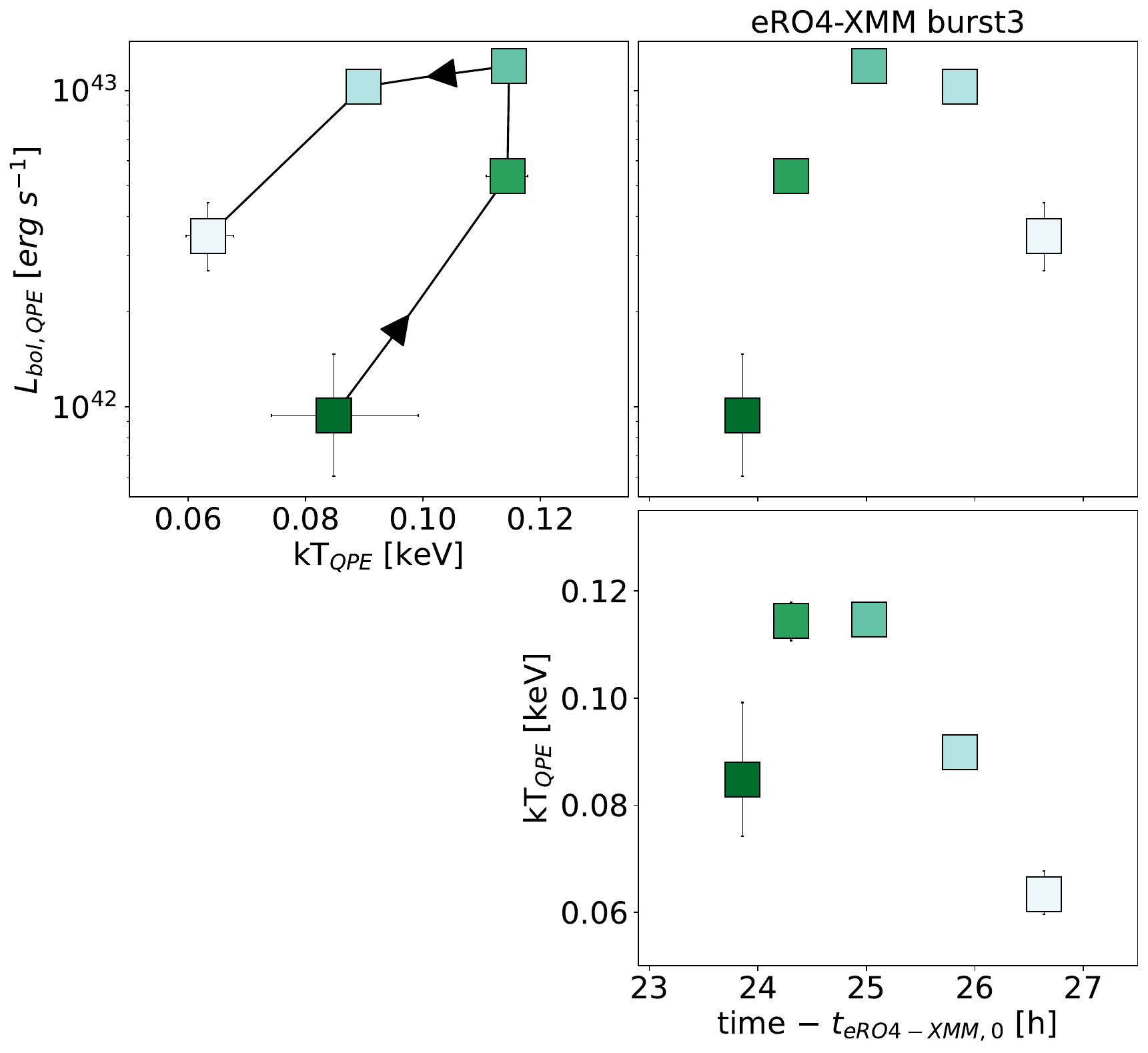}
		\caption{Same as Fig.~\ref{fig:energyevol}, but for the eruption of the eRO3-XMM2 observation (\emph{top}) and the third of the eRO4-XMM observation (\emph{bottom}).
  }
		\label{fig:energyevol_backup}
	\end{figure}
    
\end{appendix}
	
\end{document}